\documentclass[final,twocolumn,12pt]{elsarticle}
\usepackage[utf8]{inputenc}
\usepackage[T1]{fontenc}
\usepackage{fixltx2e}
\usepackage{graphicx}
\usepackage{longtable}
\usepackage{float}
\usepackage{wrapfig}
\usepackage{rotating}
\usepackage[normalem]{ulem}
\usepackage{amsmath}
\usepackage{stackengine}
\usepackage{textcomp}
\usepackage{marvosym}
\usepackage{wasysym}
\usepackage{amssymb}
\usepackage{hyperref}
\usepackage{amsthm}
\usepackage{dblfloatfix} 
\usepackage{subcaption}
\usepackage{tabularx}
\theoremstyle{definition}

\theoremstyle{plain}

\usepackage{mdframed}
\mdtheorem{h0}{H0 (EMH):}
\mdtheorem{h1}{H1 (Critique):}
\mdtheorem{oresq}{Overarching research question:}
\mdtheorem{resq}{Research question:}
\usepackage{todonotes}
\usepackage{cuted}

\newfloat{eqfloat}{h}{eqflts}
\floatname{eqfloat}{Equation}
\makeatletter
\let\c@eqfloat\c@equation
\makeatother

\begin{document}

\begin{frontmatter}

\title{\vspace{-2cm}Privacy-enhancing Aggregation of Internet of Things Data via Sensors Grouping}

\author[institute1]{Stefano Bennati\corref{cor1}}
\ead{sbennati@ethz.ch}
\cortext[cor1]{Corresponding author}
\author[institute1]{Evangelos Pournaras}
\ead{epournaras@ethz.ch}
\address[institute1]{Professorship of Computational Social Science\\
ETH Zurich, Zurich, Switzerland
}

\begin{abstract}
  Big data collection practices using Internet of Things (IoT) pervasive technologies are often privacy-intrusive and result in surveillance, profiling, and discriminatory actions over citizens that in turn undermine the participation of citizens to the development of sustainable smart cities. Nevertheless, real-time data analytics and aggregate information from IoT devices open up tremendous opportunities for managing and regulating smart city infrastructures in a more efficient and sustainable way. The privacy-enhancing aggregation of distributed sensor data, such as residential energy consumption or traffic information, is the research focus and challenge tackled in this paper. Citizens have the option to choose their privacy level by reducing the quality of the shared data at a cost of a lower accuracy in data analytics services. A baseline scenario is considered in which IoT sensor data are shared directly with an untrustworthy central aggregator. A grouping mechanism is introduced that improves privacy by sharing data aggregated first at a group level compared to a baseline scenario in which each individual shares data directly to the central aggregator. Group-level aggregation obfuscates sensor data of individuals, in a similar fashion as differential privacy and homomorphic encryption schemes, thus inference of privacy-sensitive information from single sensors becomes computationally harder compared to the baseline scenario.
    The proposed system and its generic applicability are evaluated using real-world data from two smart city pilot projects. Privacy under grouping increases, while preserving the accuracy of the baseline scenario. Intra-group influences of privacy by one group member on the other ones are measured and fairness on privacy is found to be maximized between group members with similar privacy choices. Several grouping strategies are compared. Grouping by proximity of privacy choices provides the highest privacy gains. The implications of the strategy on the design of incentives mechanisms are discussed.
\end{abstract}


\end{frontmatter}

\section{Introduction}\label{sec-2}

Cyber-physical smart city infrastructures, such as smart grids and traffic systems, are becoming more and more data intensive. Operational services over such complex infrastructures require the collection and aggregation of citizens' data \cite{gaur2015smart}, e.g. total power demand for preventing blackouts or average speed of vehicles for mitigating traffic congestion.
On the one hand, aggregation over citizens' sensitive data raises concerns about privacy \cite{medaglia10_overv_privac_secur_issues_inter_thing}.
On the other hand, these data have a great potential to improve the performance and sustainability of smart cities~\cite{fang12_smart_grid,pellicer2013global}. Therefore, privacy-enhancing aggregation mechanisms can play a paramount role in the development and adoption of data intensive smart city applications.

In the scenario studied in this paper, collective sensing is required to provide a service. Assume a population of Internet of Things (IoT) sensors, referred to as \emph{data suppliers}, that are associated with or owned by citizens, e.g. smart meters, smart phones, wearables. Data suppliers disclose their measurements to a planner, a system operator or another smart city actor that is referred to as \emph{data consumer}. The data consumer executes a data analytics algorithm in order to make collective measurements available -- i.e. computing aggregation functions such as the summation, average, maximum, top-k, etc. Its interest is to provide high quality of service in terms of high computational accuracy. The data consumer is assumed to be \emph{honest but curious}~\cite{goldreich05_found_crypt_primer} i.e. it may run privacy-intrusive algorithms on the data received from data suppliers to infer, for instance, daily routines and habits, identity of appliance owners, health status, for the purpose of personalized pricing and/or advertising. In the context of smart cities, honest but curious data consumers may correspond to actors that use citizens' data to serve corporate interests, to profile/discriminate citizens~\cite{zhang2016source}, or increase political influence and power, for instance by nudging. This scenario is also relevant as a public goods game: disclosing data to the data consumer entails a privacy cost for the citizen, and at the same time it improves the service, providing every citizen with a benefit~\cite{ledyard1997public,diekmann1985volunteer}. A trade-off between privacy and accuracy is empirically observed in earlier work~\cite{rajagopalan2011smart}.

This paper introduces the \emph{Internet of Things Privacy-enhancing Group Aggregation} (IoT-PGA) mechanism. IoT-PGA increases the individual privacy of data suppliers, while preserving a certain level of service quality. The proposed mechanism is bottom-up, i.e. it can be applied without requiring the collaboration of the data consumer, in contrast to top-down approaches that require key management systems, e.g. homomorphic encryption~\cite{fontaine2007survey,erkin2013privacy}. Earlier work introduces different ways of increasing privacy in IoT networks in a bottom-up way: by decreasing the quality of the data, e.g. by reducing its granularity \cite{pournaras16_self_regul_infor_sharin_partic_social_sensin}, by obfuscation \cite{ardagna11_obfus_based_approac_protec_locat_privac} or by adding noise \cite{eibl2017differential}. This paper looks at a new way of increasing privacy: changing the physical or logical organization of the network by varying the topology, i.e. by grouping data suppliers. Note that this is an open challenge identified in earlier work~\cite{Haddadi2012}. Grouping is a complementary approach to other existing bottom-up privacy-preserving mechanisms in the sense that the underlying performance of the mechanism can be enhanced via grouping \cite{lu2012eppa,finster2013elderberry}.

The general applicability of IoT-PGA is empirically studied and evaluated using real-world data from two smart city pilot projects that concern the following two application scenarios:

\noindent\textbf{Mitigation of traffic congestion}: Assume a city planner who manages a data consumer for the collection and the analysis of traffic data, such as GPS and velocity traces, in order to reduce traffic congestion, to improve commute times and to design any further infrastructural expansion required~\cite{Wan2016}. The data consumer is interested in the precise locations in which traffic is the slowest, so it requires high accuracy in data analytics.
On the contrary, data suppliers may not be willing to disclose their precise location and speed, which could reveal sensitive information, e.g. infractions of the traffic laws, information about time and path of the daily commute. In this scenario, the IoT-PGA mechanism is applied to real-world data from the Regional Transportation Commission of Southern Nevada (RTCSNV), containing GPS traces of cars~\cite{nrel}.

\noindent\textbf{Optimization of power consumption}:
Assume an energy utility company that plays the role of a data consumer. It collects and analyzes the energy consumption of residential customers in order to optimize energy prices and power grid usage via load reduction or load shifting~\cite{nambi2016temporal,Pilgerstorfer2017}.
Residential consumers might not wish to disclose their consumption history in detail as it might reveal sensitive information, such as house occupancy and residential activities~\cite{Greveler2012}. Real-world data are used from the Electricity Customer Behavior Trial with residential power consumption records~\cite{ecbt}.

\noindent The illustration of both application scenarios unfolds more universal insights on privacy for smart cities that are relevant beyond domain experts: the future smart city policy makers tackling a highly inter-disciplinary and data-driven policy-making, i.e. inter-dependent infrastructural networks.
The proposed system does not require infrastructural interventions when applied to the information systems supporting application services and in this sense, it is highly relevant for service providers such as power utilities companies. In contrast, integration in the control infrastructure requires interoperation with the physical infrastructure of system operators, which is out of the scope of this paper.

The main finding of this work is the following: when data suppliers physically or logically organize themselves in groups to aggregate sensor data before sharing them with a data consumer, the privacy of data suppliers improves without compromising the quality of service, i.e. accuracy of the computed aggregation functions.
Changing the number of groups in the system, introducing heterogeneous group sizes or assigning arbitrary privacy choices to data suppliers, does not influence the main finding: individual privacy increases and accuracy remains constant.
Nevertheless, all of these variables affect the efficiency of the method, thus the effect of different parameter configurations at the population level and at the group level are studied.
Finally, several grouping strategies are compared. Grouping data suppliers according to their individual privacy choices is found to be the strategy that offers the highest level of privacy. Implications of these results on incentive mechanisms are discussed.

The contributions of this paper are summarized as follows:

\begin{itemize}
\item A new bottom-up approach for enhancing privacy by changing the network topology, i.e. grouping data suppliers, without reducing the accuracy of aggregation.
\item New privacy and accuracy metrics in the group setting.
\item Measured trade-offs between privacy and accuracy for networks performing group-level data sharing.
\item Different grouping strategies and trade-offs comparisons between privacy and accuracy.
\item A proof-of-concept on two real-world smart city datasets, confirming the general applicability independently of the type of input data used.
\end{itemize}

The rest of the paper is organized as follows: In Section \ref{sec:model} the proposed model is introduced, together with the measures of accuracy and privacy. Section \ref{sec:exp} outlines the experimental methodology, a discussion of design choices, and the datasets used in the experiments. Section \ref{sec:res} discusses the results of the experiments. Section \ref{sec:disc} summarizes the results, their implications for system and policy design and possible directions for future work. Section \ref{sec:related_work} illustrates related work and positions this paper to literature. Finally, Section \ref{sec:concl} concludes this paper.

\section{Privacy-enhancing Grouping}
\label{sec-3}
\label{sec:model}

Table \ref{tbl:notation} illustrates the mathematical notation used in the paper, listed in the order they appear in the text.

\begin{table}[!htb]
\centering
\caption{Mathematical notation.}
\label{tbl:notation}
{\footnotesize
\begin{tabularx}{\columnwidth}{@{}lX@{}}
Math symbol & Description\\
\hline
$\mathcal{G} = (\mathcal{I} \cup \{A\},E)$ & the network graph\\
$\mathcal{I}$ & the set of $n$ data suppliers\\
$A$ & the data consumer\\
$E$ & set of network connections\\
$T_e$ & number of measurements for epoch $e$\\
$r_{i,e,t}$ & record of raw data of supplier $i$ at epoch $e$ and time $t$\\
$R_{i,e}$ & raw data of supplier $i$ in epoch $e$\\
$\mathcal{R}$ & the domain of the raw data\\
$\alpha$ & aggregation function\\
$f_S : \mathcal{R}^{T_e} \rightarrow \mathcal{S}^{T_e}$ & summarization function\\
$s_{i,e,t}$ & record of summarized data\\
$S_{i,e}$ & summarized data of supplier $i$ at epoch $e$\\
$\mathcal{S}$ & the domain of the summarized data\\
$\epsilon_{e,t}$ & local error at epoch $e$ and time $t$\\
$\varepsilon_{e,t}$ & global error at epoch $e$ and time $t$\\
$G \subseteq \mathcal{I}$ & a group of data suppliers\\
$m$ & the number of groups \\
$\alpha_{e,t}^G$ & intra-group aggregation\\
$a,a_1,a_2$ & data suppliers \\
$\epsilon^G_{e,t}$ & local group error for group $G$\\
$\varepsilon^G_{e,t}$ & total group error for group $G$\\
\hline
\end{tabularx}
}
\end{table}

This paper studies a collective sensing system, for smart cities and beyond, consisting of $n$ data suppliers and a single data consumer, which is in charge of computing statistical information, i.e. aggregation functions, on the data generated by the data suppliers.

The system is a graph $\mathcal{G} = (V,E)$, where $V=\mathcal{I} \cup \{A\}$, $\mathcal{I}=\{1,\ldots,n\}$ is the set of data suppliers, $A$ is the data consumer and $E=\{ e_{i,j}: \forall i \ne j \in V\}$ is the set of connections between data suppliers. The data consumer is assumed connected to all data suppliers, i.e. $E \supseteq \{e_{i,A}, \forall i \in \mathcal{I}\}$. Data suppliers are also connected with each other in non-overlapping groups formed by network \mbox{(self-)organization} algorithms~\cite{jelasity06_t_man,Shaikh2009}. Each supplier in the group can interact to each other group member through a communication network such as the Internet, either assuming a fully connected network, e.g. via a lookup server for this purpose, or via distributed routing protocols, e.g. gossiping communication~\cite{jelasity2007gossip}.

The aggregation functions are computed at each epoch, e.g. each day. Each data supplier produces $T_e$ measurements during epoch $e$, one for each time step $1 \le t \le T_e$, e.g. every hour.
The sequence of measurements produced by data supplier $i$ during epoch $e$ is defined as $R_{i,e}=(r_{i,e,t})_{t=1}^{T_e}, ~r \in \mathcal{R}$. These data are referred to as \emph{raw data} and they are treated as privacy-sensitive information. The data consumer collects individual measurements and uses them to perform data analysis with an arbitrary algorithm $\alpha$, referred to as \emph{aggregation function}, for instance, the average, i.e. $\alpha(X)=\mathbb{E}(X)$, and the sum, i.e. $\alpha(X)=\sum{(X)}$.

The data consumer is assumed \emph{honest but curious}~\cite{goldreich05_found_crypt_primer}: an adversary that passively collects privacy-sensitive data about citizens and can perform privacy-intrusive operations over these data.
For this reason each measurement shared to the data consumer entails a privacy cost for the citizen, which should be minimized. Data suppliers increase their privacy by applying a \emph{summarization function} $f_S$ to the data before sharing. The summarization function $f_S : \mathcal{R}^{T_e} \rightarrow \mathcal{S}^{T_e}$ transforms a vector of $T$ values in the domain $\mathcal{R}$ to $T$ values in the domain $\mathcal{S}$. By definition $|\mathcal{S}| \le |\mathcal{R}|$: the number of possible discrete values in $\mathcal{S}$ is lower than the number of possible discrete values in $\mathcal{R}$.

The summarization function obfuscates, i.e. reduces the quality of, the information shared to the data consumer, by reducing the granularity of data, in order to make it computationally harder to infer characteristics of the raw data. The obfuscated data are referred to as \emph{summarized data} and defined as $S_{i,e}=f_S(R_{i,e})$. The \emph{summarization level} of a data supplier is defined as $1/|\mathcal{S}|$. A higher level of obfuscation corresponds to a lower $|\mathcal{S}|$ meaning higher privacy for the data suppliers but lower accuracy in the aggregation.

In the scenario of power consumption, obfuscation can be applied by reducing the temporal granularity of the smart meter readings, by either reducing the frequency of measurements or by computing the average load across measurements.
In the traffic scenario, obfuscation can be implemented by decreasing either the spatial granularity or the temporal granularity of location reports.

This paper makes use of privacy and accuracy measures from earlier work~\cite{pournaras16_self_regul_infor_sharin_partic_social_sensin}, which are general enough to be applied to different aspects of the smart city application scenarios. A discussion of these measures is presented in Appendix \ref{sec:model:corr}. Privacy is measured by the \emph{local error} that is defined as follows:

\begin{equation}
\epsilon_{e,t}=\frac{1}{n}\sum_{i=1}^n \epsilon_{i, e,t}; ~ \epsilon_{i, e,t}=\frac{|r_{i,e,t}-s_{i,e,t}|}{|r_{i,e,t}|+|s_{i,e,t}|},
\label{eq:sym_le}
\end{equation}

\noindent where each term is the difference between the raw and summarized data of supplier $i$. A higher local error provides higher privacy. Note that the local error does not depend on the aggregation function. Accuracy, measured by the \emph{global error} is defined as follows:

\begin{equation}
\varepsilon_{e,t}=\frac{|\alpha(R_{e,t})-\alpha(S_{e,t})|}{|\alpha(R_{e,t})|+|\alpha(S_{e,t})|},
\label{eq:sym_ge}
\end{equation}

\noindent which is the average difference between the raw data $R_{e,t}=(r_{i,e,t})_{i=1}^n$ and the summarized data $S_{e,t}=(s_{i,e,t})_{i=1}^n$ collected by the data consumer. A higher global error corresponds to lower accuracy.

\subsection{Grouping}
\label{sec-3-1}
\label{sec:model:grouping}

\begin{figure}[!htb]
\centering
\includegraphics[width=.9\linewidth]{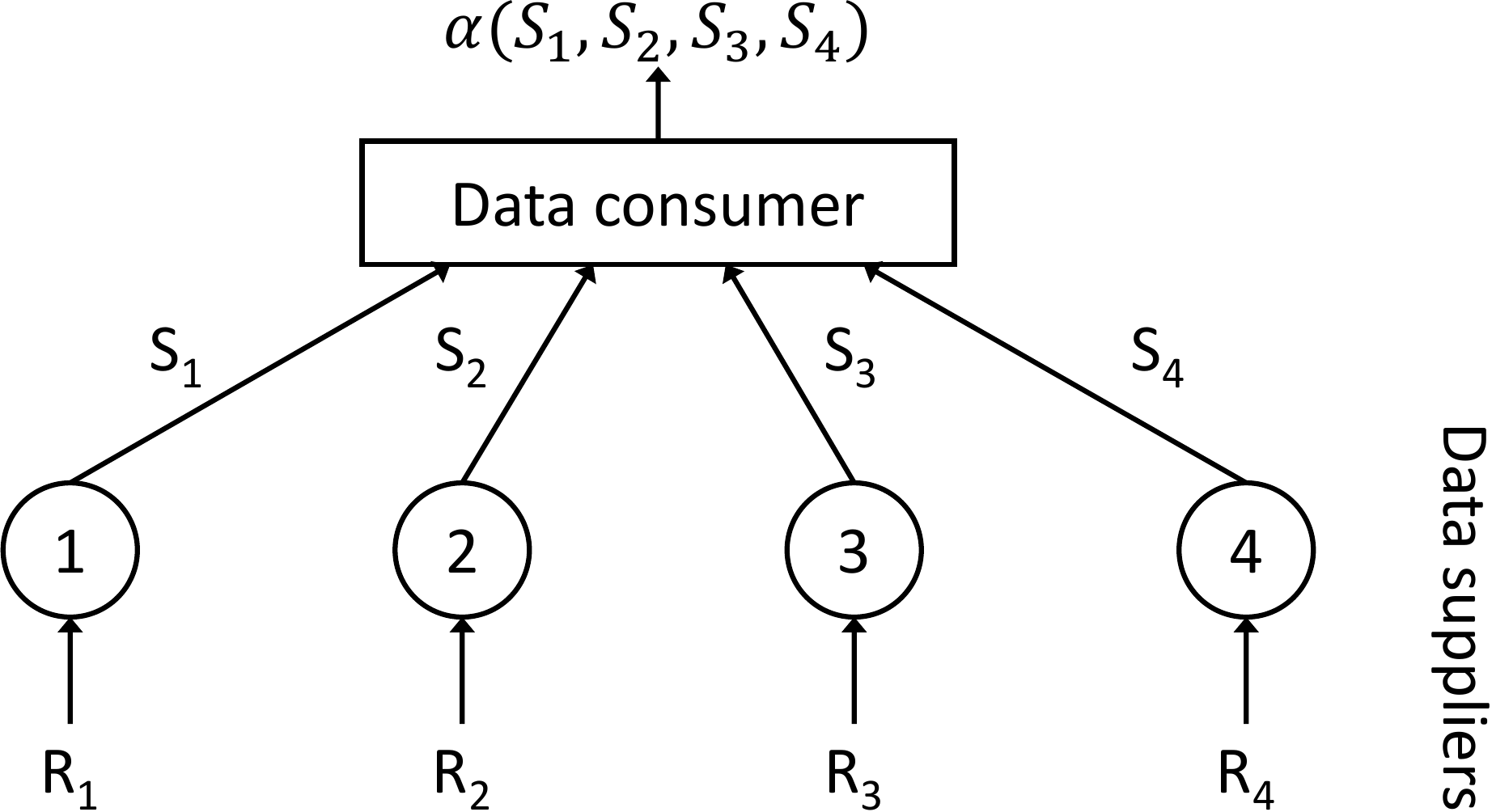}
\caption{\label{fig:centr}An illustration of the model for four data suppliers. In this setting the data consumer computes $\alpha(S_{1,e},...,S_{n,e})=\alpha(f_S(R_{1,e}),\ldots,f_S(R_{n,e}))$.}
\end{figure}

\begin{figure}[!htb]
  \begin{minipage}[t]{\linewidth}
    \centering
    \includegraphics[width=.9\linewidth]{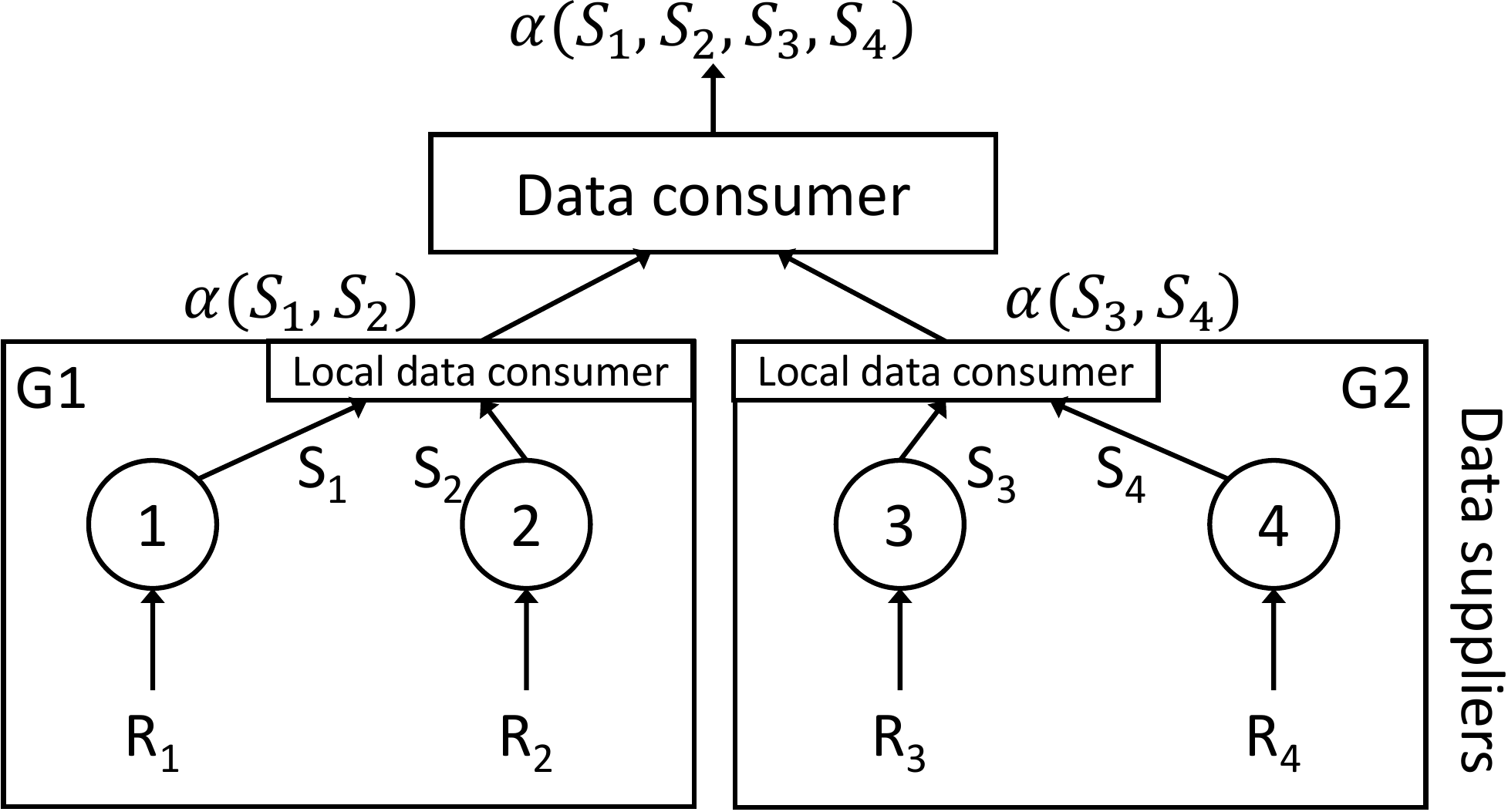}
    \subcaption{\label{fig:hier}Hierarchical organization. Each local aggregator transmits the locally-aggregated data to the consumer.}
  \end{minipage}
  \begin{minipage}[t]{\linewidth}
    \centering
    \includegraphics[width=.9\linewidth]{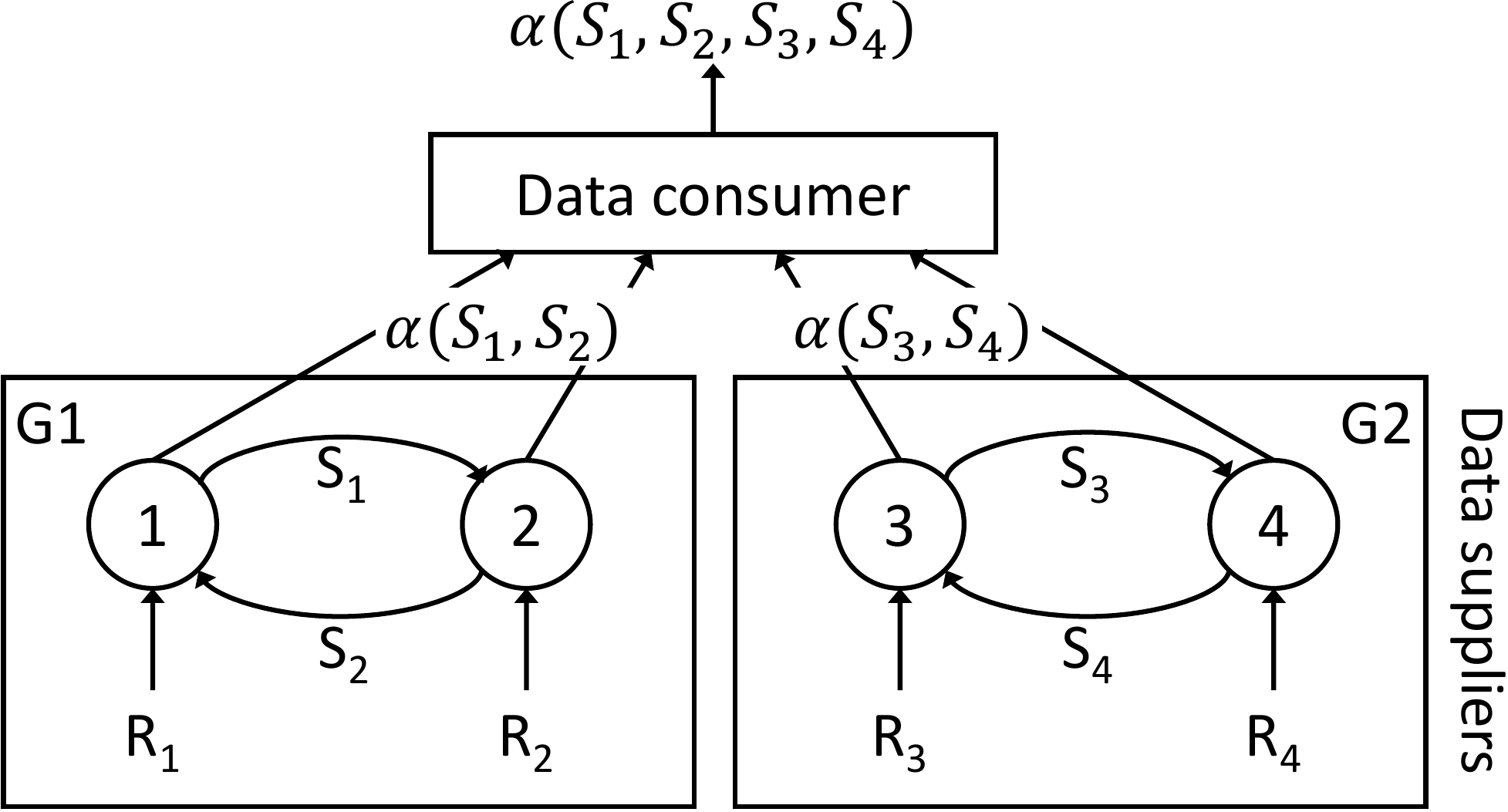}
    \subcaption{\label{fig:distr}Equivalent distributed organization. Each supplier transmits a fraction of the locally-aggregated data such that the aggregation at the consumer level of the group reports produces the locally-aggregated data.}
  \end{minipage}
\caption{An illustration of a collaborative intra-group aggregation for improving privacy-preservation. In this setting the data consumer computes $\alpha_{e,t}(\alpha_{e,t}(G_1),...,\alpha_{e,t}(G_n))$, where $G_i$ is the local aggregated data generated by group $i$.}
\end{figure}

So far data suppliers are assumed to transmit their summarized values directly to the data consumer. This setting corresponds to a centralized organization (Figure \ref{fig:centr}), in which the aggregation is performed at a central level. The equivalent of distributed aggregatio~\cite{jelasity05_gossip_based_aggreg_large_dynam_networ,pournaras17_engin_democ_inter_thing_data_analy} is a setting where data suppliers are (self-)organized~\cite{jelasity06_t_man} in groups within which they perform local aggregation in a privacy-preserving way \cite{vaidya2003privacy,mohammed2011anonymity} (Figure \ref{fig:hier}).

Groups in a smart city scenario may correspond to sets of citizens or devices that are in some close proximity that is measurable, for instance, via the Euclidean distance.
An example of a proximity measure is physical proximity, where groups could contain neighbouring households connected to the same smart-grid infrastructure e.g. sharing power feeders, plants~\cite{pudjianto07_virtual_power_plant_system_integ} or smart meters. Groups may also represent citizens waiting at the same crossroad or riding the same public transport vehicle \cite{thiagarajan2010}.
The physical proximity though may not be a practical option when it rapidly varies. Alternative grouping criteria that cover the connectivity of IoT devices can be applied to create virtual groups, e.g. software defined networks \cite{li2013software} or overlay networks \cite{lua2005survey}.
An integration of grouping strategies in the power software infrastructure of the power grids is out of the scope of this work, yet it is feasible as earlier shown for the applicability of multi-agent systems in the SCADA systems of power grids \cite{davidson2006applying}.
An example of a grouping criterion not based on physical constraints is semantic proximity, where groups could be formed according to some similarity between users such as the type of energy demand e.g. residential vs industrial, the final destination of the journey, or social network similarity e.g. participating to the same online communities.

A group $G$ is defined as a set of data suppliers in $\mathcal{I}$ that are connected with each other:\\

\[
  G \ne \emptyset \subseteq \mathcal{I}: \Shortunderstack[l]{$\underbrace{\forall i \ne j \in G ~ \exists e_{i,j}  \in E \vphantom{(y)}}$
    \scriptsize\rlap{pairwise connected data suppliers}}
\]

\noindent Data suppliers are assumed to have a channel for secure communication with other data suppliers, such that a malicious eavesdropper, e.g. the data consumer, cannot know or infer the content of the information exchange. For example, group communication could be encrypted or could be performed over a separate network, e.g. a peer-to-peer Bluetooth network, to which the eavesdropper does not have physical access~\cite{yao1982protocols}.

Data suppliers have the option to cooperate with the members of their group to increase their privacy: every group elects a data consumer that computes the aggregation function $\alpha$ using the group data, before sharing them to the central data consumer (Figure \ref{fig:hier}). For convenience of notation, the output of the data consumer within group $G$, i.e. the aggregation over the data of each data supplier, is denoted as $\alpha_{e,t}^G:= \alpha(s_{i,e,t},\forall i \in G)$.

The election of a data consumer at a group-level can be performed via distributed election protocols~\cite{Kuhn2010} in the hierarchical scenario (Figure \ref{fig:hier}). An alternative approach is a more distributed information exchange (Figure \ref{fig:distr}) in which every data supplier within a group can play the role of data consumer and send \footnote{This equation might change depending on the chosen aggregation function.} the mean value of the computed aggregation function $\alpha_{e,t}^G/|G|$. The summarized data used as input in the aggregation reach every group member via network communication by means of a routing protocol, e.g. \cite{jelasity05_gossip_based_aggreg_large_dynam_networ,paruchuri2003optimal}, at either the network layer or at the application layer \cite{nunes2014survey}. Both the aggregation function and the group size can be computed in a distributed fashion~\cite{jelasity05_gossip_based_aggreg_large_dynam_networ,pournaras17_engin_democ_inter_thing_data_analy} using such protocols.

Given that each group-level data consumer computes the same aggregation function computed by the central consumer, the output of the aggregation function in each scenario is expected to be equivalent: aggregating the supplier data directly or aggregating the data that the group-level data consumers earlier aggregated. This aspect is empirically studied in Section \ref{sec:res}.

\subsubsection{Privacy cost}

If data consumers at a group-level are not assumed to be honest-but-curious, for instance groups built based on a level of trust~\cite{Shaikh2009}, then the privacy cost at the group level is minimized. If local consumers are assumed to be honest-but-curious, then the privacy problem moves from the central to the group level. Countermeasures for decreasing the privacy cost in this case is the frequent change of (i) groups and (ii) elected data consumers that share group-level data in the hierarchical scenario. In this way, each data supplier is limited to coarser accumulated data about its group \cite{zang11_anony,eibl15_influen_data_granul_smart_meter_privac}.

\subsubsection{Local group error}
\label{sec-3-1-1}
\label{sec:model:le}

The data aggregated at a group-level are computed from the summarized data of the group members.
Compared to the baseline scenario, it is computationally harder for the central aggregator to infer the raw data of each data supplier using the data aggregated at a group-level. As experimentally confirmed in Figure \ref{fig:strategic}, for any given summarization level $1/|\mathcal{S}|$, the local error, i.e. the privacy, of data suppliers in a group is higher when data suppliers summarize and share their data directly with the central data consumer.

Group-level aggregation has two phases: (i) exchange of summarized data within the group and (ii) exchange of aggregated data at a group-level with the central data consumer. In the first phase, the local error measures the privacy cost for sharing summarized data with other group members. It is equivalent to the local error in the case of having no groups (cf. Equation \ref{eq:sym_le}). In the second phase, the \emph{local group error} is introduced computed by the difference between the raw data and the result of the group-level aggregation:

\begin{equation}
\epsilon^G_{ e,t}=\frac{|r_{i,e,t}-\alpha_{e,t}^G|}{|r_{i,e,t}|+|\alpha_{e,t}^G|}, i \in G.
\label{eq:group_le}
\end{equation}

Note that the local group error becomes equivalent to the local error $\epsilon^G_{e,t}= \epsilon_{e,t}$ if $G_j = \{ j \} \Rightarrow \alpha_{e,t}^{G_j}=S_{j,e,t} ~ \forall j: ~1 \le j \le n$ i.e. each individual becomes its own group.

\subsubsection{Total group error}
\label{sec:model:groupe}

The \emph{total group error} measures the difference between the summarized data and the data aggregated at a group-level.

\begin{equation}
\epsilon^G_{e,t}=\sum_{i \in G}\frac{|s_{i,e,t}-\alpha_{e,t}^G|}{|s_{i,e,t}|+|\alpha_{e,t}^G|}.
\label{eq:groupe}
\end{equation}

In contrast to the local group error that measures the privacy gain of the data supplier by being member in a group, i.e. the difference between the supplier's raw data and the group-level aggregated data, the total group error measures the overall privacy within a group, i.e. the difference between the summarized data shared within the group and the group-level aggregated data shared out of the group, to the central data consumer. If the group-level aggregated data and the summarized data of a group member are similar, it is easier for the data consumer to infer information about the data suppliers.

\subsubsection{Global error}
\label{sec-3-1-3}
\label{sec:model:ge}

Compared to the baseline scenario, grouping to $m$ disjunct groups results in the same global error in the following cases: (i) in sum, due the the associative property and (ii) in mean due to the property of the grand mean for which $\mathbb{E}_{e,t}(\mathbb{E}_{e,t}(G_1),\ldots,\mathbb{E}_{e,t}(G_m))=\mathbb{E}_{e,t}(S_{e,t})$ under groups of the same size.

In contrast, accuracy of the grouping mechanism varies from the accuracy of the baseline scenario if groups have heterogeneous sizes. In this case data suppliers in groups of lower size are weighted higher than data suppliers in larger groups as expressed in Equation \ref{eq:grandmean}. This accuracy difference is empirically investigated in Section \ref{sec:res1} for the two smart city application scenarios.

\begin{figure*}[!htb]
  \begin{align}
    \begin{split}
\mathbb{E}_{e,t}(\mathbb{E}_{e,t}(G_1),\ldots,\mathbb{E}_{e,t}(G_m)) &= \mathbb{E}_{e,t}(\frac{\sum_{j=1}^{|G_1|}s_{j,e,t}}{|G_1|},\ldots,\frac{\sum_{j=1}^{|G_m|}s_{j,e,t}}{|G_m|}) \\
& = \mathbb{E}_{e,t}(\frac{s_{1,e,t}}{|G_1|},\ldots,\frac{s_{a,e,t}}{|G_j|},\ldots,\frac{s_{|\mathcal{I}|,e,t}}{|G_m|}). \\
& a \in G_j, ~ 1 \le j \le m
    \end{split}
    \label{eq:grandmean}
  \end{align}
\end{figure*}

\begin{eqfloat*}[!htb]
\begin{flalign}
&\mbox{Given:}~ s_1=s_2=s_3=10, s_4=20.&
\end{flalign}
  \begin{flalign*}
& \mbox{No groups:~} & \mathbb{E}(s_1,s_2,s_3,s_4) & =(10+10+10+20)/4& =25/2 & \nonumber \\
& \mbox{Same size:~} & \mathbb{E}(\mathbb{E}(s_1,s_2),\mathbb{E}(s_3,s_4)) & =((10+10)/2+(10+20)/2)/2 &=25/2 & \nonumber  \\
& \mbox{Different size:~} & \mathbb{E}(\mathbb{E}(s_1,s_2,s_3),s_4) & =((10+10+10)/3+20)/2 &=15/2 & \nonumber
  \end{flalign*}
\addtocounter{equation}{-1}
\caption{Example of computing the mean for different configurations of groups. If groups have the same size, the result is equivalent to a simple mean.}
  \label{eq:grandmean_ex}
\end{eqfloat*}


\section{Experimental Methodology}
\label{sec-4}
\label{sec:exp}

The goal of the experimental evaluation is to compare the baseline scenario in which data suppliers send summarized data directly to the data consumer with the grouping scenario in which data suppliers form groups whose maximum size is defined by the parameter $N$. Note that varying the group size varies also the number of groups, as the population has a fixed size.

In the two compared scenarios two types of effects are studied: (i) \emph{macro-level} and (ii) \emph{micro-level}. Macro-level effects denote the changes seen at the population level, e.g. average privacy, accuracy. Experiments with varying group sizes, and thus varying number of groups, are performed while the summarization levels are kept constant during each experiment. Group sizes are sampled at the start of each experiment from distributions with parameter $N$: (i) a uniform distribution from 2 to $N$, (ii) a power law distribution from 2 to $N$ biased towards lower values, (iii) a bipolar distribution where 2 and $N$ have both 50\% probability, as well as (iv) a deterministic function returning $N$. Micro-level effects denote the changes seen at the group level, e.g. difference in privacy between group members. Experiments with varying summarization levels and fixed groups size are performed.

\subsection{Summarization function}\label{sec-4-1}

\begin{figure}[!htb]
\centering
\includegraphics[width=.9\linewidth]{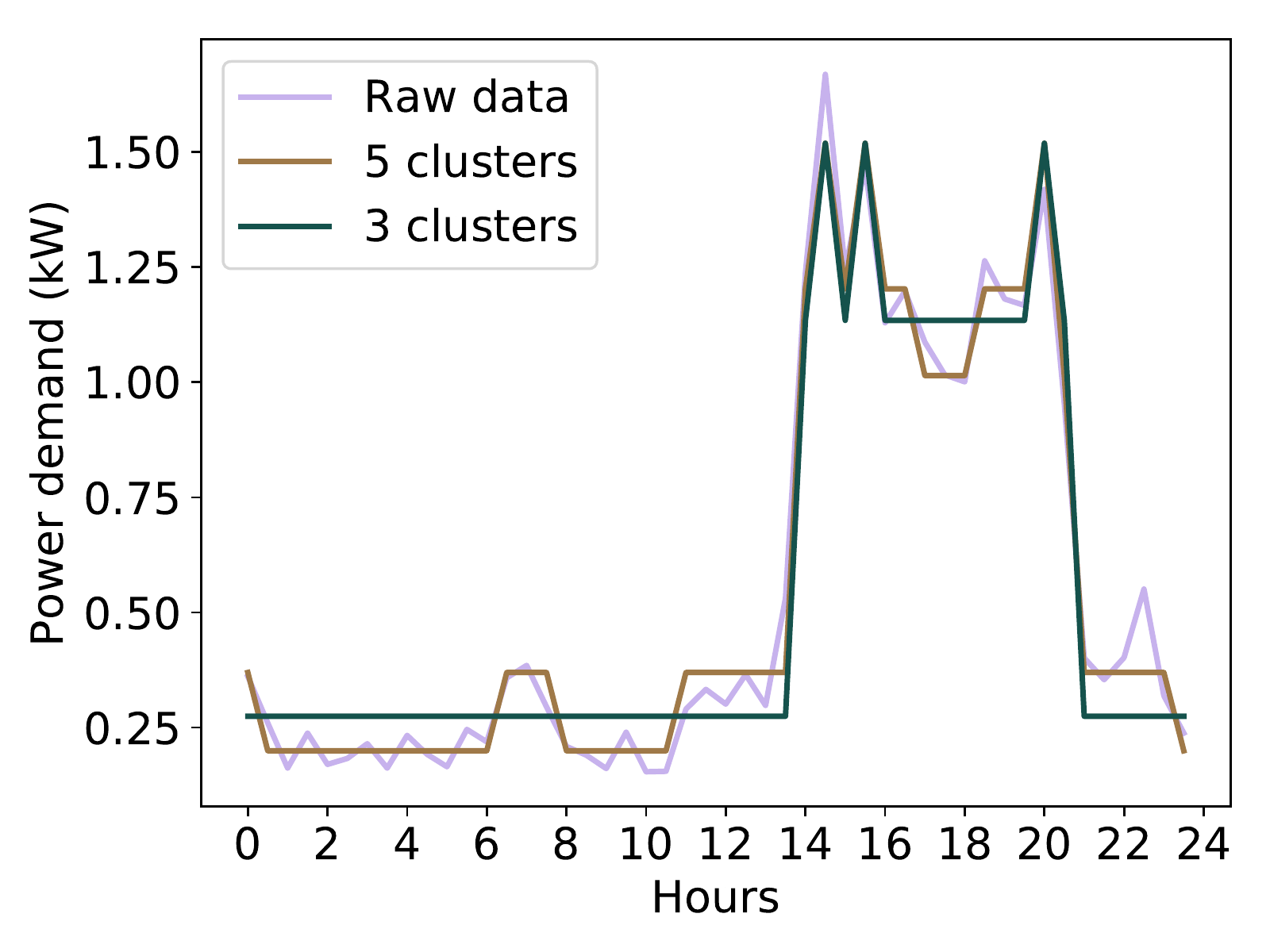}
\caption{\label{fig:data_example}An example of smart meter reading for user 3182 at day 06.12.2009, showing the effect of clustering on the signal with 3 and 5 clusters respectively.}
\end{figure}

The clustering algorithm \emph{k}-means with $k$ clusters is chosen as the summarization function. Clustering is a versatile machine learning technique, broadly used in data mining~\cite{jain12_survey_recen_clust_techn_data_minin}. Clustering can also work as a privacy-preserving mechanism~\cite{laszlo05_minim_spann_tree_partit_algor_microag,panagiotakis13_succes_group_selec_microag}, for example it is used to achieve \emph{k}-anonimity~\cite{domingo-ferrer05_ordin_contin_heter_k_anony_throug_microag} and t-closeness~\cite{soria-comas15_t_closen_throug_microag}, to improve the efficiency of a differential privacy mechanism~\cite{soria-comas14_enhan_data_utilit_differ_privac} and to improve privacy of IoT data \cite{pournaras17_engin_democ_inter_thing_data_analy}. Moreover the concept of clustering maps intuitively to the idea of information reduction~\cite{defays98_maskin_microd_using_micro_aggreg}: centroids, representative points of a cluster, substitute the original values in the cluster, thus obfuscating the data (see Figure \ref{fig:data_example}). Figure~\ref{fig:data_example} illustrates how a power consumption signal is obfuscated using k-means clustering.

IoT-PGA is independent of the summarization function, so clustering can be substituted by any other function that offers better privacy guarantees, for instance adding Laplace noise to the group aggregates as in earlier work~\cite{kellaris13_pract_differ_privac_via_group_smoot,Asikis2017}.

A summarization level of $1/|\mathcal{S}|$ corresponds to summarization with $k=|\mathcal{S}|$ clusters.
Summarization is applied on a sequence of measurements, e.g. the 48 half-hour power consumption records of the ECBT data, that are input to \emph{k}-means. Clustering divides the data points into $k$ clusters and returns the centroids, the representative points that define each cluster. Once the centroids are obtained, data suppliers generate the summarized data by substituting every point in the raw data with the representative centroid, the one at the minimum Euclidean distance, converting a sequence of length $|\mathcal{S}|$ with $K \le |\mathcal{S}|$ distinct values, to a sequence of the same length and $k \le K$ distinct values. The number of possible distinct values determines how high the information content of the summarized data is: If $k=K$, the raw and summarized data are equivalent. If $k=1$ the summarized data contain only one distinct value, the mean.

\subsection{Group formation}\label{subsec:group=formation}

The performed experiments start with all data suppliers having a summarization level of 1/10. The distribution of summarization levels gradually varies by randomly pairing data suppliers and moving summarization units from one member of the pair to the other. Practically, a fixed number of centroids is transferred within every group, from one data supplier to the other. Assume two randomly paired data suppliers, $a_1$ and $a_2$, with summarization levels $1/k_1$ and $1/k_2$. As a result of this process the first data supplier could gain $k$ centroids and the second looses $k$ centroids, ending up with new summarization levels of $1/(k_1+k)$ and $1/(k_2-k)$, or vice versa with $1/(k_1-k)$ and $1/(k_2+k)$. The experiment is repeated for different distributions of summarization levels, identified by their \emph{standard deviation}: a low standard deviation corresponds to a population with uniform summarization levels, whereas a high standard deviation corresponds to a highly dispersed population with a large number of extreme summarization values.

The effect of three different grouping strategies on the privacy and accuracy is studied, for different number of groups and distribution of summarization levels.
The strategies are generic in the context of smart cities applications in the sense that they do not rely on the physical proximity as a grouping criterion, nor on any critical assumption in power grids and transport systems.
In contrast, groups are logically linked within virtual topologies, i.e. overlay networks, that can be instantiated using information systems that support an application service, without changing the control infrastructure.
In other words, grouping is decoupled from the underlying physical network as no physical controllers or actuators are defined.
This practice is extensively applied in smart grid infrastructures~\cite{Pilgerstorfer2017,jelasity06_t_man} and other application domains.
This paper studies the following grouping strategies:

\begin{itemize}
\item \emph{Random}: At every epoch data suppliers are randomly assigned to groups of uniform size.
\item \emph{Data proximity}: At every epoch data suppliers are grouped according to the similarity between their raw data.
\item \emph{Summarization proximity}: Data suppliers are grouped according to the similarity between their summarization levels.
\end{itemize}

The strategy of data proximity groups together citizens with similar measurements, e.g. all passengers of a vehicle report the same speed of travel, or neighbors equipped with similar energy production facilities produce similar amounts of energy.
This strategy can also be used to form semantically interrelated groups in case there are different types of users, e.g. residential vs. industrial households, car drivers vs. pedestrians.
The strategy of summarization proximity groups together citizens who value the privacy of their data in a similar way. Random grouping represents a large set of grouping strategies, whose criterion does not consider either individual measurement nor individual preferences, e.g. physical proximity as in cars waiting at the same traffic light. Note that raw data changes at every epoch as data supplier receive new measurements, while summarization choices of each individual are assumed to be constant across epochs with varying standard deviation among individuals.

\subsection{Smart city datasets}\label{sec:datasets}

The general applicability of IoT-PGA in the domain of smart city is validated using real-world data from two pilot projects of two critical sectors of cities: (i) \emph{energy} and (ii) \emph{traffic}.

Experiments are performed using the data of the ``Electricity Customer Behavior Trial'' (ECBT) project~\cite{ecbt}, a collection of electricity consumption profiles of both residential households and small-medium enterprises, for a total of 6435 users, collected for a period of 52 weeks from 2009 to 2010. The measurements are collected with a frequency of 30 minutes and aggregated daily.
The dataset is preprocessed to improve its quality, excluding all data suppliers with less than 95\% of data availability. The polished dataset includes 68.42\% of the original data. Less than 1\% of missing data is interpolated.

The same experiments are repeated on the ``NREL Regional Transportation Commission of Southern Nevada'' (RTCSNV) dataset~\cite{nrel}, a travel survey performed in 2014. The dataset contains the GPS sensor data of the study, which comprises wearable GPS sensors for a total of 2293 people. The data cover a total of three days, and every day has a variable number of measurements (trips). The average speed of each trip is considered to be privacy-sensitive information. A measurement is defined as the collection of average speed values of every trip in a day.

This paper assumes the records in the dataset to be the ground truth data, meaning that detecting and removing sampling or malfunction errors is out of the scope of this work. In any case, such errors increase the obfuscation level and therefore the privacy of users as measured in this paper.

\begin{figure}[!htb]
\centering
\includegraphics[width=.9\linewidth]{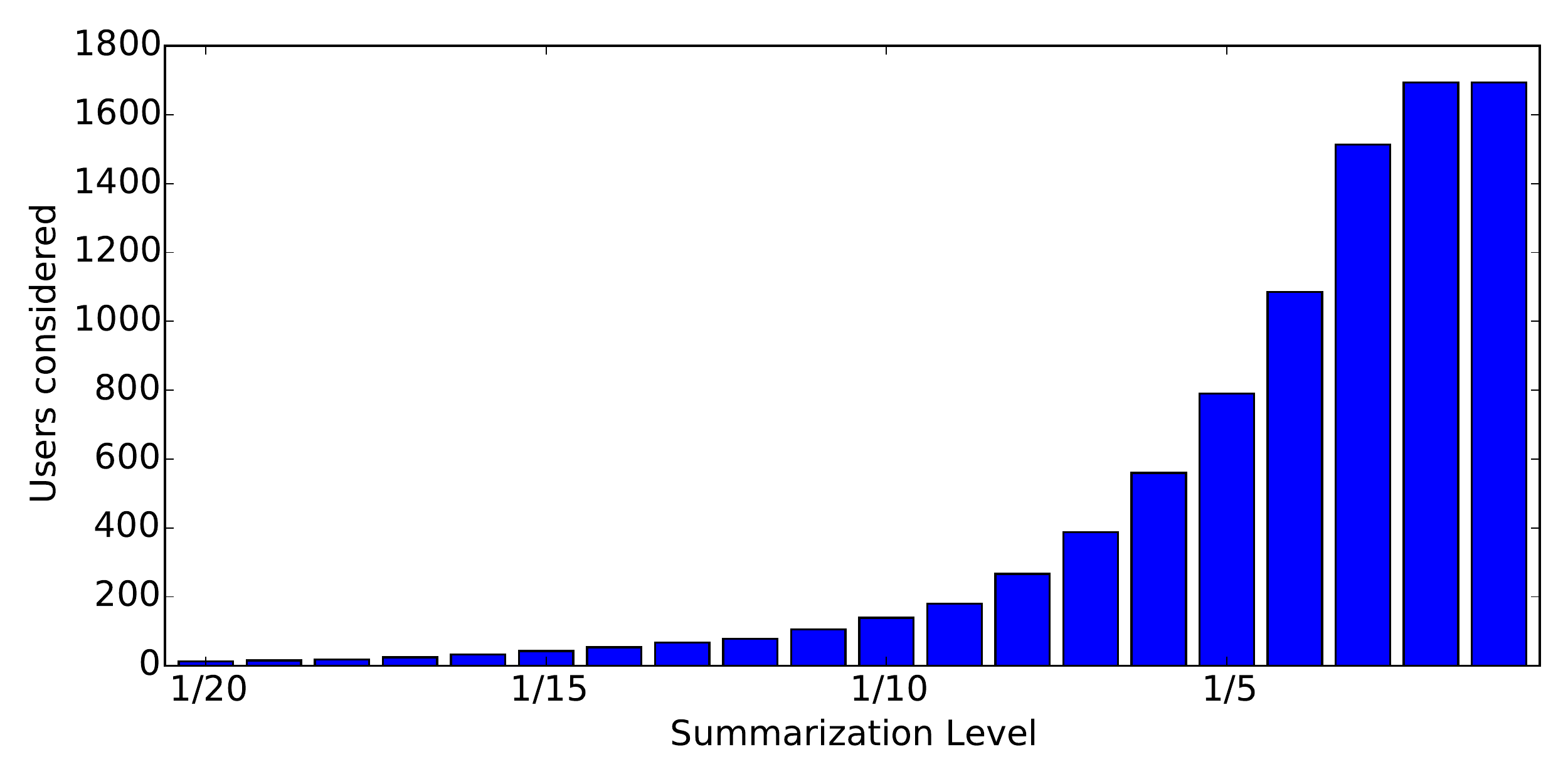}
\caption{\label{fig:data_counts}Number of data suppliers in the NREL dataset considered in the experiments for each summarization level.}
\end{figure}

The main drawback of the NREL dataset, compared to the ECBT dataset, is the lower number of epochs (3 vs. 365 days) and the lower number of data suppliers (variable and lower than 1600 vs. higher than 6000 for the ECBT dataset). In the ECBT dataset each epoch has 48 records. In contrast, the NREL dataset has a variable number of trips for each data supplier and epoch, therefore choosing a fixed number of non-overlapping clusters for every data supplier is not an option as it may result in a larger number of clusters than the number of input data points (see also~\cite{arabie1981overlapping}). One way to overcome this technical issue is to leave out data suppliers with fewer than $k$ points, given a summarization level of 1/$k$. Figure \ref{fig:data_counts} illustrates how the number of suitable data suppliers in the NREL dataset declines with a decreasing summarization level: the data supplier population decreases to half at summarization level of 1/5, and at summarization level of 1/10 fewer than 200 data suppliers remain. Taking this into consideration, all experiments are limited to summarization levels lower than 1/10, which allow for statistical purposes a large enough population of data suppliers.

\subsection{Aggregation function}\label{sec-4-3}

The sum, together with the mean, is part of a large class of operations that can be computed in a distributed way, therefore the results are applicable in the context of the decentralized group management and aggregation shown in this paper~\cite{pournaras17_engin_democ_inter_thing_data_analy,jelasity05_gossip_based_aggreg_large_dynam_networ}.

The mean is used as aggregation function in the performed experiments. The sum has applicability in smart grids, for example when computing the total load of the network. This paper focuses on mean as it encodes the sum given the size of the groups, which makes it more interesting in the setting where groups have heterogeneous size (cf. Equation \ref{eq:grandmean_ex}).


\section{Experimental Evaluation}\label{sec:res}

\begin{figure}[!htb]
\begin{minipage}[b]{.5\linewidth}
\centering
\includegraphics[width=\linewidth]{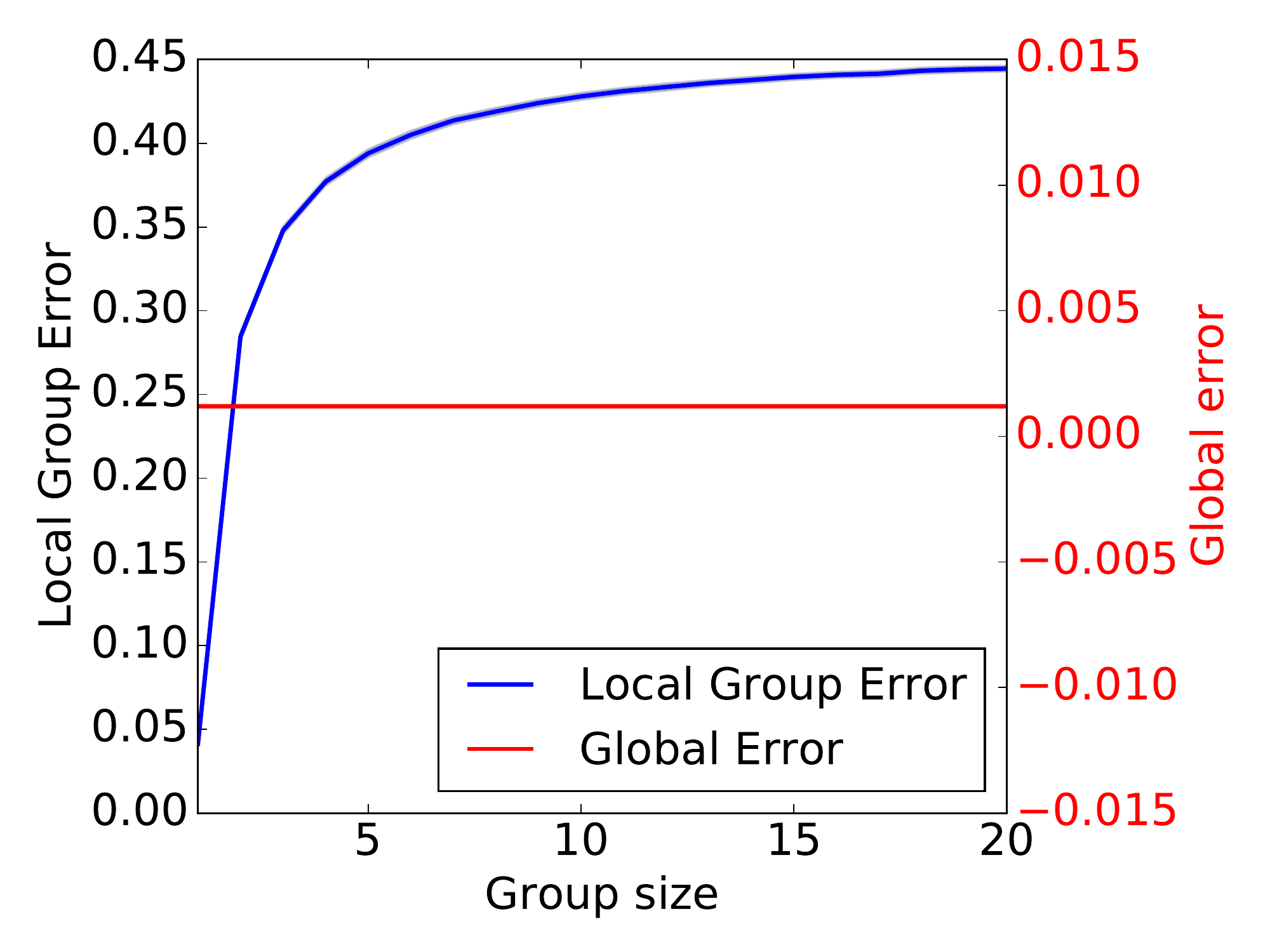}
\subcaption{ECBT dataset}\label{fig:1a}
\end{minipage}%
\begin{minipage}[b]{.5\linewidth}
\centering
\includegraphics[width=\linewidth]{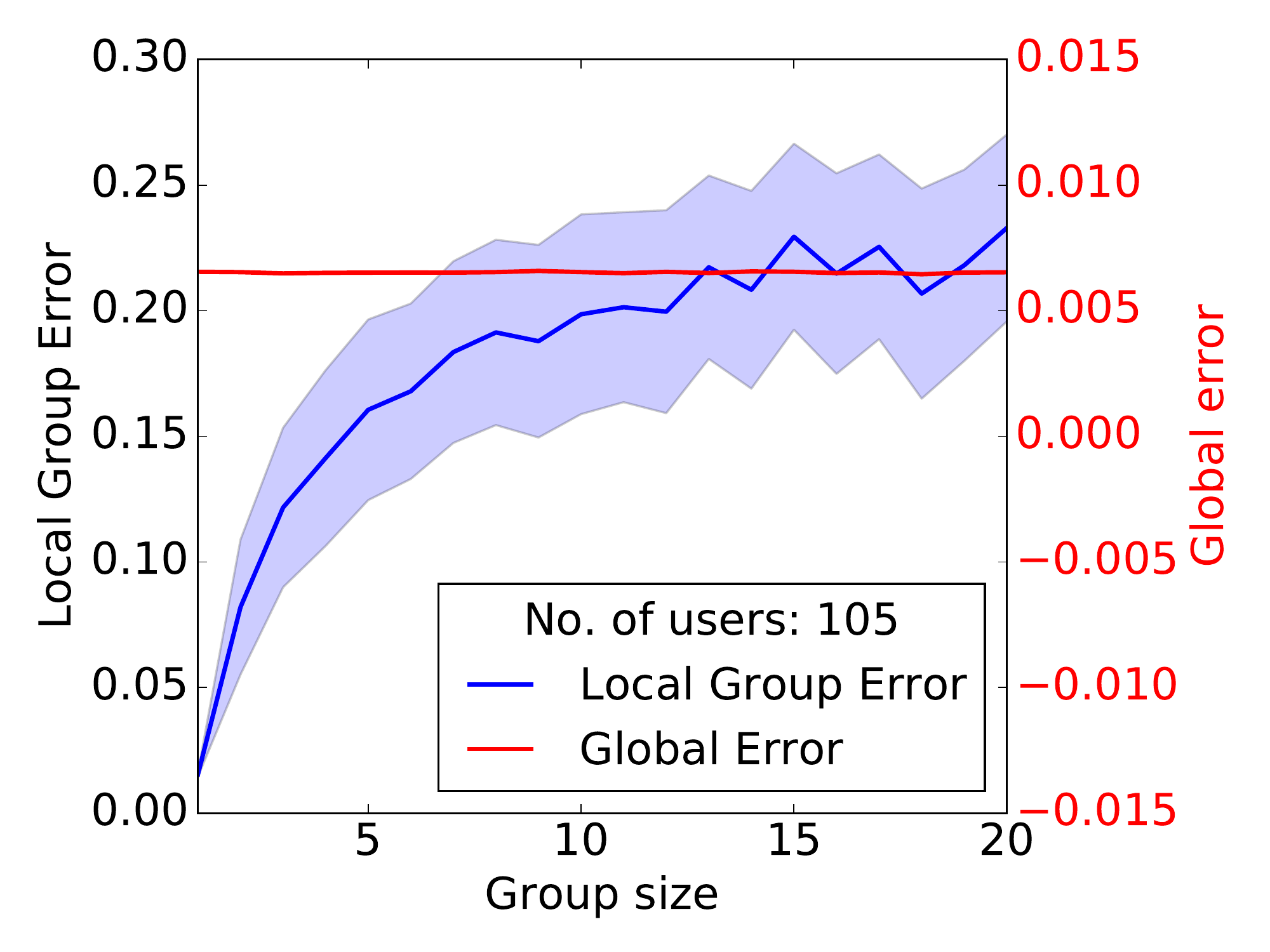}
\subcaption{NREL dataset}\label{fig:1b}
\end{minipage}
\caption{Average local group error and global error for varying group size. The results are obtained for a summarization level of 1/10, but they are qualitatively similar for  different summarization levels.}\label{fig:1}
\end{figure}

Figure \ref{fig:1} shows that the local group error increases by increasing the group size, while the global error remains constant. The largest gain in privacy is observed by moving from no groups to groups of size 2, with an increase of around 600\%, while increasing the group size even further increases the local group error around 50\% the level for group size 2. Qualitatively similar results are obtained for different values of summarization level (cf. different rows of Figure \ref{fig:strategic}). This result confirms the hypothesis that, given a global accuracy objective, grouping data suppliers increases privacy against the data consumer.

\subsection{Non-uniform grouping}
\label{sec:res1}

One assumption so far is that all groups have the same size. This property makes the aggregation function of average over data suppliers (baseline scenario) equals to the average among the group averages (grouping scenario, see discussion in Section \ref{sec:model:ge}), which ensures that the global error remains constant under grouping. This assumption is relaxed by studying different distributions of group sizes defined by the parameter $N$, which determines the maximum size any group can take.

\begin{figure}[!htb]
\begin{minipage}[b]{.5\linewidth}
\centering
\includegraphics[width=\linewidth]{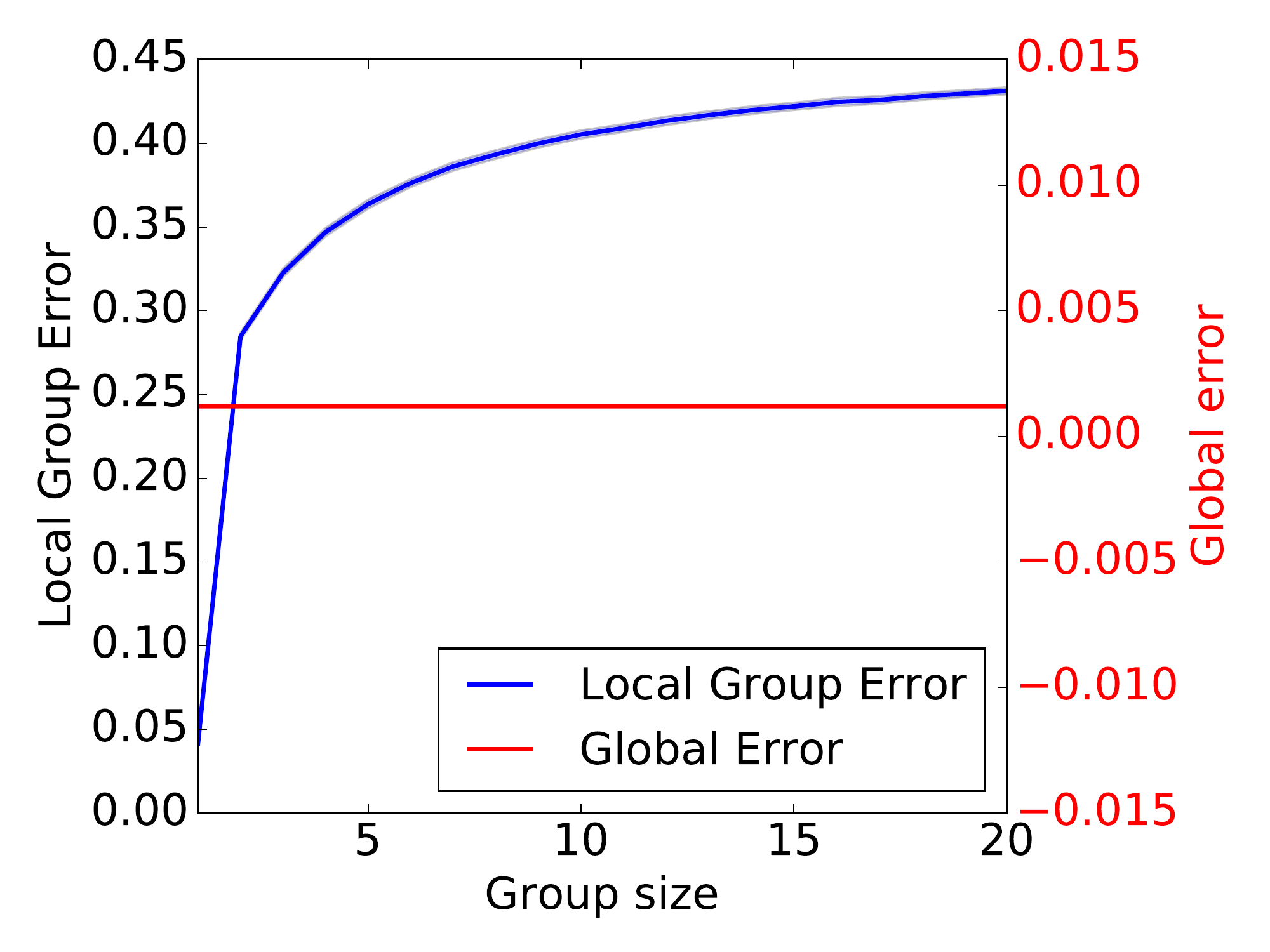}
\subcaption{ECBT dataset}\label{fig:unif_a}
\end{minipage}%
\begin{minipage}[b]{.5\linewidth}
\centering
\includegraphics[width=\linewidth]{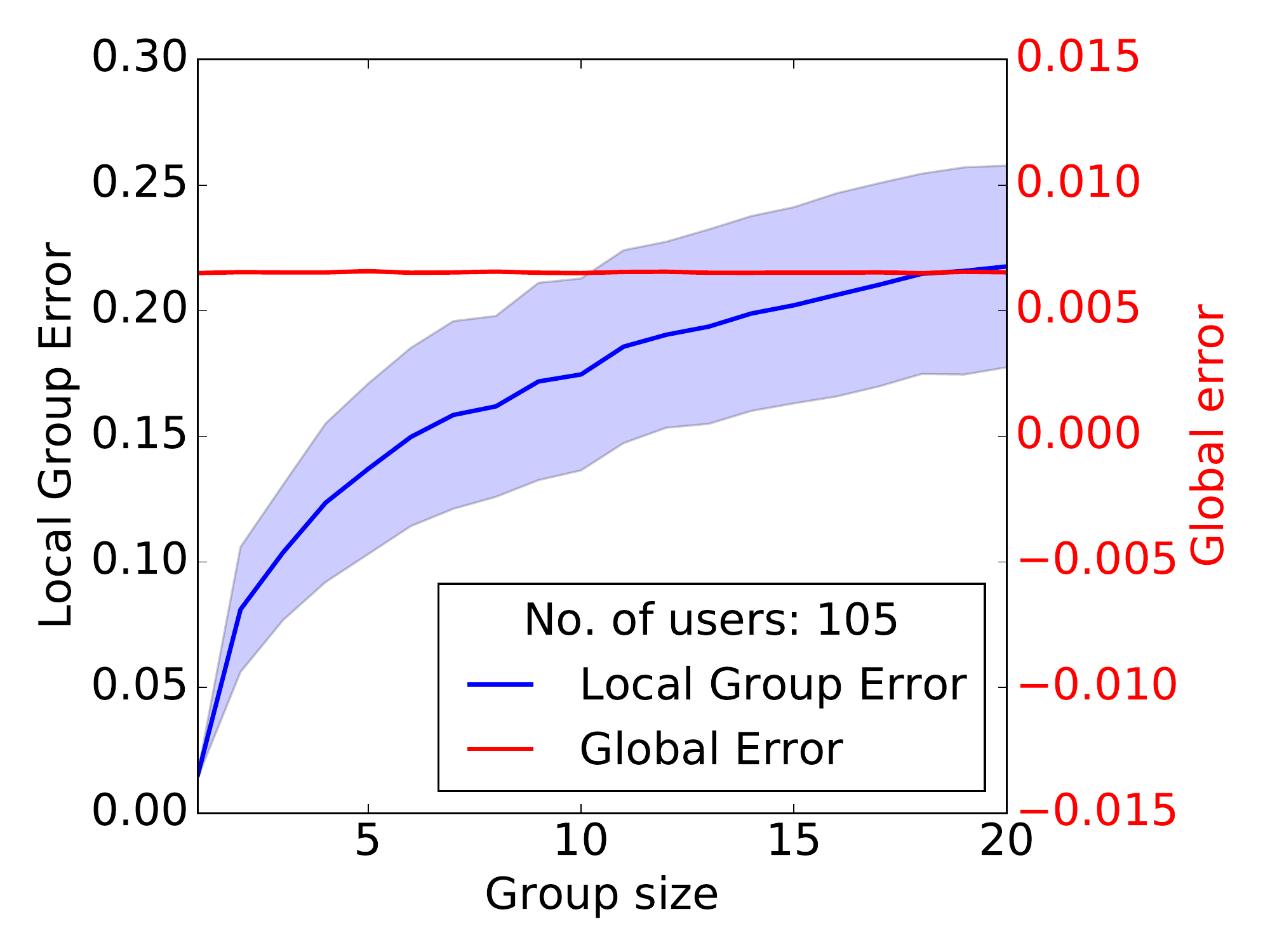}
\subcaption{NREL dataset}\label{fig:unif_b}
\end{minipage}
\caption{Average local group error and global error for varying group size. Group sizes generated randomly by sampling from a uniform distribution.}\label{fig:unif}
\begin{minipage}[b]{.5\linewidth}
\centering
\includegraphics[width=\linewidth]{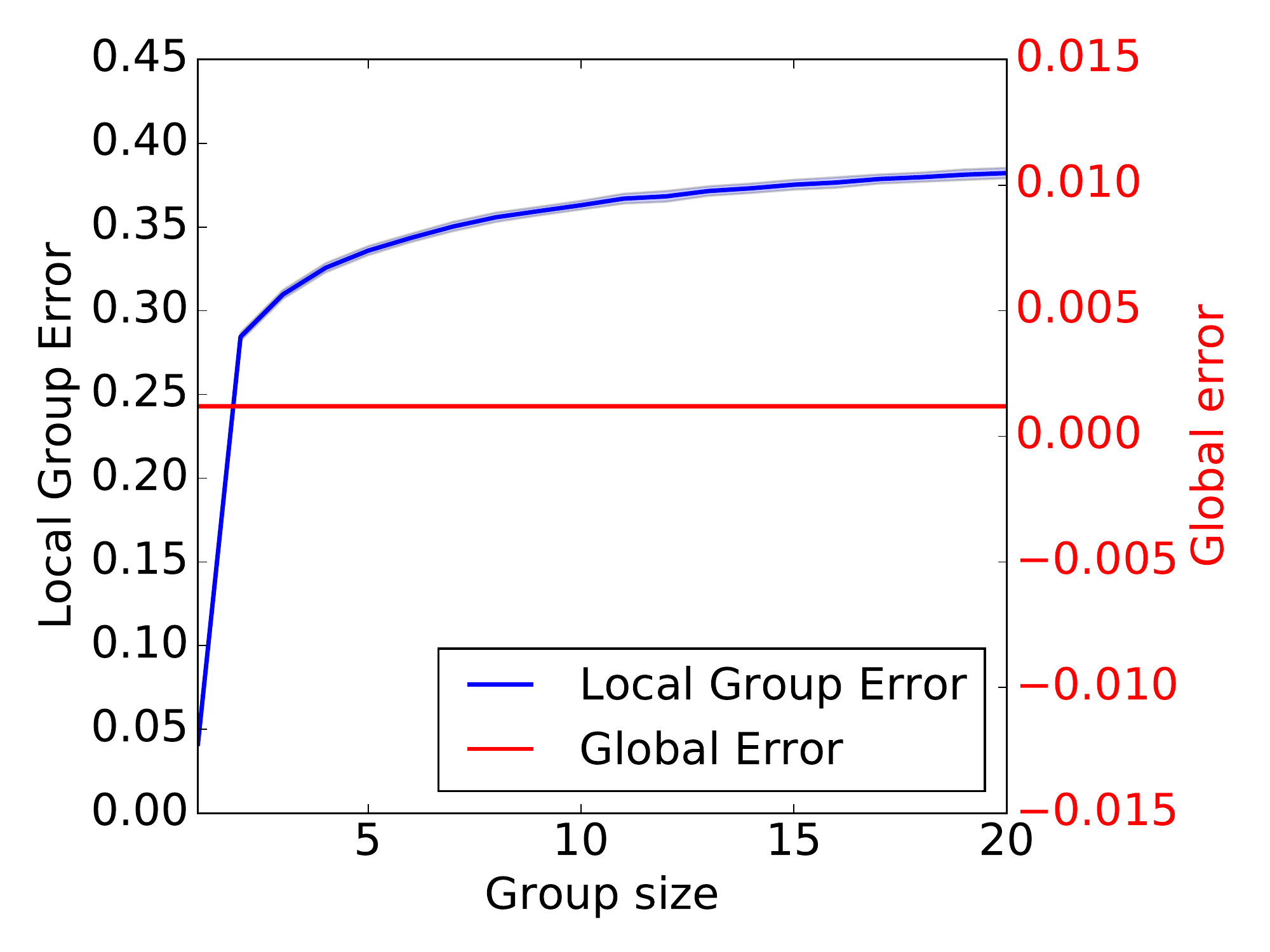}
\subcaption{ECBT dataset}\label{fig:plaw_a}
\end{minipage}%
\begin{minipage}[b]{.5\linewidth}
\centering
\includegraphics[width=\linewidth]{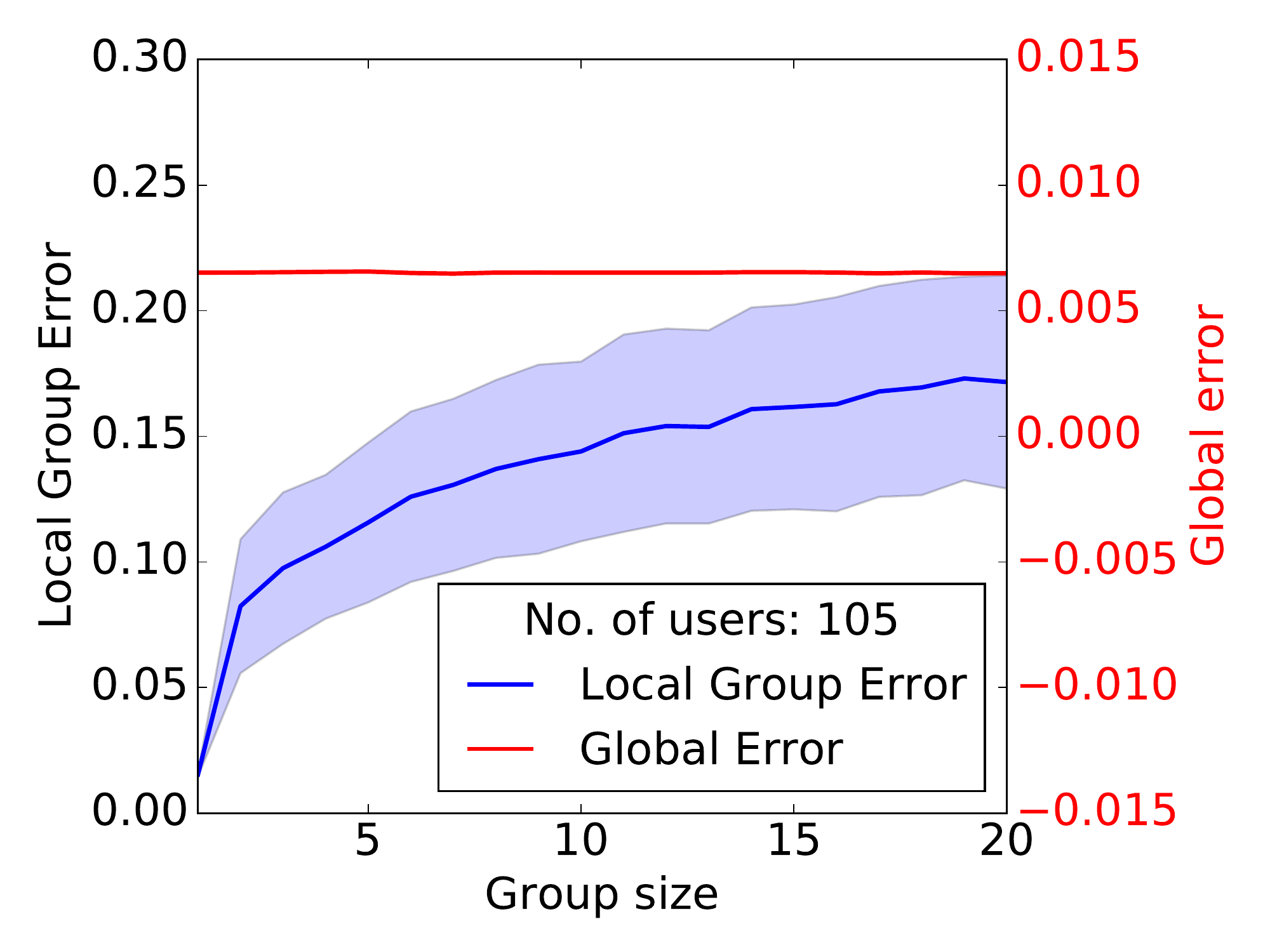}
\subcaption{NREL dataset}\label{fig:plaw_b}
\end{minipage}
\caption{Average local group error and global error for varying group size. Group sizes generated randomly by sampling from a power law distribution.}\label{fig:plaw}
\begin{minipage}[b]{.5\linewidth}
\centering
\includegraphics[width=\linewidth]{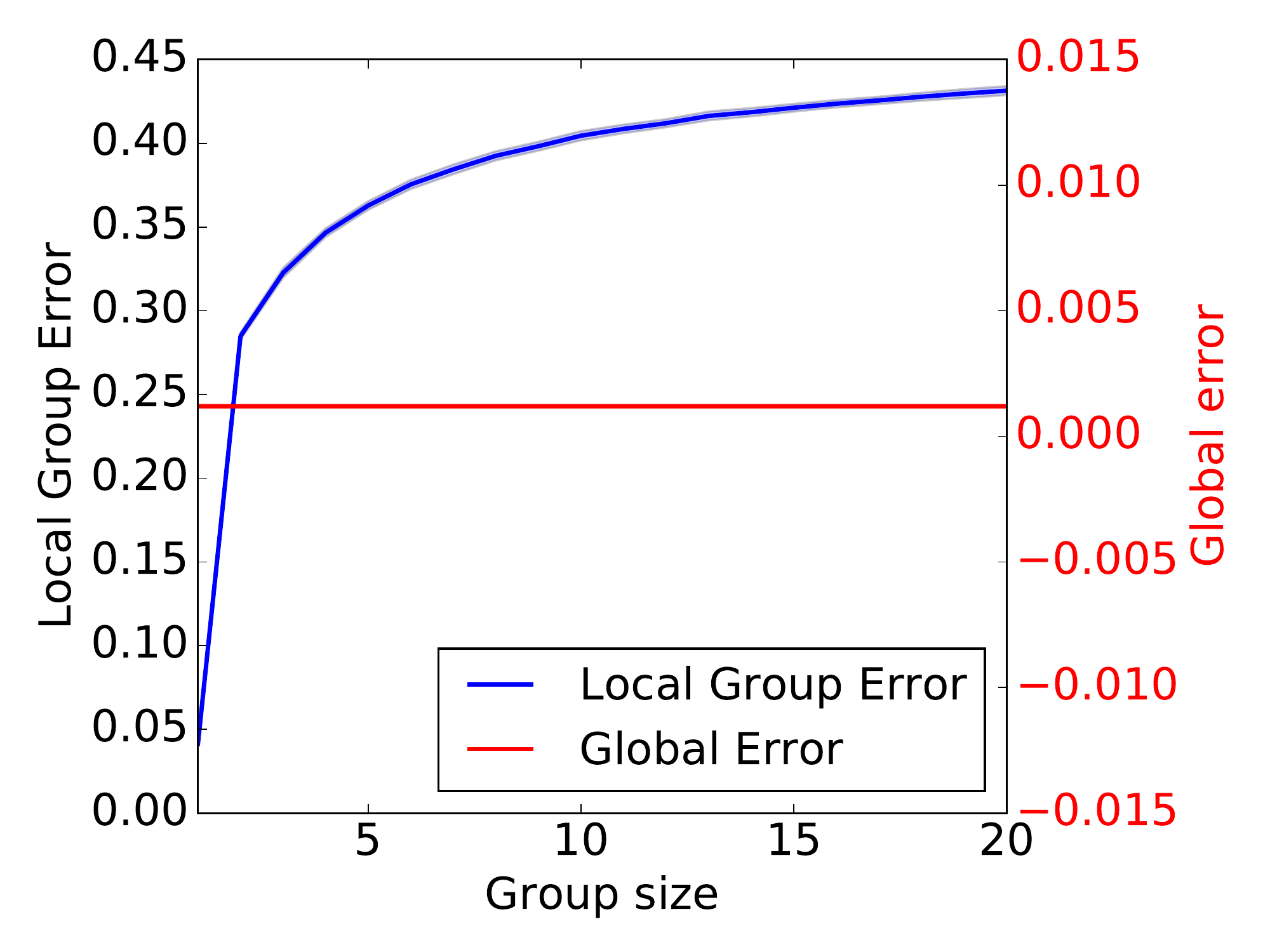}
\subcaption{ECBT dataset}\label{fig:step_a}
\end{minipage}%
\begin{minipage}[b]{.5\linewidth}
\centering
\includegraphics[width=\linewidth]{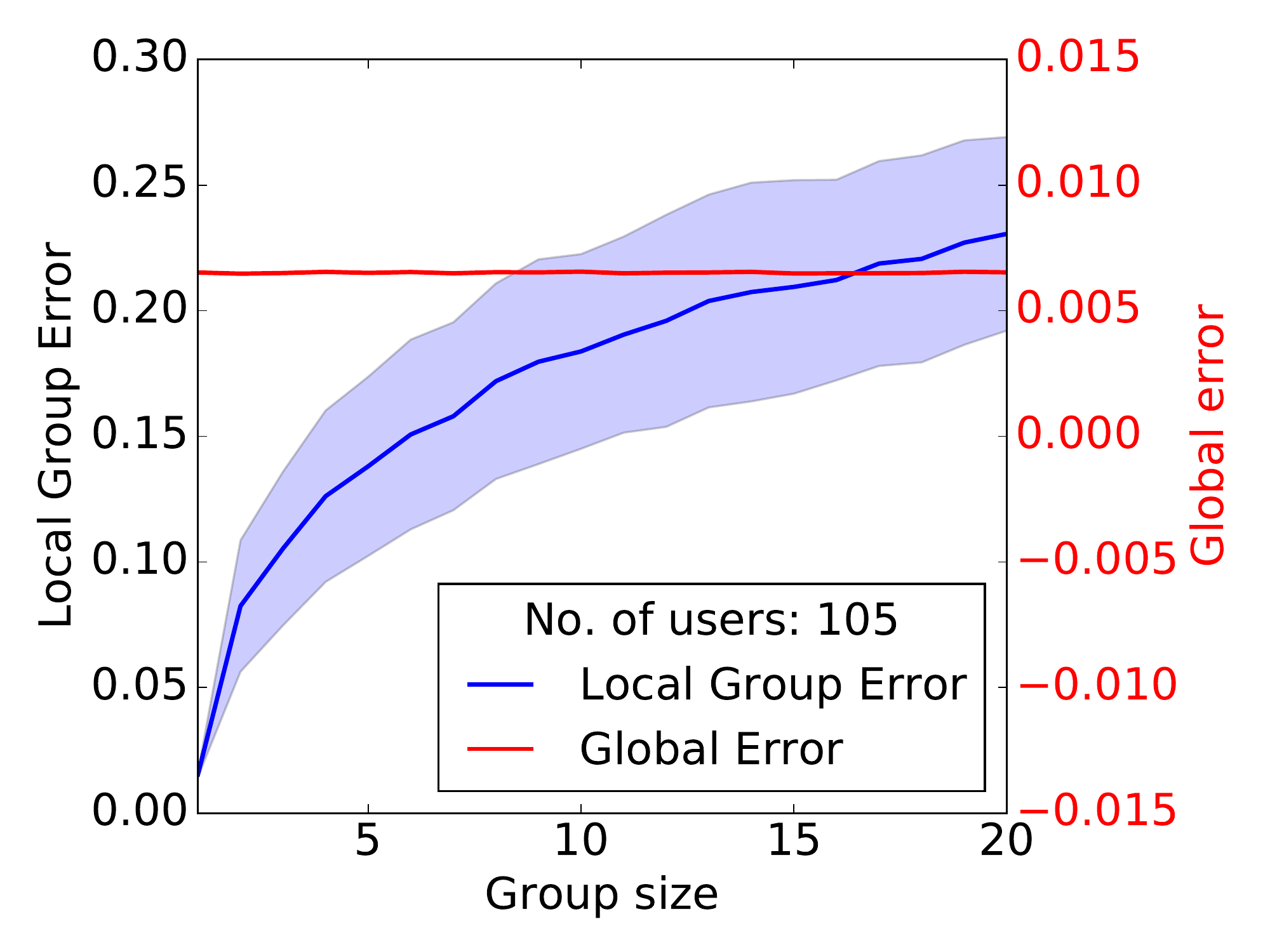}
\subcaption{NREL dataset}\label{fig:step_b}
\end{minipage}
\caption{Average local group error and global error for varying group size. Group sizes generated randomly by sampling from a step function.}\label{fig:step}
\end{figure}

In the first experiment the group sizes are chosen uniformly at random between 2 and $N$ (Figure \ref{fig:unif}). Results are comparable to fixed group sizes, in particular there is no change in the global error.

In the second experiment a bias is introduced towards smaller groups: group sizes are sampled from a power-law distribution that generates a higher number of small groups than large groups. If groups are sampled from a power law distribution (Figure \ref{fig:plaw}) there is no distinguishable change in global error, but there is a decrease of about 10\% in the local group error.

In the third experiment the bias changes to extreme group sizes: samples are taken from a step function that returns either 2 or $N$ with 50\% probability each. In this case, the global error remains at the same level as in all previous experiments (Figure \ref{fig:step}).

Concluding, this section confirms that grouping increases privacy without decreasing accuracy, independently of the group characteristics of groups. The actual increase in privacy is influenced though by group characteristics. The latter is further investigated in Section \ref{sec:res3}

\subsection{Inner group dynamics}\label{sec:res2}

\begin{figure}[t!]
  \begin{minipage}[t]{1.0\linewidth}
    \begin{minipage}[b]{.5\linewidth}
\centering
\includegraphics[width=\linewidth]{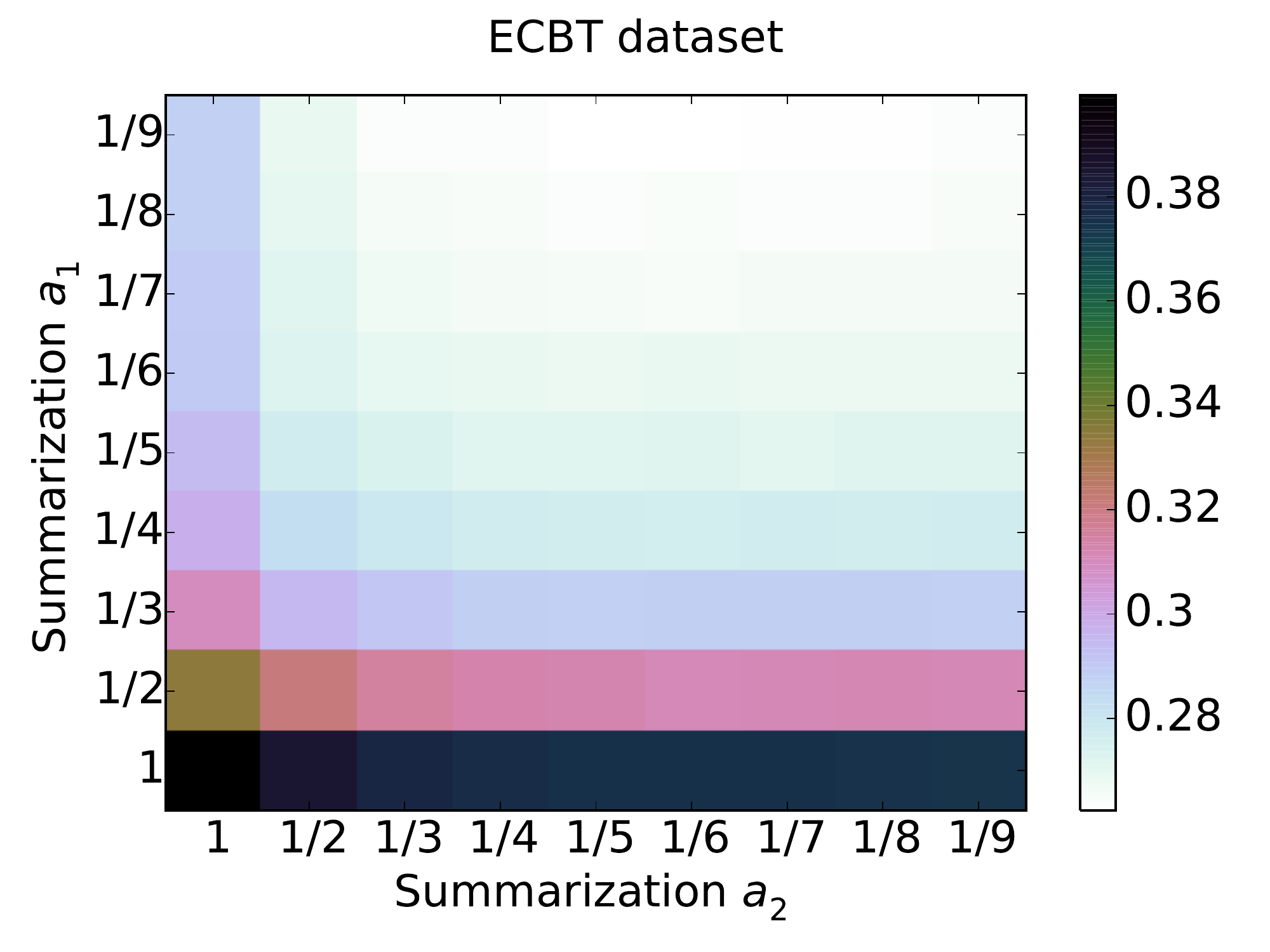}
\end{minipage}%
\begin{minipage}[b]{.5\linewidth}
\centering
\includegraphics[width=\linewidth]{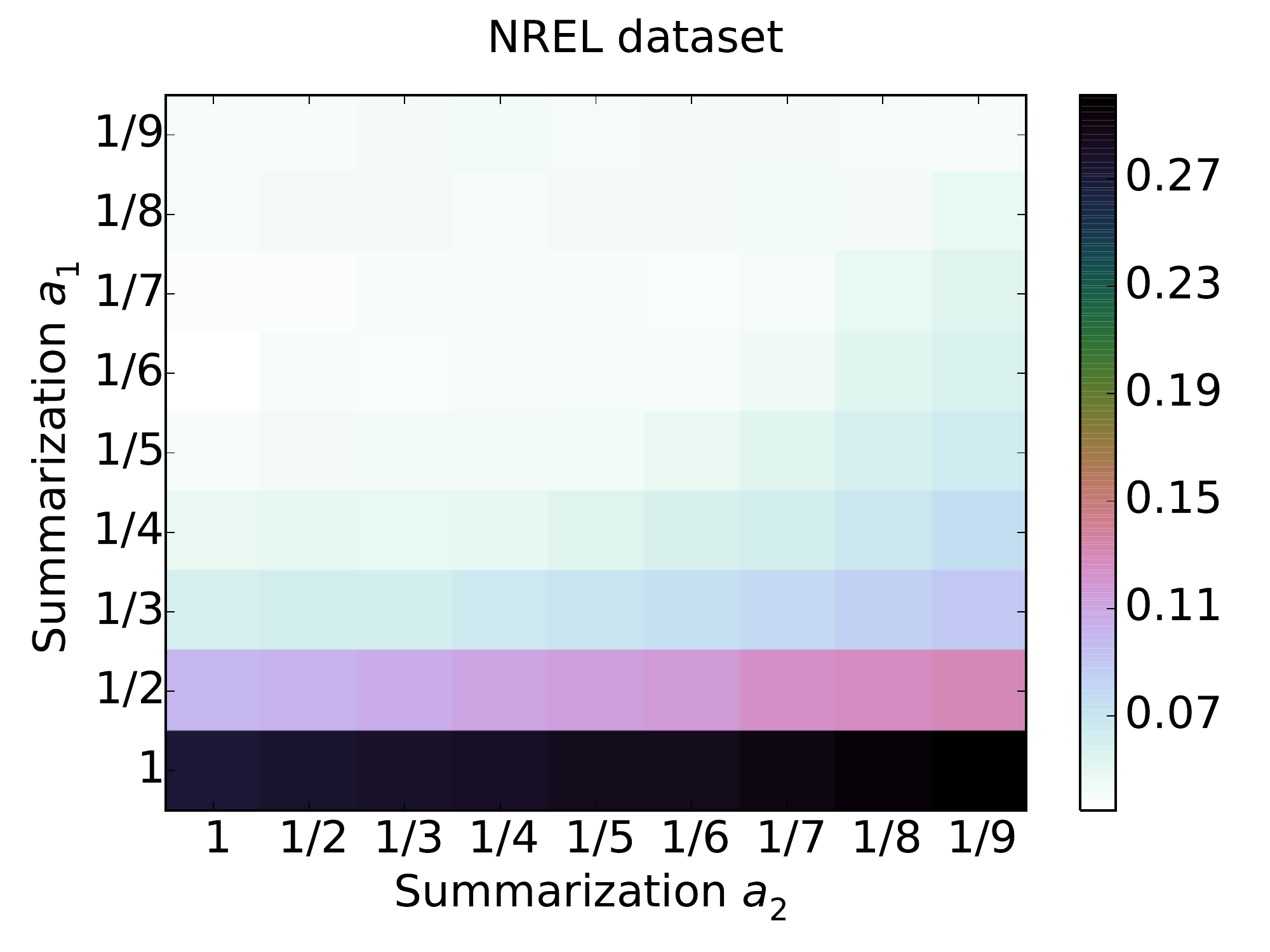}
\end{minipage}
\subcaption{Local group error of $a_1$.}\label{fig:lerr_a1}
  \end{minipage}
  \begin{minipage}[t]{1.0\linewidth}
    \begin{minipage}[b]{.5\linewidth}
\centering
\includegraphics[width=\linewidth]{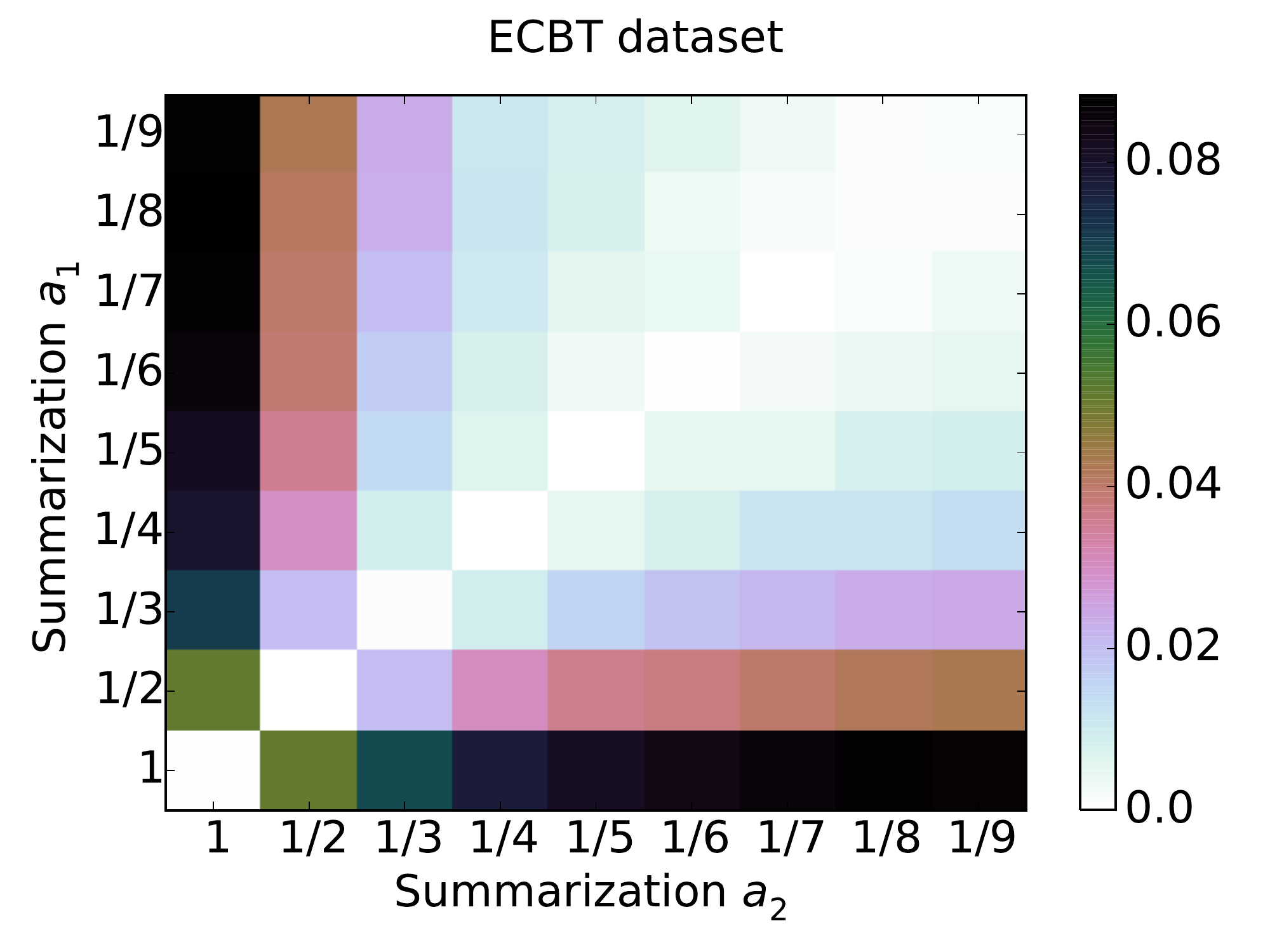}
\end{minipage}%
\begin{minipage}[b]{.5\linewidth}
\centering
\includegraphics[width=\linewidth]{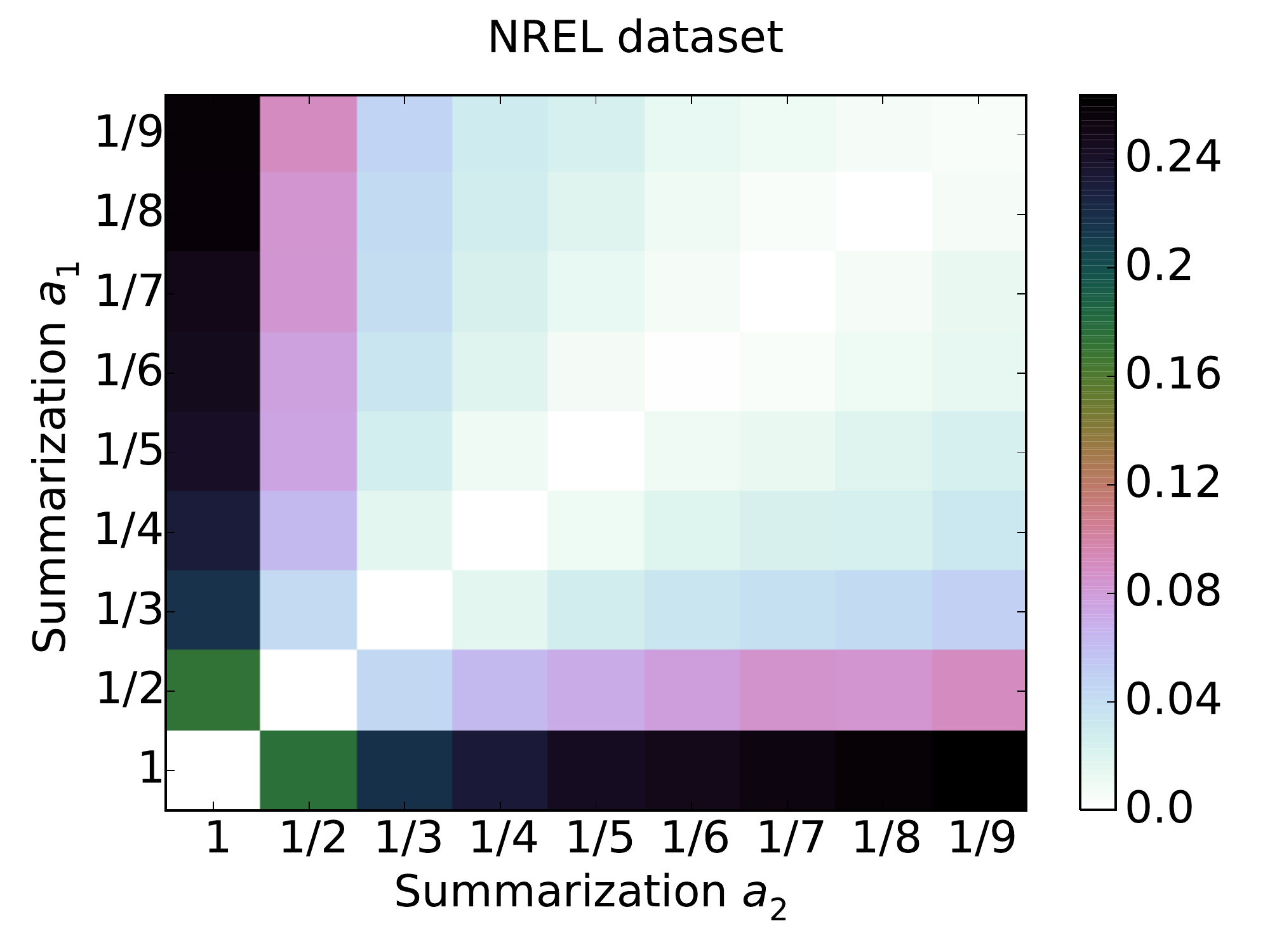}
\end{minipage}
\subcaption{Absolute difference between the local group errors of $a_1$ and $a_2$.}\label{fig:diff_lerr}
  \end{minipage}
  \begin{minipage}[t]{1.0\linewidth}
    \begin{minipage}[b]{.5\linewidth}
\centering
\includegraphics[width=\linewidth]{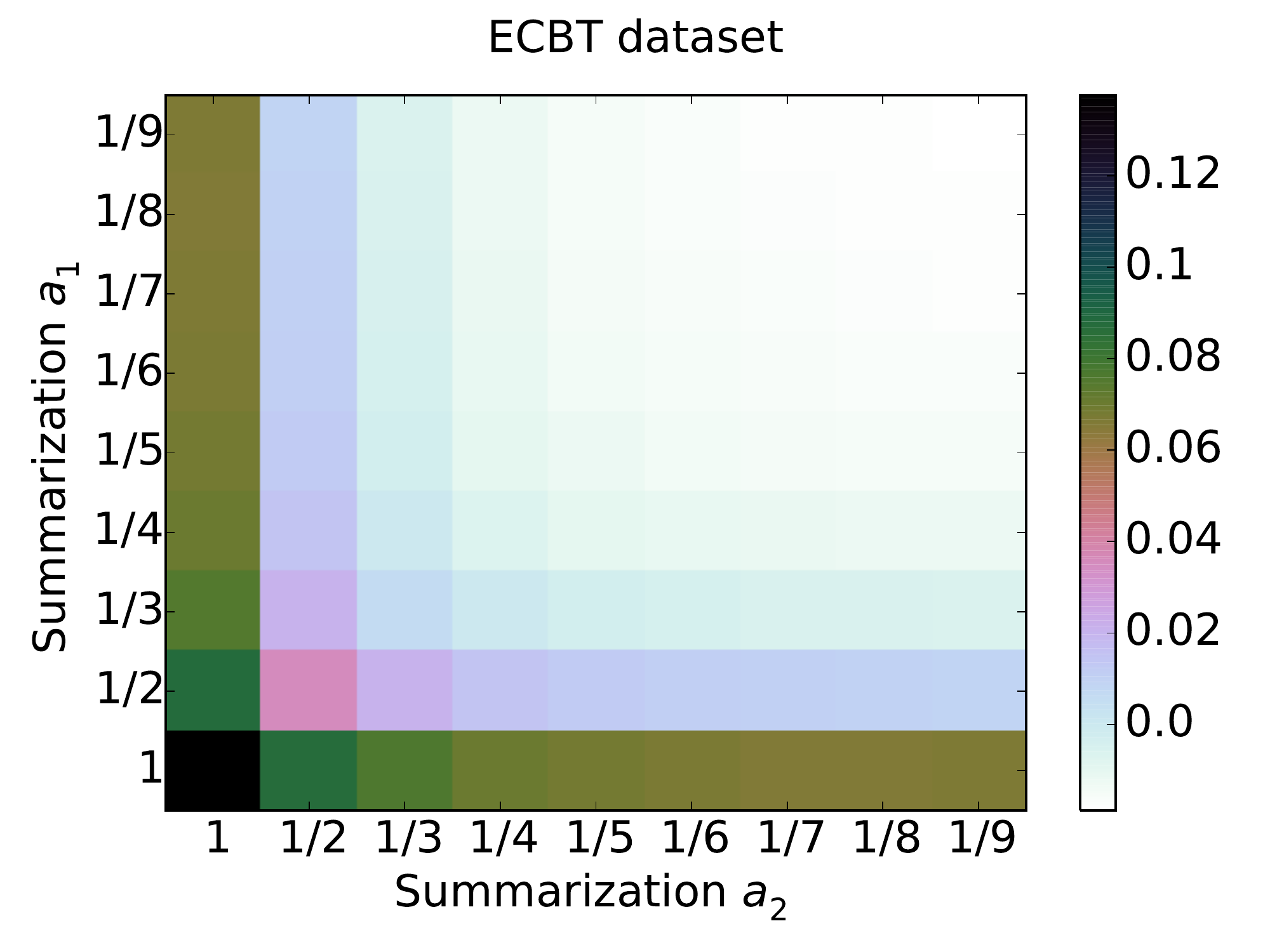}
\end{minipage}%
\begin{minipage}[b]{.5\linewidth}
\centering
\includegraphics[width=\linewidth]{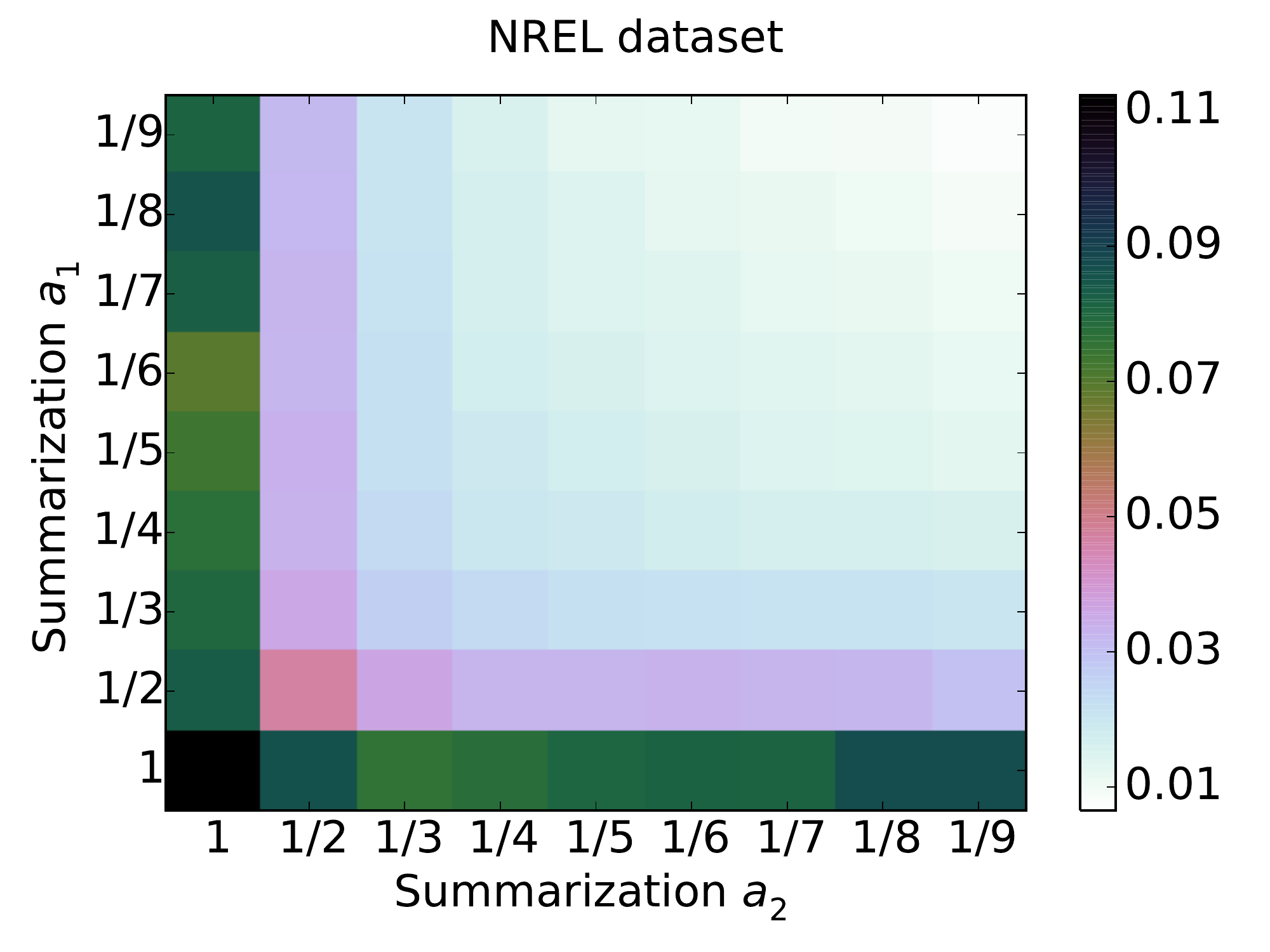}
\end{minipage}
\subcaption{Average global error of groups.}\label{fig:mean_gerr}
  \end{minipage}
\caption{Errors for varying summarization levels of $a_1$ and $a_2$ . The y axis shows the summarization level of $a_1$, the x axis the summarization level of $a_2$.}
\end{figure}

Data suppliers are grouped in pairs\footnote{Further results for groups of size 3 have the same effect. Due to space limitations they are not shown in this paper and they are available upon request.} with two members of an arbitrary group denoted as $a_1$ and $a_2$, and each of them is assigned with a summarization level between 1 and 1/9. Experiments cover in total 81 combinations of summarization levels. Looking at the local group error of each data supplier, the privacy of one data supplier is influenced by the summarization choice of the other one: Figure \ref{fig:lerr_a1} shows that the higher the summarization of $a_1$, the higher its increase of privacy. The summarization choice of $a_2$ influences positively the privacy of $a_1$. Further privacy measures in Appendix \ref{sec:model:corr} confirm this observation.

Figure \ref{fig:diff_lerr} shows the absolute difference between the local group errors of the two data suppliers. Both data suppliers have the same error when they summarize at the same level. The higher the difference between the summarization levels of the two data suppliers, the higher the difference between the local group errors.

Figure \ref{fig:mean_gerr} shows the trade-off between summarization and accuracy: The higher the summarization, the higher the global error. The plots also reveal that the greatest reduction in accuracy happens for high summarization levels, while accuracy is almost constant for summarization levels below 1/6.

\subsection{Grouping as incentivization}\label{sec:res3}

So far, experiments show that the privacy of a data supplier is maximized for high summarization levels, yet it does not vary significantly for summarization levels lower than 1/4. Reward mechanisms can steer data suppliers to a regime in which privacy is high while accuracy is at an acceptable level. A grouping mechanism can act as an incentive mechanism by rewarding with group membership data suppliers that choose to summarize at low levels. The size of the group and the collective group member choices of the summarization level influence the final privacy gain (cf. Figure \ref{fig:1}).

\begin{figure}[!htb]
\begin{minipage}[b]{.5\linewidth}
\centering
\includegraphics[width=\linewidth]{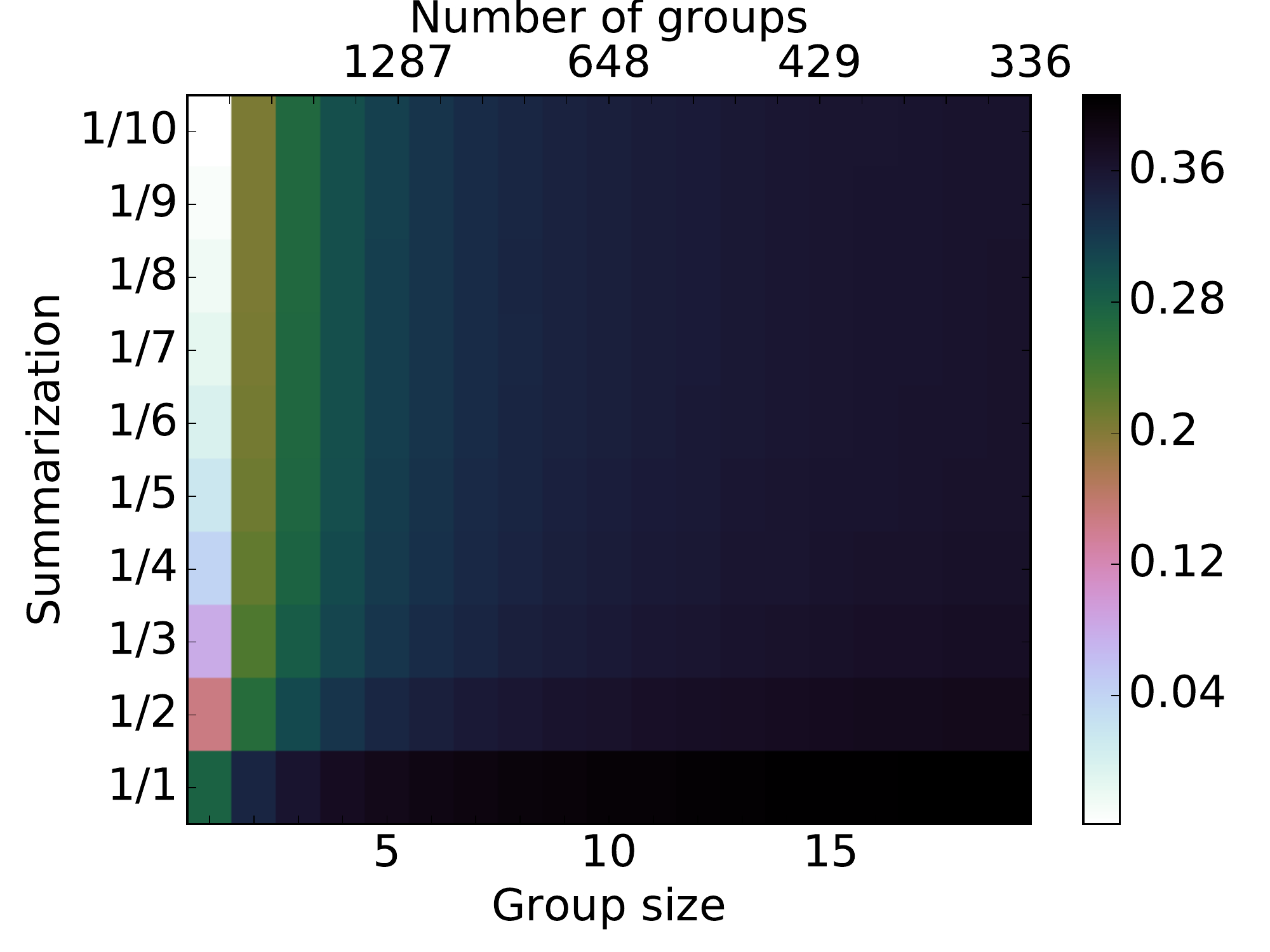}
\subcaption{ECBT dataset}\label{fig:strategic_a}
\end{minipage}%
\begin{minipage}[b]{.5\linewidth}
\centering
\includegraphics[width=\linewidth]{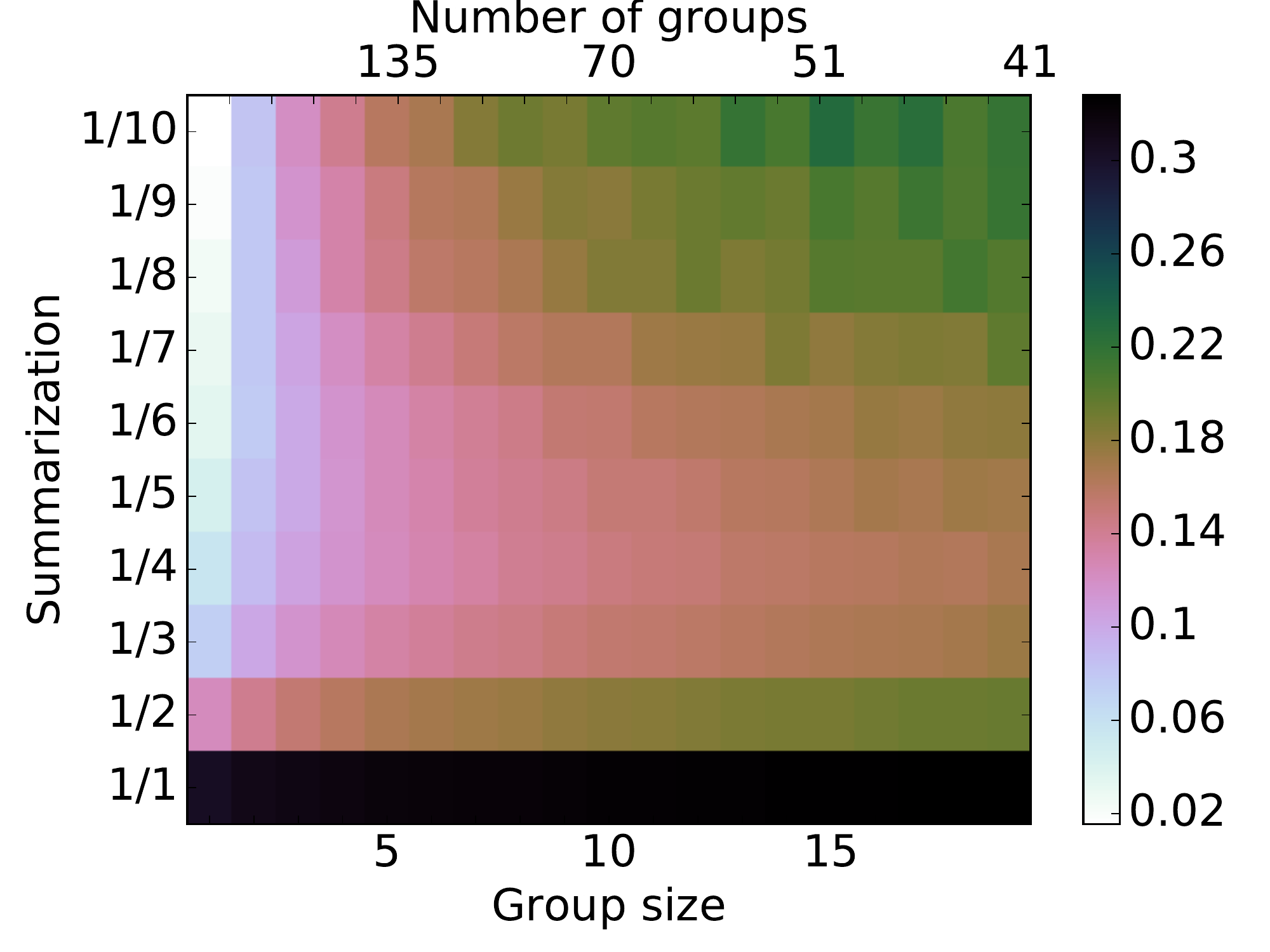}
\subcaption{NREL dataset}\label{fig:strategic_b}
\end{minipage}
\caption{Average local group error across groups for a given group size and a summarization level.}\label{fig:strategic}
\end{figure}

\begin{figure}[t!]
\begin{minipage}[b]{.5\linewidth}
\centering
\includegraphics[width=\linewidth]{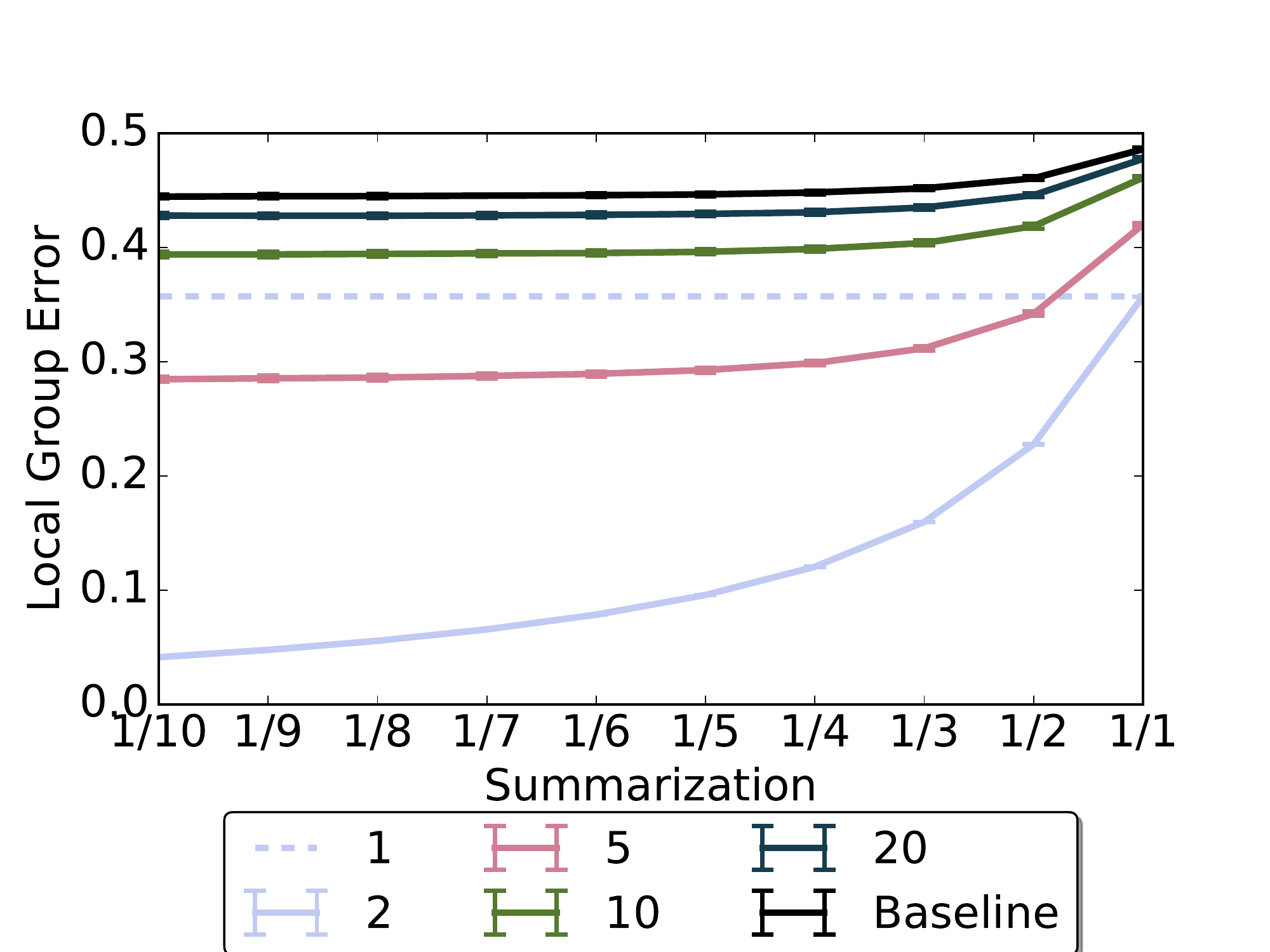}
\subcaption{ECBT dataset}\label{fig:strategic_line_a}
\end{minipage}%
\begin{minipage}[b]{.5\linewidth}
\centering
\includegraphics[width=\linewidth]{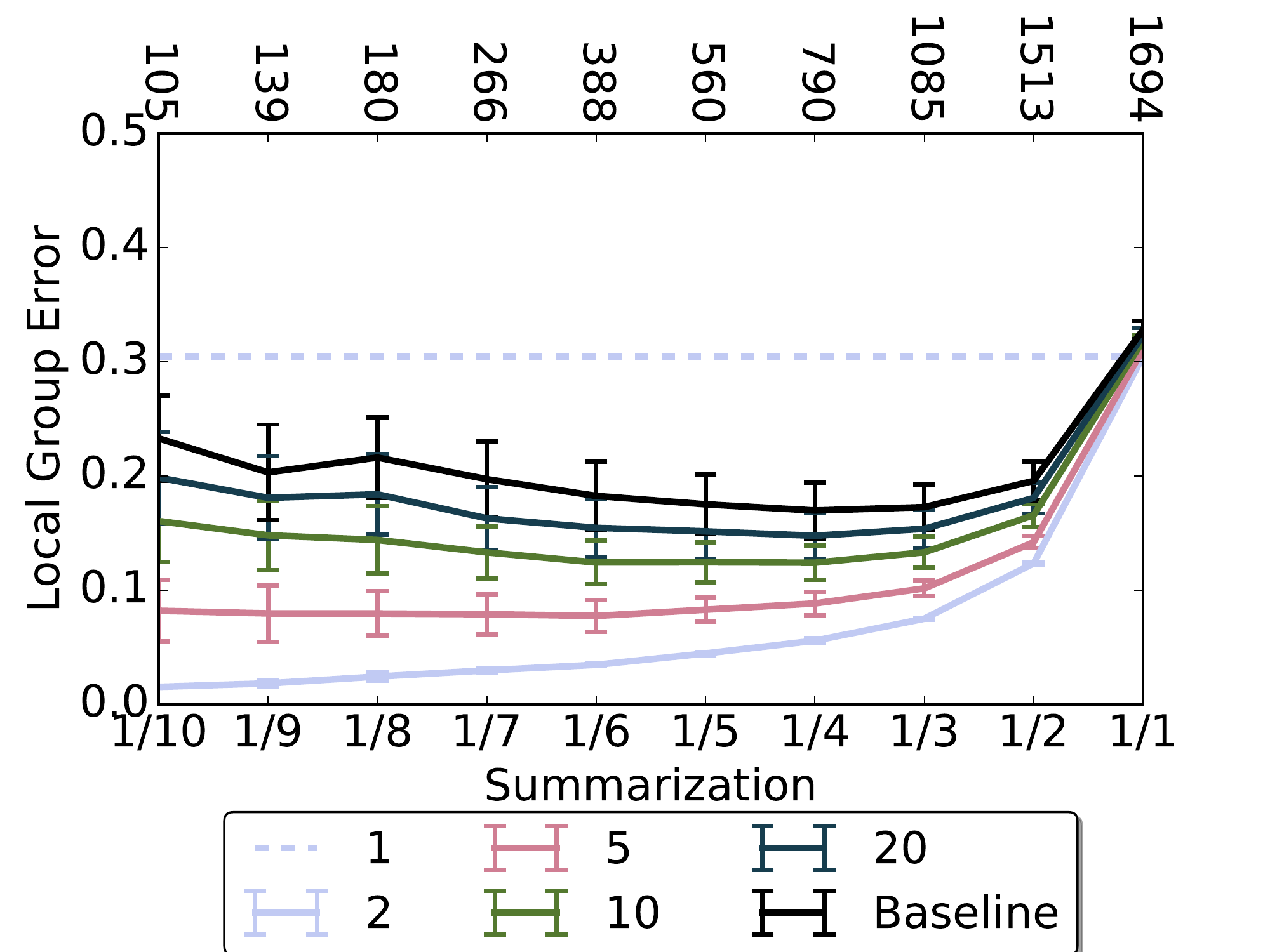}
\subcaption{NREL dataset}\label{fig:strategic_line_b}
\end{minipage}
\caption{Local group error for a given group size and a summarization level. The legend indicates the size of groups. Values above the dashed line (combinations of group size and summarization level) indicate a higher local group error compared to the the highest local group error in the baseline scenario.}\label{fig:strategic_line}
\end{figure}

The benefit of participating in a group compared to sharing data directly to the data consumer is studied in this section. An experiment is performed, in which all groups have the same size and the summarization level of all data suppliers is fixed. Figure \ref{fig:strategic} shows that the average local group error increases with increasing group size, and decreases with decreasing summarization level. Note that the first column in the plots represents the local group error of data suppliers in the baseline scenario, i.e. no groups. In this scenario, the privacy of data suppliers decreases with decreasing summarization level, but there is a group size after which the local group error, at each summarization level, is higher than the privacy value of a data supplier summarizing at level 1 (depicted by a dashed line in Figure \ref{fig:strategic_line_a}). In this case a data supplier benefits from being in a group to higher extent for a given group size, even if it is required to reduce the summarization level. For example, in the ECBT dataset, a data supplier summarizing at level 1 in the baseline scenario increases its local group error up to around 0.35, while within a group larger than 5 the local group error increases up to around 0.4, even when summarizing at level 1/10.

The same effect, though at a smaller scale, is observed in the NREL dataset (Figure \ref{fig:strategic_line_b}). Grouping does not increase the privacy of data suppliers above the privacy level they reach individually, although from the trend in the data it is reasonable to expect that this can potentially happen for larger groups\footnote{Recall that groups size in the NREL dataset cannot be increased above 20 as the number of data suppliers decreases with the summarization level (cf. Section \ref{sec:datasets}).}.

Moreover, Figure \ref{fig:strategic_line} shows that grouping has a positive effect on the individual privacy, regardless of the individual choices on summarization level: the larger the groups the higher the local group error, for any summarization level. This effect varies across datasets, but the trend holds.
In conclusion it is feasible to incentivize data suppliers to summarize at lower levels by means of the grouping mechanism, but the parameters of the incentive mechanism depend on the characteristics of the data and  can be chosen empirically.

\subsection{Total group error}

\begin{figure}[t!]
\begin{minipage}[b]{.5\linewidth}
\centering
\includegraphics[width=\linewidth]{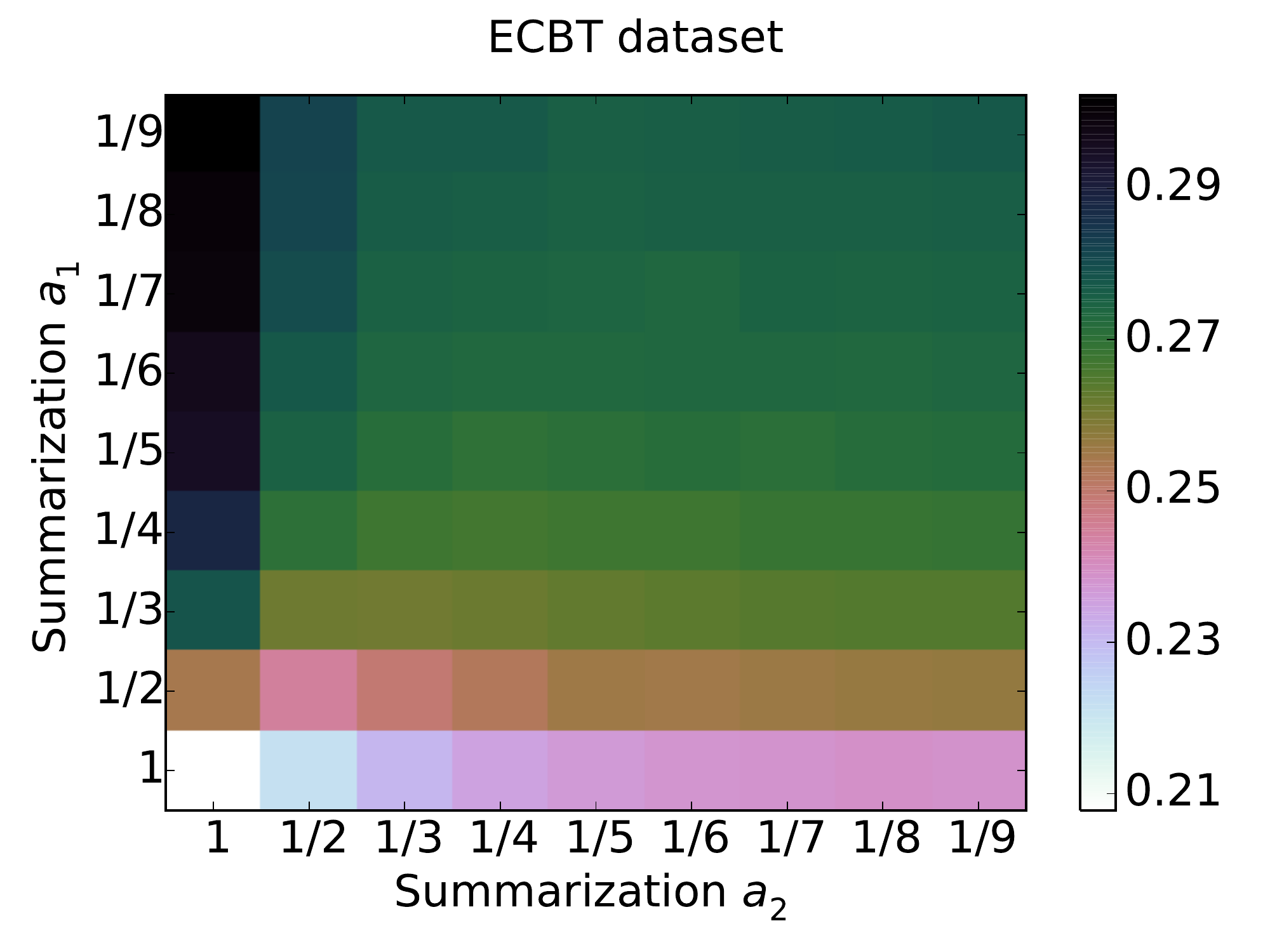}
\end{minipage}%
\begin{minipage}[b]{.5\linewidth}
\centering
\includegraphics[width=\linewidth]{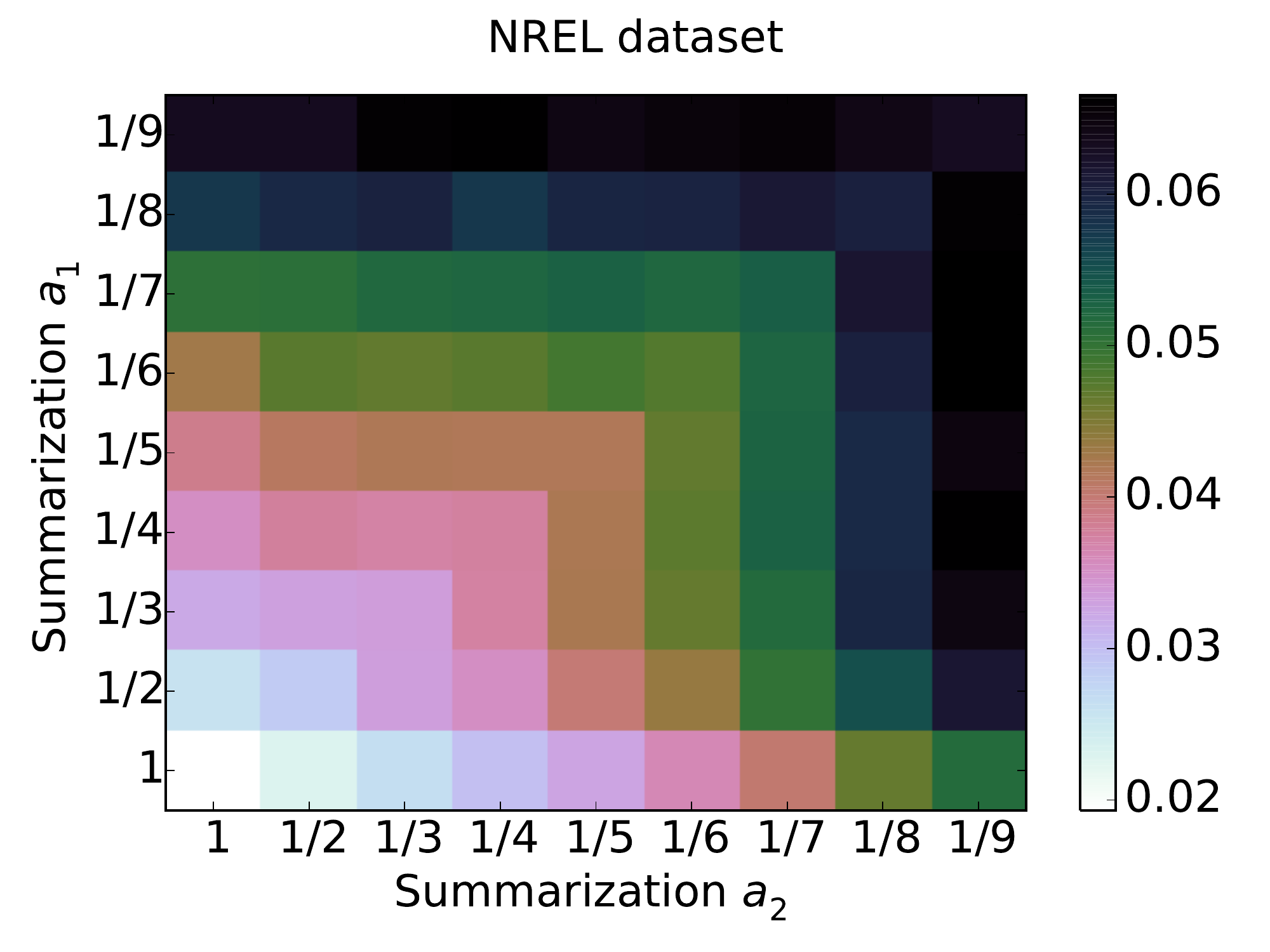}
\end{minipage}
\caption{Average total group error of $a_1$ across groups of size 2 for varying summarization levels. The y axis shows the summarization level of $a_1$, the x axis the summarization level of $a_2$.}\label{fig:g_lerr_a1}
\end{figure}

If the total group error is low, each data supplier's summarized data are similar to the data aggregated at the group-level, therefore being in a group does not improve privacy significantly. The measure of total group error can be interpreted as the efficiency of the grouping mechanism. In both datasets, the total group error increases when decreasing the individual summarization level (Figure \ref{fig:g_lerr_a1}), thus data suppliers who summarize at low levels have a higher incentive for grouping.

\subsection{Grouping strategies}\label{sec-10-2}

\begin{figure}[!htb]
  \begin{minipage}[t]{1.0\linewidth}
\begin{minipage}[b]{.5\linewidth}
\centering
\includegraphics[width=\linewidth]{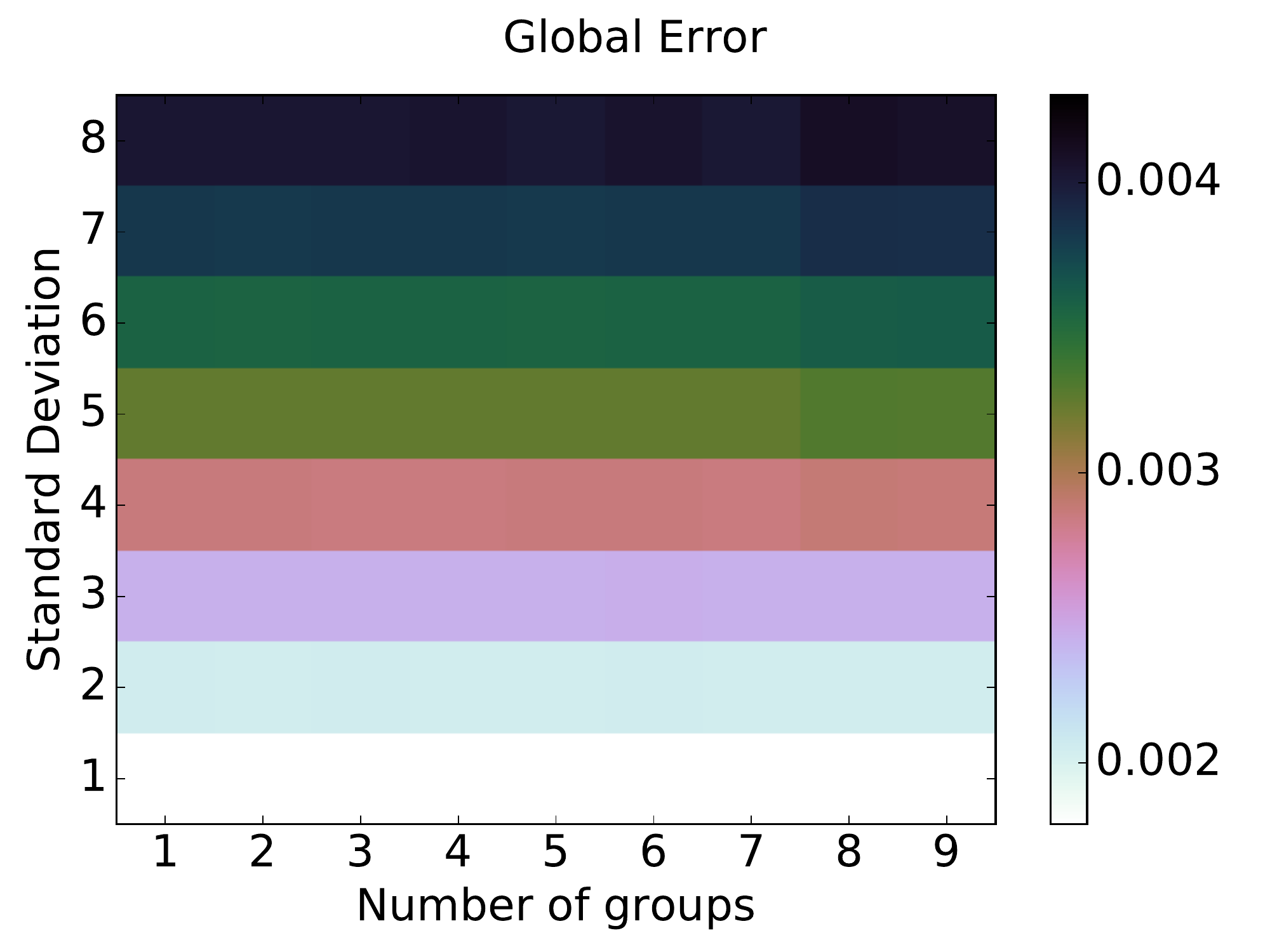}
\end{minipage}%
\begin{minipage}[b]{.5\linewidth}
\centering
\includegraphics[width=\linewidth]{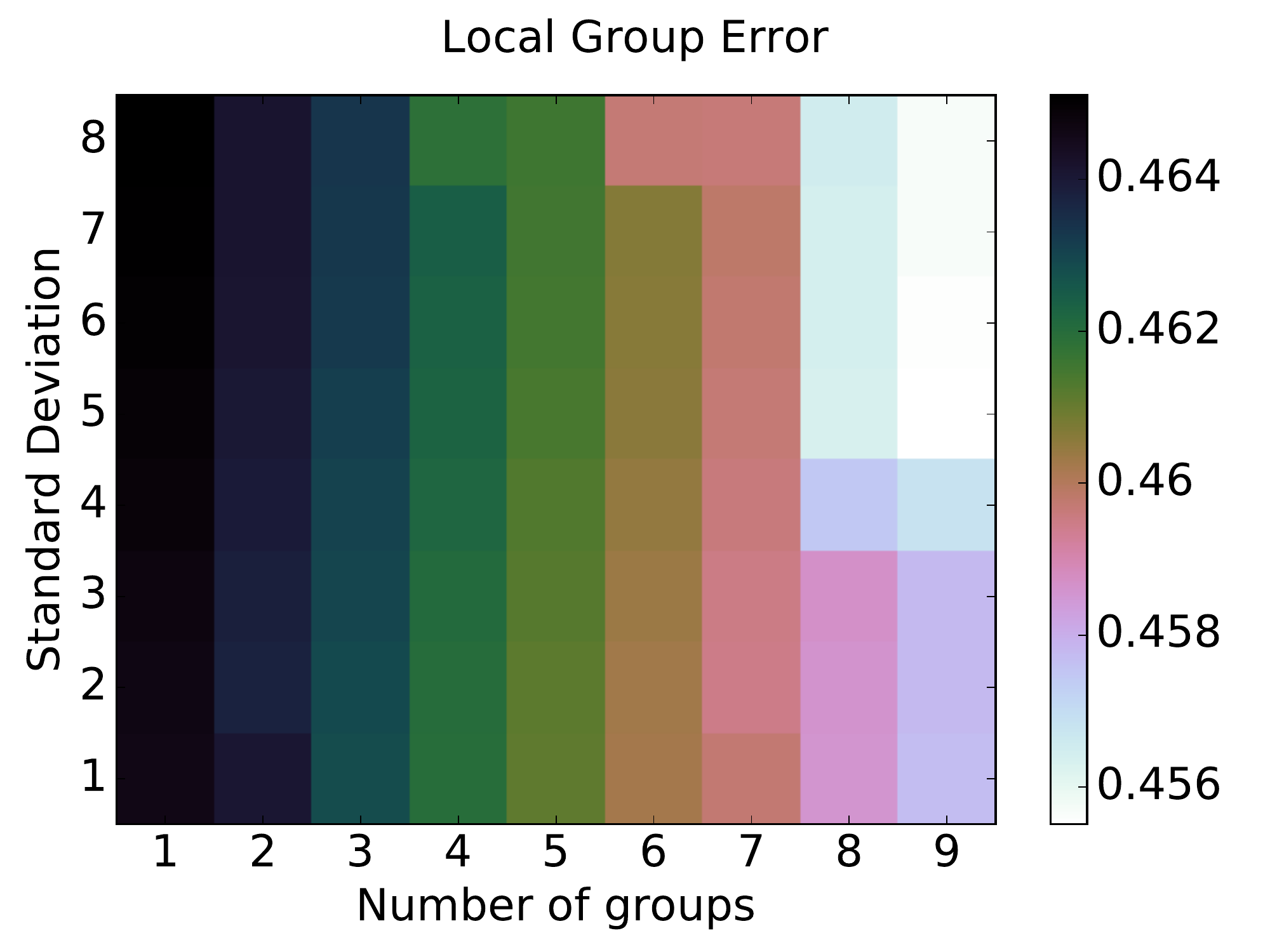}
\end{minipage}
\subcaption{Data suppliers are grouped randomly.}\label{fig:rnd}
  \end{minipage}
  \begin{minipage}[t]{1.0\linewidth}
    \begin{minipage}[b]{.5\linewidth}
\centering
\includegraphics[width=\linewidth]{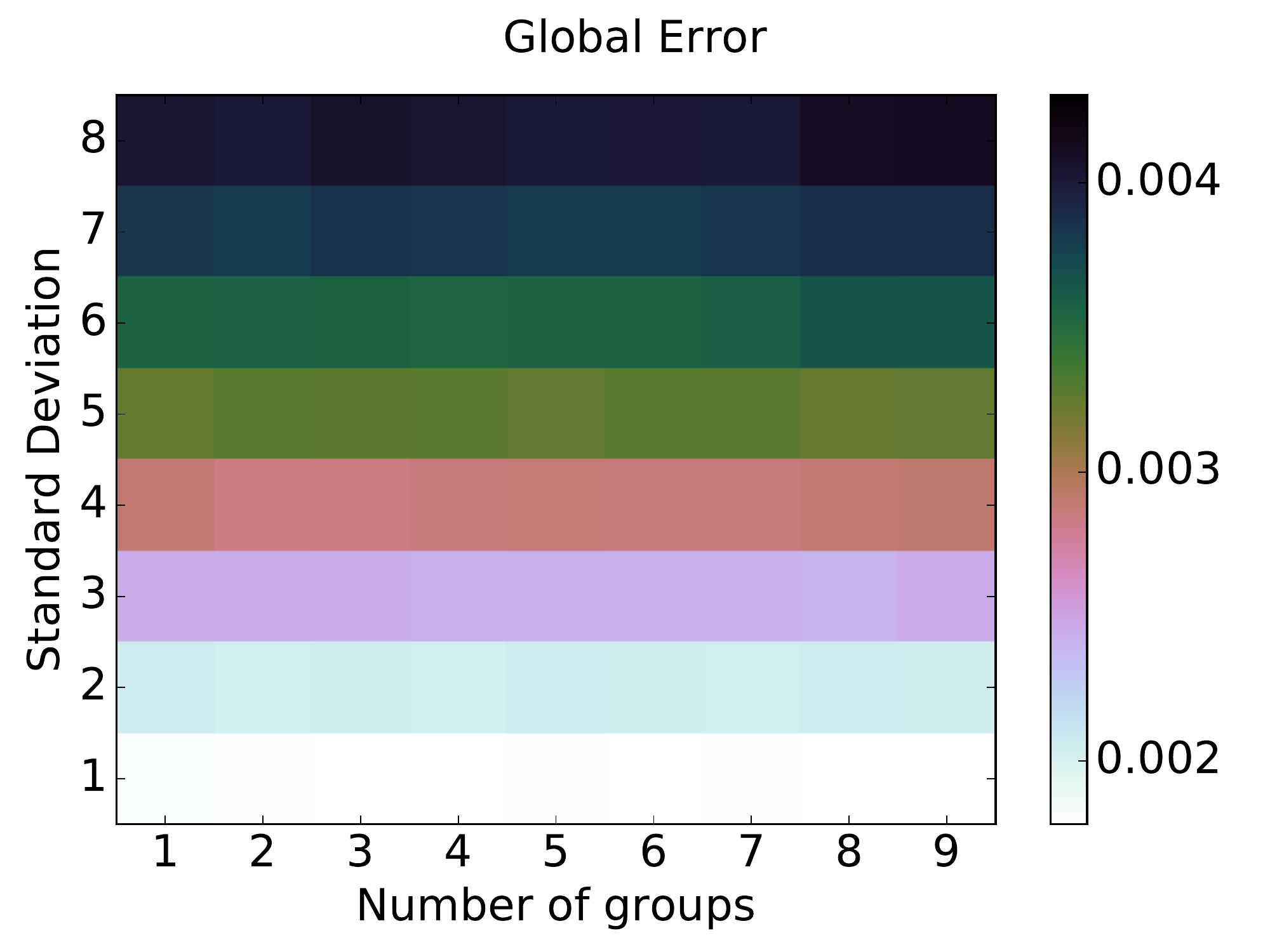}
\end{minipage}%
\begin{minipage}[b]{.5\linewidth}
\centering
\includegraphics[width=\linewidth]{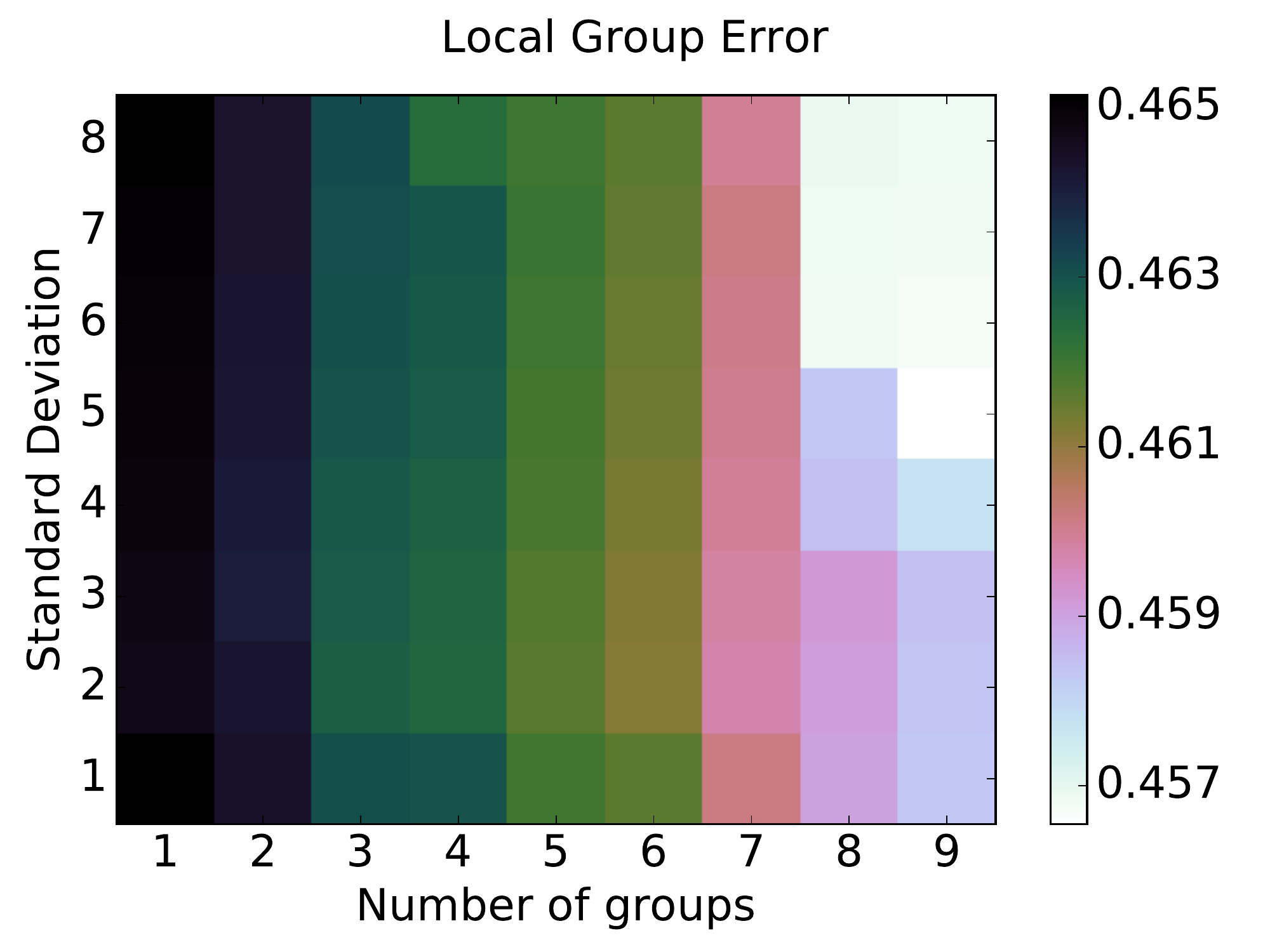}
\end{minipage}
\subcaption{Data suppliers are grouped according to the similarity of their raw data.}\label{fig:data}
  \end{minipage}
  \begin{minipage}[t]{1.0\linewidth}
    \begin{minipage}[b]{.5\linewidth}
\centering
\includegraphics[width=\linewidth]{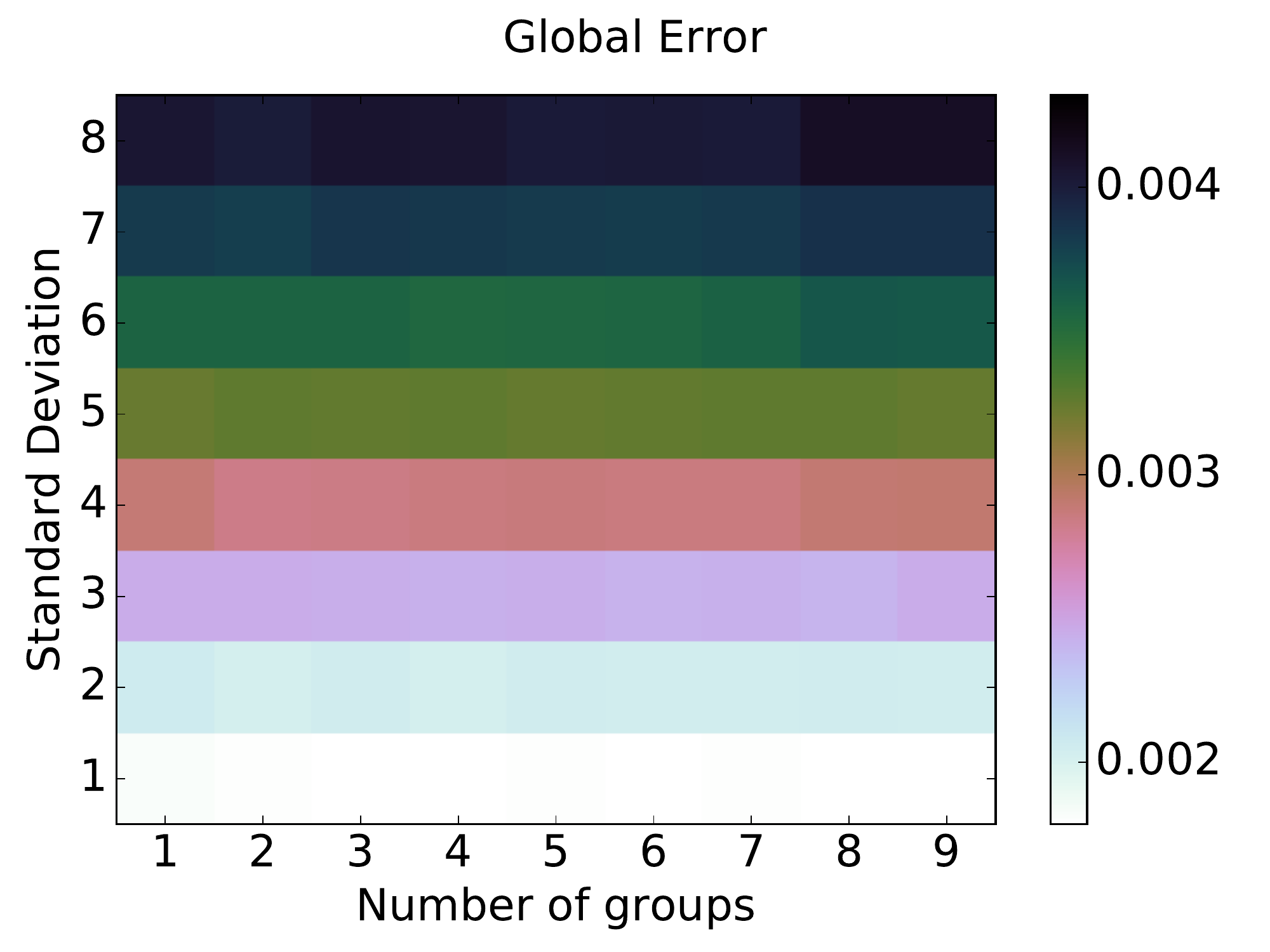}
\end{minipage}%
\begin{minipage}[b]{.5\linewidth}
\centering
\includegraphics[width=\linewidth]{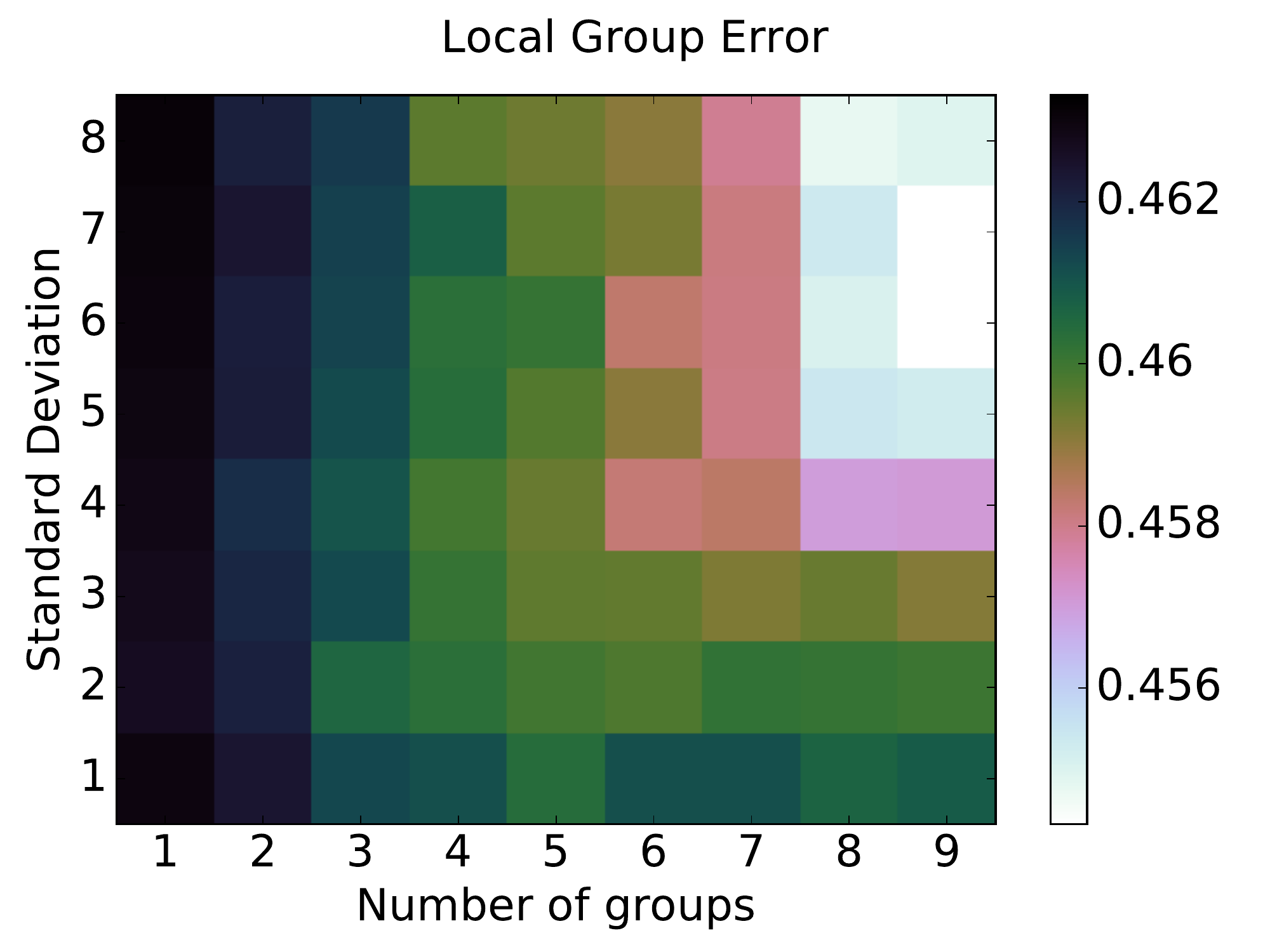}
\end{minipage}
\subcaption{Data suppliers are grouped according to the similarity of their summarization choice.}\label{fig:summ}
  \end{minipage}
\caption{Errors for a given number of groups and standard deviation of summarization levels. ECBT dataset.}
\end{figure}

This section studies the effect of different grouping strategies on the trade-off privacy vs. accuracy. The first finding is that, independently of the grouping strategy, accuracy decreases with an increasing standard deviation of group sizes, while it is not influenced by the number of groups (global error in Figures~\ref{fig:rnd},~\ref{fig:data},~\ref{fig:summ}). This result is consistent with previous results as it shows that the grouping strategies do not influence the global error (cf. Sections \ref{sec:res}). The effect of standard deviation on global error can be explained as follows: (i) The negative influence on accuracy by a data supplier that increases its summarization level is higher than the positive influence by a data supplier that decreases its summarization level (Figure \ref{fig:orig}). (ii) A higher value of standard deviation corresponds to more extreme summarization choices.

\begin{figure}[!t]
\centering
\includegraphics[width=\linewidth]{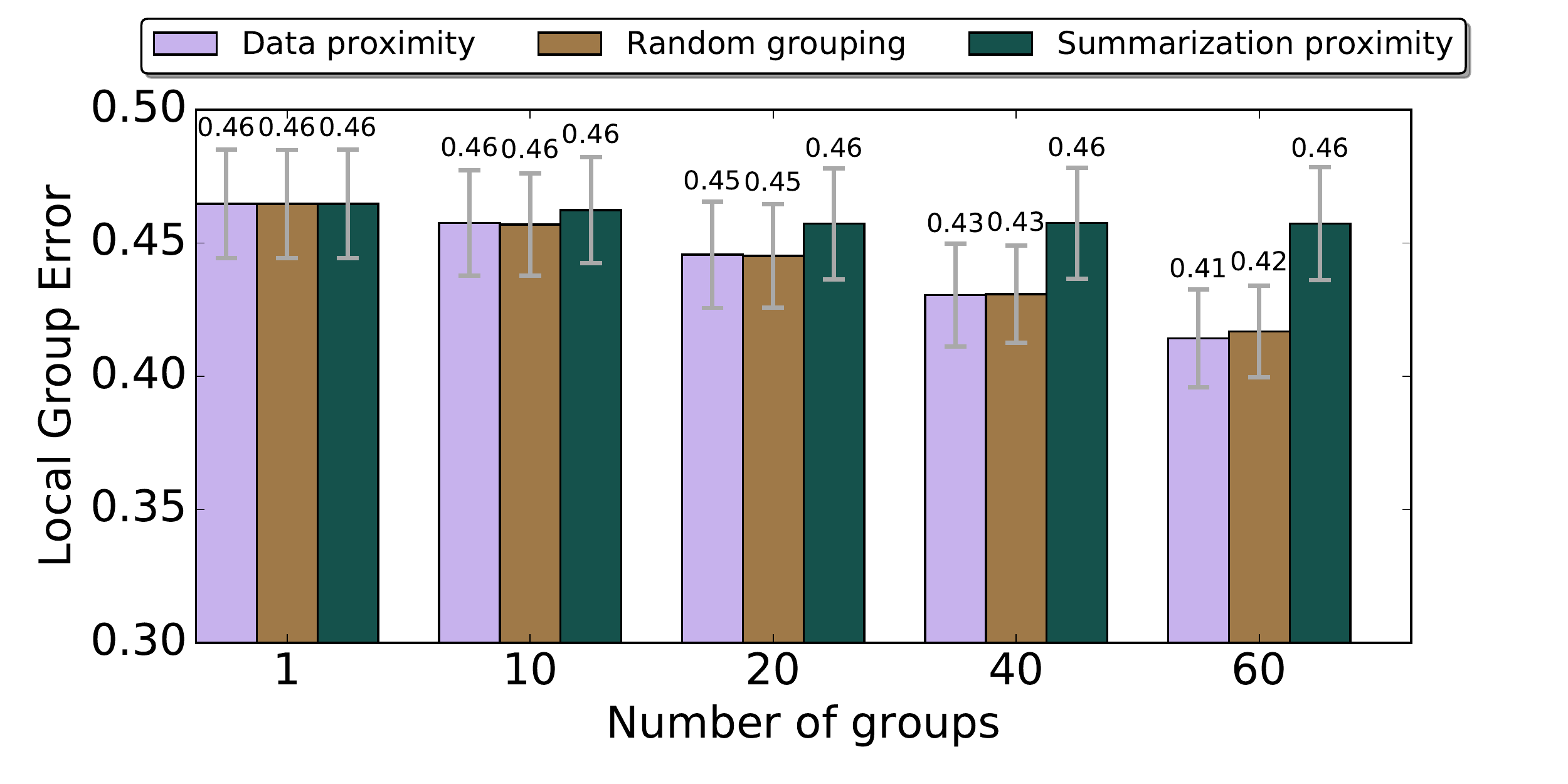}
\caption{\label{fig:stat_sig} Comparison of local group error for different grouping strategies, standard deviation of 2. Error bars represent the standard deviation across simulations. ECBT dataset.}
\end{figure}

In contrast, privacy shows a dependency on both the grouping mechanism and the summarization choices. There is no substantial difference between grouping randomly (Figure \ref{fig:rnd}) and grouping by data proximity (Figure \ref{fig:data}), while there is a difference when grouping by summarization proximity (Figure \ref{fig:summ}, for a high number of groups and low levels of standard deviation). This difference is investigated in more detail in Figure \ref{fig:stat_sig}, which shows the privacy for a standard deviation of 2. The data proximity strategy performs as the random, while the summarization proximity strategy results in a local group error of around 10\% higher than random for more than 60 groups. 

The illustrated results have several implications: (i) The summarization proximity strategy groups data suppliers with similar summarization levels together, and this promotes fairness within groups (cf. Figure \ref{fig:diff_lerr}). (ii) Grouping can be optimized using knowledge about the individual summarization choices. (iii) More fair incentive mechanisms can be designed if data suppliers are rewarded based on their summarization level and they are grouped by the summarization proximity strategy. Incentive mechanisms\footnote{The concept can be extended by punishing data suppliers in the less cooperative groups by preventing them from being in a group, thus reducing their privacy level. Further research on such an extension is part of future work.} that group data suppliers with the same criteria have been proven to support cooperation~\cite{gunnthorsdottir10_near_effic_equil_contr_based_compet_group,nax14_stabil_welfar_ofmer_based_match}.


\section{Discussion and Future Work}\label{sec:disc}

The experimental results illustrated in this paper show the following: (i) IoT-PGA increases the individual privacy of data suppliers without compromising system accuracy. (ii) The result of (i) holds across different summarization levels, numbers of groups, groups sizes, and distributions of group sizes, which suggests that local aggregation is independent of groups sizes and composition. (iii) Larger groups improve privacy. (iv) The same trade-off between privacy and accuracy in the baseline scenario is also relevant within groups: Uniform levels of summarization promote fair treatment, as the difference in privacy among group members is minimized if all members summarize at the same level. (v) An increase in summarization choices increases the privacy and decreases the accuracy. An incentive system is required to guarantee a certain level of accuracy. (vi) Data suppliers gain more privacy by being in a group, independently of their summarization level, compared to sharing data directly to the data consumer. This result holds if the size of groups is larger, or the number of groups is lower, than a threshold, which depends on the dataset. (vii) Grouping data suppliers by the similarity of the chosen summarization levels increases privacy by approximately 10\% over random grouping. (viii) The choice of grouping strategy does not influence the accuracy of the system.

These results have implications on system design choices: (i) Data suppliers belonging to groups can reduce their summarization level in exchange for (monetary) rewards. Their privacy is higher for each summarization level compared to the baseline scenario. (ii) Group membership itself can be a reward, if groups are large enough, as the privacy level within the group is higher than the privacy level outside the group, at any summarization level. (iii) Grouping according to the summarization choices promotes fair treatment within the groups and delivers an increase in privacy by 10\% compared to random grouping.

A citizen may join a group using the (Internet) connectivity offered by the IoT devices, for the following reasons: (i) Improve the quality of service, i.e. accuracy, by sharing data without compromising privacy. (ii) Improve privacy without degrading the service quality. (iii) Allow other citizens to improve their own privacy, even if the privacy of this citizen is not a concern. (iv) Improve privacy without other concerns.

Formal investigation of privacy guarantees, i.e. anonymity~ \cite{domingo-ferrer05_ordin_contin_heter_k_anony_throug_microag,soria-comas15_t_closen_throug_microag} and differential privacy, is outside of the scope of this paper. Instead, this paper focuses on the empirical evaluation of the IoT-PGA with real-world data from two smart city pilot projects. Future work towards the direction of more formal privacy guarantees concerns the integration of micro-aggregation~\cite{defays98_maskin_microd_using_micro_aggreg,fayyoumi10_survey_statis_discl_contr_micro} and differential privacy mechanisms~\cite{eibl2017differential,kellaris13_pract_differ_privac_via_group_smoot} in IoT-PGA. Moreover, grouping strategies that encode trust models~\cite{Shaikh2009} and security against malicious attackers that eavesdrop data~\cite{medaglia10_overv_privac_secur_issues_inter_thing} are also subject of future work.


\section{Related Work}\label{sec:related_work}

Privacy-preserving mechanisms that operate on groups improve fault-tolerance \cite{chan12_privac_preser_stream_aggreg_fault_toler} and allow more distributed and privacy-preserving computations \cite{eibl2017differential}. The concept of group is also relevant in Big Data analytics \cite{jain16_big_data_privac} and the Map-Reduce paradigm. Privacy preservation is performed at each node (mapper) in order to provide privacy guarantees on the local data \cite{xu15_privac_preser_machin_learn_algor}. The Big Data scenario differs from the collective sensing scenario in that each node operates on a large database, as opposed to local data records, on which standard anonymization techniques can be applied. Nevertheless, mechanisms designed to run on databases could be deployed on a group of IoT devices.

The design of IoT-PGA draws parallels to micro-aggregation~\cite{defays98_maskin_microd_using_micro_aggreg,fayyoumi10_survey_statis_discl_contr_micro}, however there are several differences: (i) In a collective sensing scenario, queries target macro-level data\footnote{Note that privacy concerns are about microdata, which could reveal privacy-sensitive information about the data suppliers.} i.e. estimators of the population characteristics, while in micro-aggregation, queries target micro-level data i.e. individual characteristics~\cite{ciriani07_microd}. (ii) Micro-aggregation is mainly an anonymization technique, while the collective sensing scenario studied focuses on obfuscation techniques. (iii) Micro-aggregation requires a grouping mechanism that maximizes inter-group data homogeneity, while IoT-PGA\footnote{If groups are at least of size $k$ and members have similar data, IoT-PGA fits the requirements of micro-aggregation.} does not require this.

Secure multiparty computation encrypts communication~\cite{yao1982protocols,Clifton2002tools}, e.g. with homomorphic encryption, which allows for performing mathematical operations on encrypted data \cite{gentry09}.
However, such techniques are computationally expensive and cannot easily satisfy high performance requirements in resource-constraint networks running Internet of Things applications~\cite{jain16_big_data_privac}.

Anonymization breaks the link between the data and the identity of the owner~\cite{sweeney02_k_anony}, e.g. mix zones, in which pseudonyms are exchanged between users in a certain spatial region \cite{beresford04_mix}. Both of these approaches are centralized, thus require a trusted management system, e.g for cryptographic key distribution or anonymization~\cite{gedik2005location,gruteser2003anonymous,Marmol2013}, as well as potential changes in the aggregation algorithm. Obfuscation can instead be adopted by individual users~\cite{duckham05_formal_model_obfus_negot_locat_privac,ardagna11_obfus_based_approac_protec_locat_privac} e.g. to reduce data granularity~\cite{shokri11_quant_locat_privac,pournaras16_self_regul_infor_sharin_partic_social_sensin} or introduce perturbations~\cite{dwork2011differential}.

When such management systems are not available or too costly to employ, obfuscation is an alternative for citizens to adopt e.g. to reduce data granularity or introduce perturbations.

Several attacks have been developed to deanonymize data by exploiting auxiliary information:
(i) episodic observation of individual behavior \cite{ma13_privac_vulner_publis_anony_mobil_traces,srivatsa12_deanon,gambs14_de_anony_attac_geoloc_data}, (ii) contextual information about the users \cite{zang11_anony,narayanan2008robust}, which allows to draw links between anonymized records and an external database~\cite{lane2012feasibility}, or (iii) statistical properties of the data, e.g. regularities in mobility traces~\cite{tsoukaneri2016inference}, even if the data are obfuscated \cite{montjoye13_unique_crowd}. Therefore, measures such as \emph{k}-anonymity~\cite{samarati01_protec_respon_ident_microd_releas} or entropy~\cite{beresford03_locat_privac_pervas_comput} do not always preserve privacy in the system.
In response to this finding, the privacy measure of \emph{expected distance error} is introduced~\cite{shokri11_quant_locat_privac}. The measures of privacy used in the present work (as well as those in \cite{pournaras16_self_regul_infor_sharin_partic_social_sensin}) are derived from the measure of the expected distance error, in which the attacker does not apply any inference function to the observed data.


\section{Conclusions}\label{sec:concl}

Previous work identified several privacy-preserving mechanisms that increase privacy by degrading the quality of shared data with several techniques discussed. This work contributes a  new opportunity to enhance the privacy of all these mechanisms:
adaptation of the network organization as a new means for increasing privacy in the Internet of Things.

A new bottom-up privacy-preserving mechanism for data aggregation, referred to as IoT-PGA, is devised and evaluated, in which data suppliers are grouped and perform group-level aggregation. New privacy and accuracy metrics are defined and used for the evaluation in the grouping scenario. Privacy and accuracy of IoT-PGA are measured using real-world data from two smart city pilot projects to evaluate its general applicability. The mechanism is found to increase the privacy of individuals, without degrading the accuracy of aggregation for several different experimental and parameter settings.

Moreover, trade-offs between privacy and aggregation accuracy are studied in the context of privacy-enhancing network grouping. Different grouping strategies are designed and experimentally evaluated. When groups are formed based on the proximity of users' choices on the summarization level (summarization proximity strategy), the highest increase in privacy is achieved, approximately 10\% above random grouping.

IoT-PGA is relevant for smart city pilot projects in energy management, traffic management, and other application scenarios, as it reduces the privacy cost for citizens to contribute data. In this context, IoT-PGA can encourage participation to smart city initiatives and ultimately the sustainability of the nowadays digital society.


\section{Acknowledgements}

The authors acknowledge support by the European Commission through the ERC Advanced Investigator Grant 'Momentum' [Grant No. 324247] and by the European Community’s H2020 Program under the scheme ‘ICT-10-2015 RIA’ [Grant No. 688364] ‘ASSET: Instant Gratification for Collective Awareness and Sustainable Consumerism’.


\section{References}
\label{sec-8}
\bibliographystyle{apalike}
\bibliography{references}

\appendix
\renewcommand{\thesection}{\Roman{section}}
\label{sec-10}
\section{Privacy and Accuracy Measures}\label{sec:model:corr}

Earlier work introduces measures of privacy and accuracy~\cite{pournaras16_self_regul_infor_sharin_partic_social_sensin}.
The local error, measuring privacy, is defined as follows:
\begin{equation}
\epsilon_{e,t}=\frac{1}{n}\sum_{i=1}^n \epsilon_{i, e,t}; ~ \epsilon_{i, e,t}=\frac{|r_{i,e,t}-s_{i,e,t}|}{|r_{i,e,t}|},
\end{equation}
\\ where each term is the difference between the raw and summarized data of supplier $i$. The global error, measuring accuracy, is defined as:
\begin{equation}
\varepsilon_{e,t}=\frac{|\alpha(R_{e,t})-\alpha(S_{e,t})|}{|\alpha(R_{e,t})|},
\end{equation}
\\ which is the average difference between the raw data $R_{e,t}=(r_{i,e,t})_{i=1}^n$ and the summarized data $S_{e,t}=(s_{i,e,t})_{i=1}^n$ collected by the data consumer.

Note that the local error is the average of the individual errors, while the global error is the error of the aggregates. Both measures are normalized on the raw data, in order to make them comparable.
Also note that the global error depends on the aggregation function, but it does not change when considering the mean and the sum as aggregation functions.
In both cases the global error is specified as follows:

\begin{equation}
\varepsilon_{e,t}=\frac{|\sum_{i=1}^n r_{i,e,t}-\sum_{i=1}^n s_{i,e,t}|}{|\sum_{i=1}^n r_{i,e,t}|},
\end{equation}
\\ as the mean divides both numerator and denominator by the same quantity $n$.

The measures of global and local error resemble the Mean Absolute Percentage Error (MAPE), and similarly suffer from the same drawbacks: they have no upper limit and they are not defined if there are zero values.
These limitations are resolved by introducing a symmetric version of the measures, based on the symmetric MAPE (sMAPE):
\begin{equation}
\epsilon_{e,t}=\frac{1}{n}\sum_{i=1}^n \epsilon_{i, e,t}; ~ \epsilon_{i, e,t}=\frac{|r_{i,e,t}-s_{i,e,t}|}{|r_{i,e,t}|+|s_{i,e,t}|},
\label{eq:sym_le}
\end{equation}
\\ which defines the symmetric local error, while the symmetric global error is defined as
\begin{equation}
\varepsilon_{e,t}=\frac{|\alpha(R_{e,t})-\alpha(S_{e,t})|}{|\alpha(R_{e,t})|+|\alpha(S_{e,t})|},
\label{eq:sym_ge}
\end{equation}
\\ where $R_{e,t}=(r_{i,e,t})_{t=1}^n$ and $S_{e,t}=(s_{i,e,t})_{i=1}^n$.

\begin{figure}[t!]
\centering
\includegraphics[width=.9\linewidth]{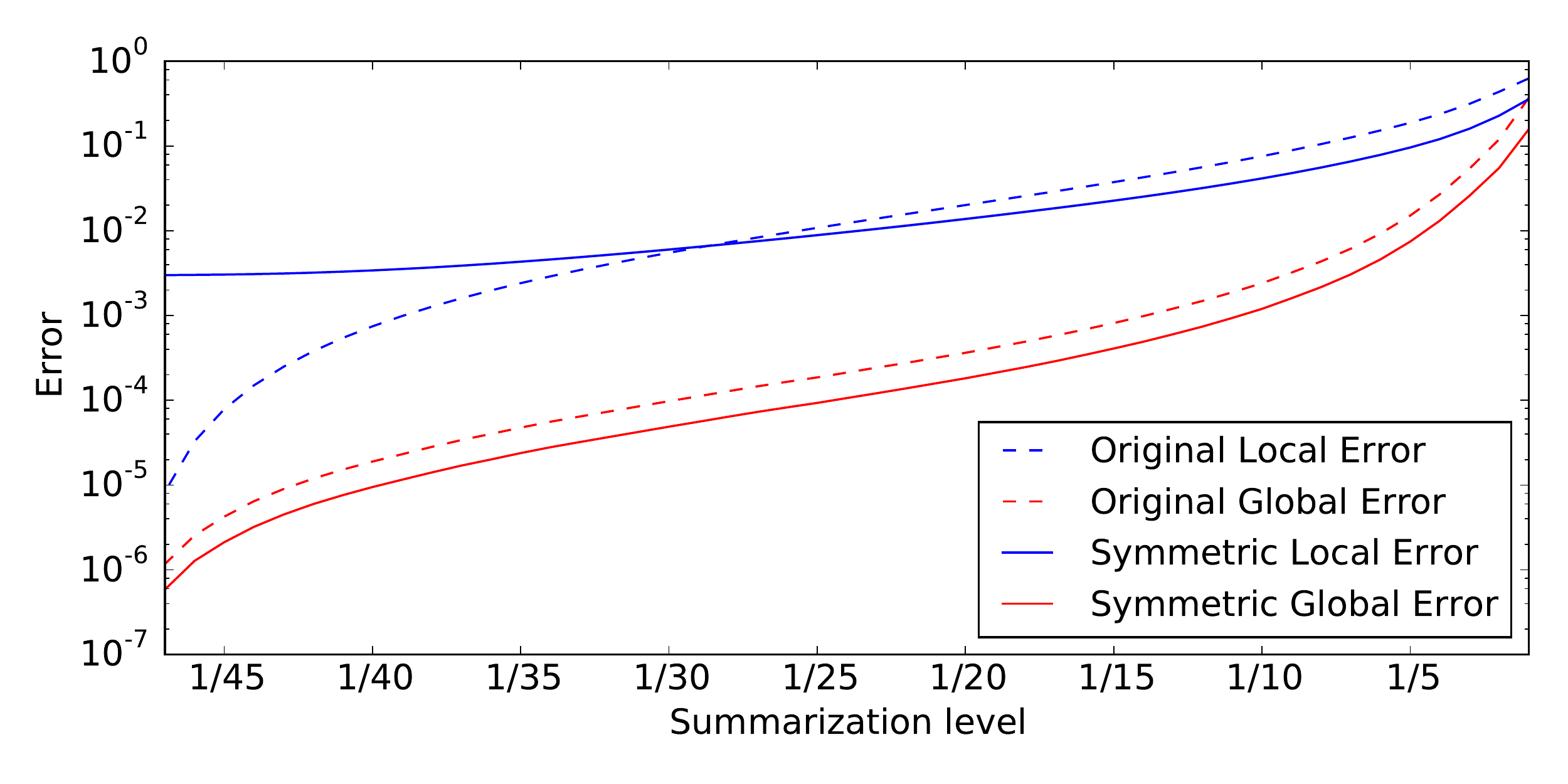}
\caption{\label{fig:orig}Comparison between the original measures, see \cite{pournaras16_self_regul_infor_sharin_partic_social_sensin}, and the symmetric measures.}
\end{figure}

\noindent This choice is validated by comparing the symmetric measures with the original measures on the ECBT dataset (Figure \ref{fig:orig}). The measures are qualitatively similar for summarization levels higher than 1/20, so in this region the two measures can be used interchangeably.

\begin{figure}[!htb]
\begin{minipage}[b]{\linewidth}
\centering
\includegraphics[width=.9\linewidth]{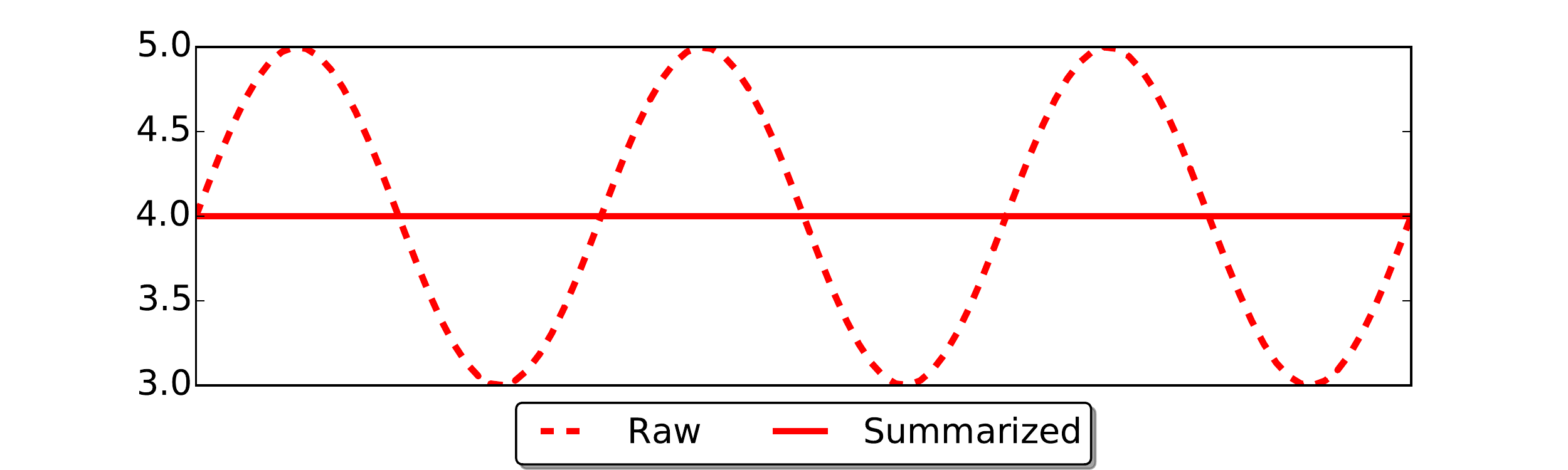}
\subcaption{Data of supplier $a_1$.}\label{fig:pc_sub1}
\end{minipage}
\begin{minipage}[b]{\linewidth}
\centering
\includegraphics[width=.9\linewidth]{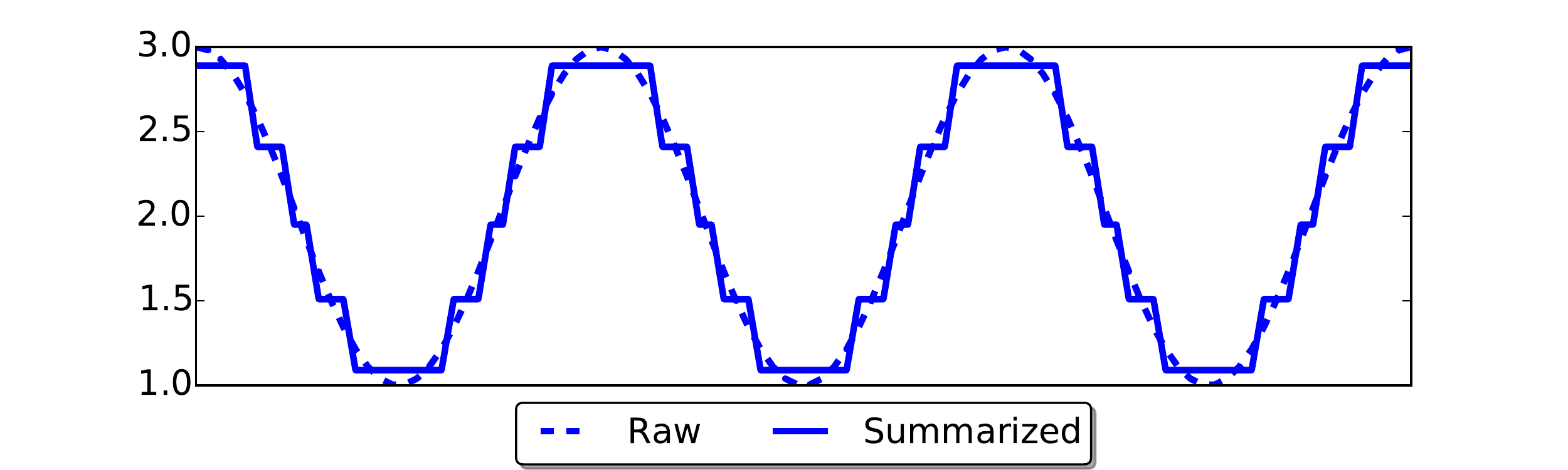}
\subcaption{Data of supplier $a_2$.}\label{fig:pc_sub2}
\end{minipage}
\begin{minipage}[b]{\linewidth}
\centering
\includegraphics[width=.9\linewidth]{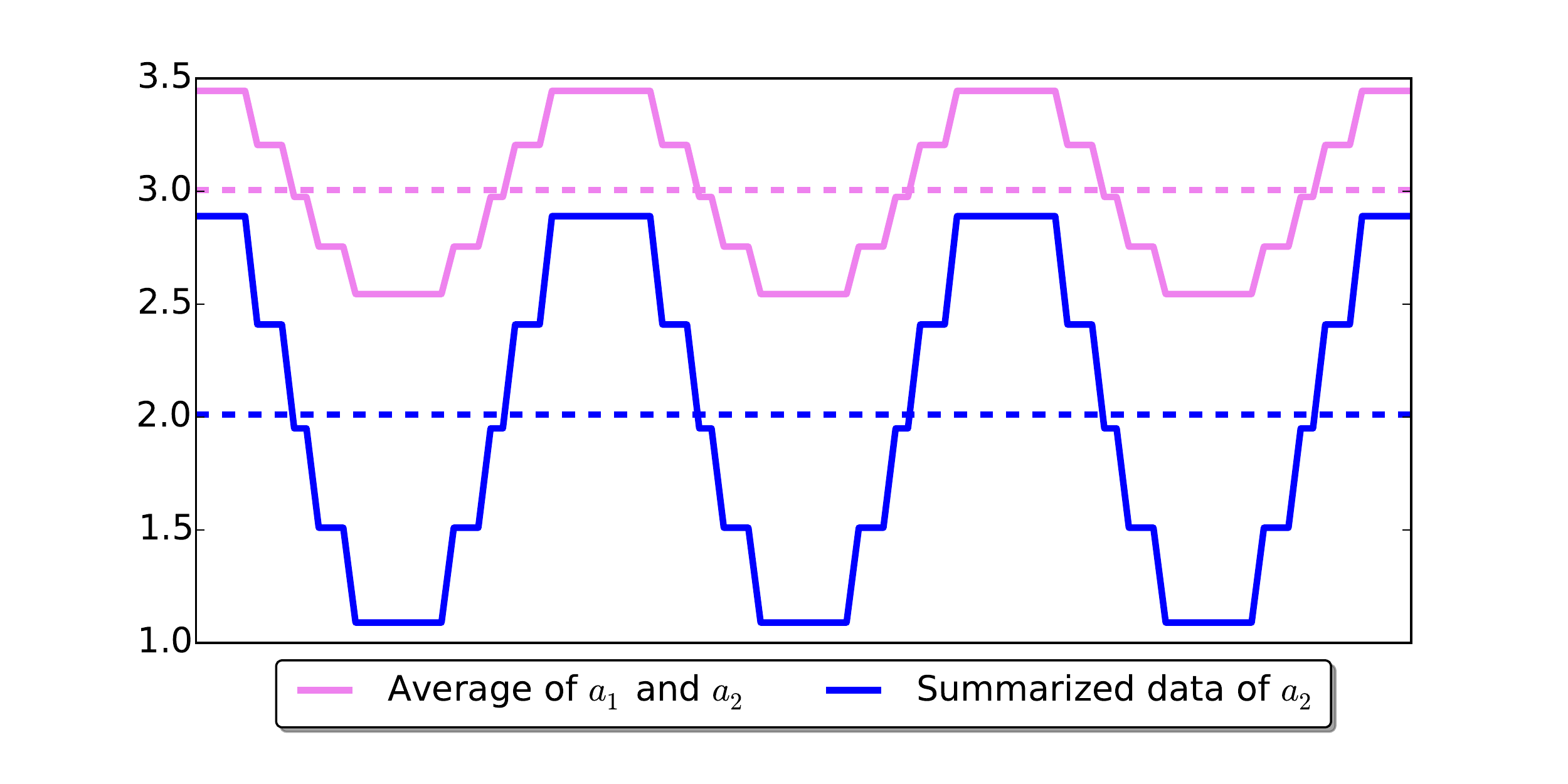}
\subcaption{Average of the summarized data of $a_1$ and $a_2$.}\label{fig:pc_sub5}
\end{minipage}
\begin{minipage}[b]{\linewidth}
\centering
\includegraphics[width=.9\linewidth]{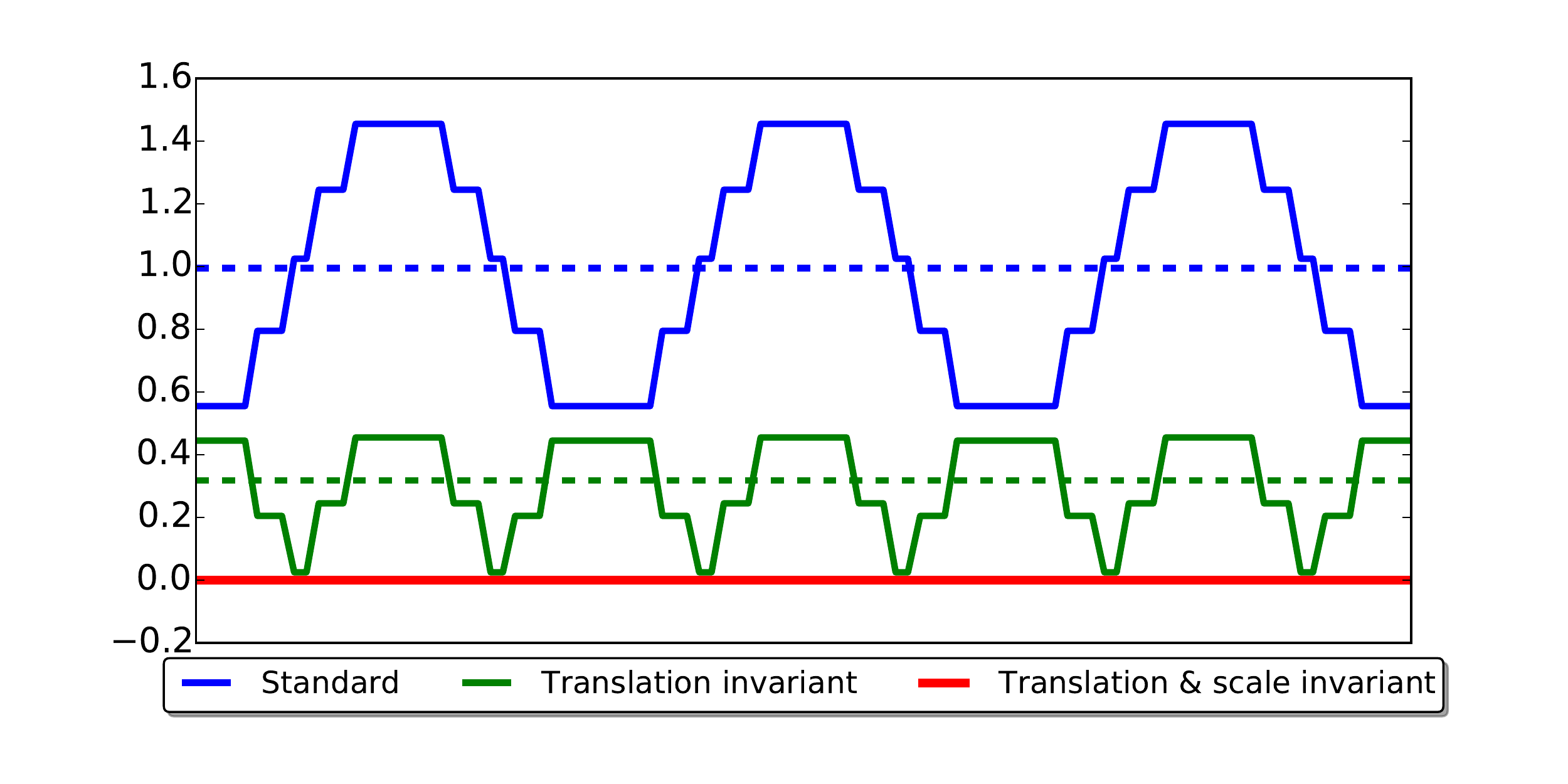}
\subcaption{Comparison of different ways of computing the error. One of the errors is constant and zero. Dotted lines represent the error means.}\label{fig:pc_sub6}
\end{minipage}
\caption{Example explaining the properties of translation- and scale-invariance for an error measure.}\label{fig:pc}
\end{figure}

A translation- and scale-invariant error measure considers the similarity between the shapes of the signals, as opposed to the similarity between individual points.
In many situation the shape of a signal can reveal as much privacy-sensitive information as the numerical values of its points, for example a sudden drop in residential energy consumption can reveal that the tenants are away.
In the group setting, the shape of the signal becomes relevant whenever a group member adopts a very high summarization and the others do not:
Assume groups have size two, so the difference between the data of one supplier and the group data is precisely the data of the other member.
Assume the first group member $a_1$, chooses a summarization level of one, thus producing a constant output greater than zero (Figure \ref{fig:pc_sub1}), and the second member $a_2$, chooses a lower summarization level (Figure \ref{fig:pc_sub2}).
In this case, the group data, the average of the member's individual data, has the same shape as the second supplier's data and a mean value equivalent to the average of the mean values (Figure \ref{fig:pc_sub5}).

In the example in Figure \ref{fig:pc} the signals have different means. A standard measure, which is neither scale nor translation invariant, indicates an error proportional to the difference of the means, because it considers the individual points (Figure \ref{fig:pc_sub6}).
The property of translation-invariance makes the measure robust against these translations: the error is computed as if the original data have zero mean.
The error is still greater than zero because the signal of $a_2$ has been rescaled during the averaging: its amplitude is reduced.
The property of scale-invariance makes a measure robust to changes in the amplitude of a signal, the error is computed on the similarity between the shapes of the signals.
The error computed by a translation- and scale-invariant measure is zero.

To address these concerns, a new error measure is required in the group setting, to compare the group-aggregated data and the data of an individual supplier: the Pearson product-moment correlation coefficient $p$ (cf. Equation \ref{eq:pearson}), which measures the correlation between two vectors, and it is both translation- and scale-invariant.

\begin{figure}[!htb]
  \centering
  \begin{equation}
p_{i,e}=\frac{\sum_{t=1}^{T_e}(r_{i,e,t}-\bar{r}_{i,e})(\alpha_{e,t}^G-\bar{\alpha}_{e}^G)}{\sqrt{\sum_{t=1}^{T_e}(r_{i,e,t}-\bar{r}_{i,e})^2}\sqrt{\sum_{t=1}^{T_e}(\alpha_{e,t}^G-\bar{\alpha}_{e}^G)^2}} 
\label{eq:pearson}
  \end{equation}
\end{figure}

\noindent The Pearson coefficient is a measure of similarity. An error measure, denoted as \emph{privacy-correlation}, is derived from the Pearson coefficient and defined as $C_{i,e}=1-p_{i,e}$.
The value of $C$ decreases with an increasing similarity between the vectors, so a higher value of $C$ indicates higher privacy-preservation.

\begin{figure}[!htb]
\begin{minipage}[b]{.5\linewidth}
\centering
\includegraphics[width=\linewidth]{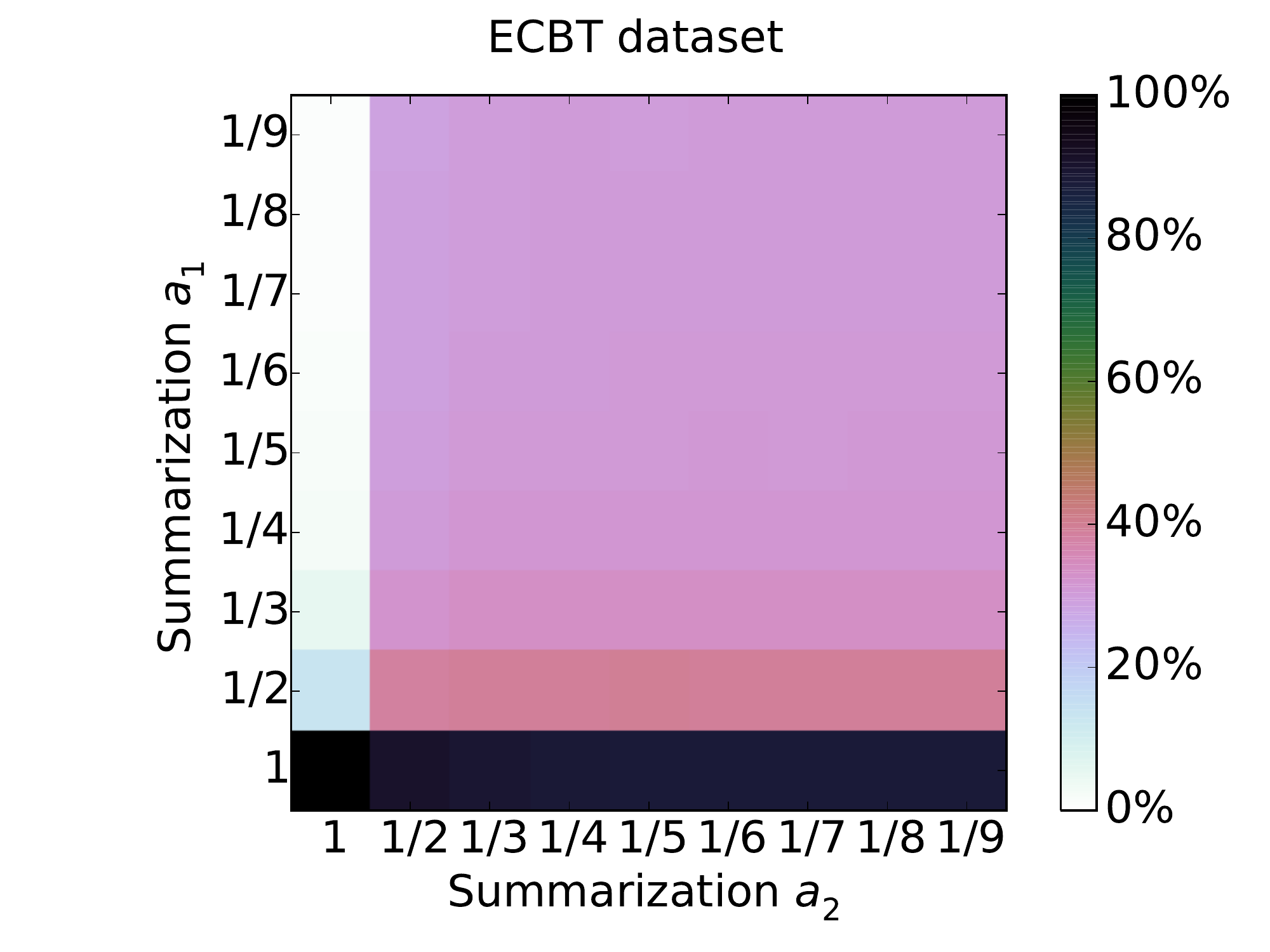}
\subcaption{ECBT dataset}\label{fig:corr_a1_a}
\end{minipage}%
\begin{minipage}[b]{.5\linewidth}
\centering
\includegraphics[width=\linewidth]{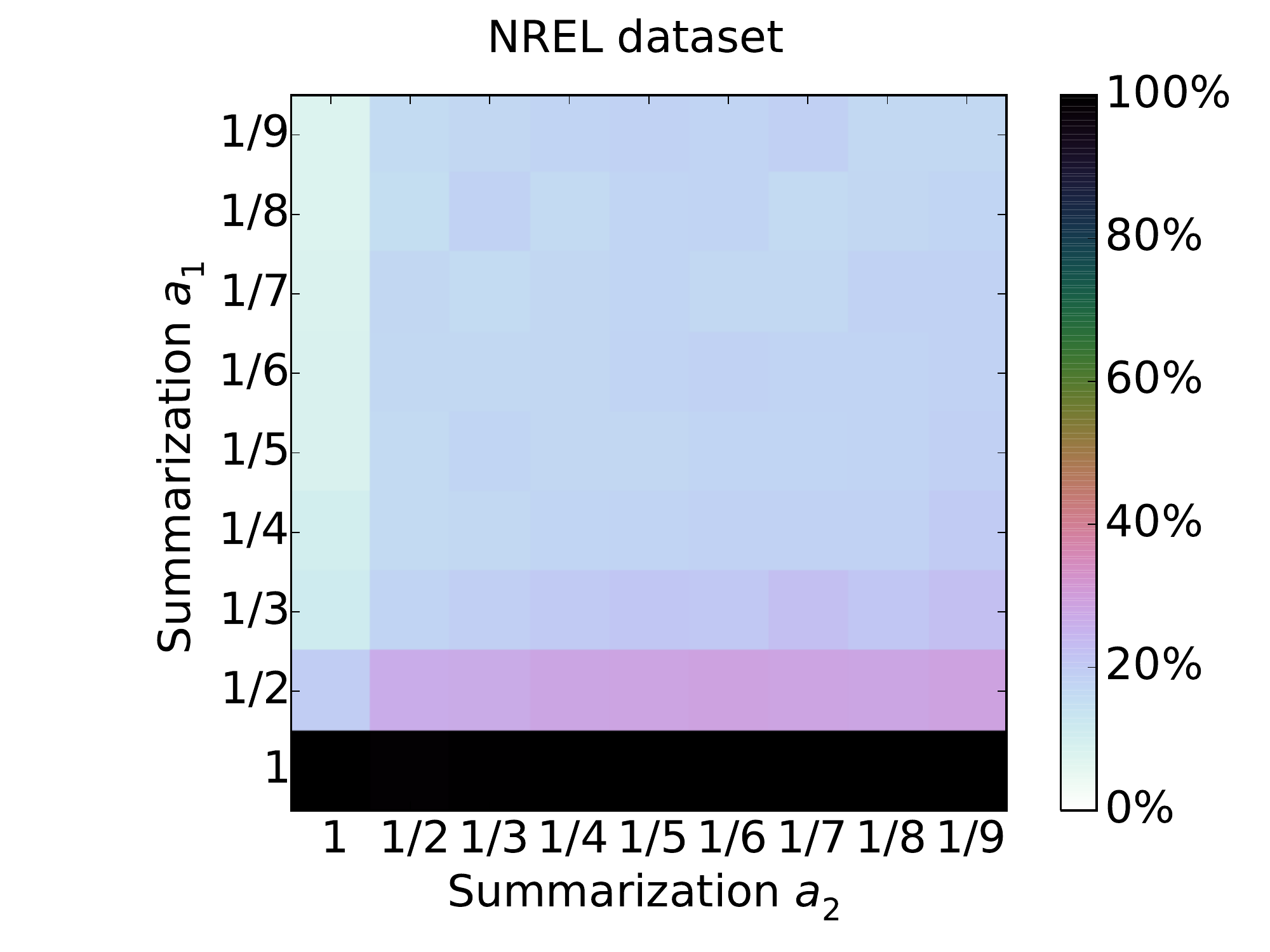}
\subcaption{NREL dataset}\label{fig:corr_a1_b}
\end{minipage}
\caption{Privacy-correlation between the summarized data of $a_1$ and the group data (summarized) for varying summarization levels. The y axis shows the summarization level of $a_1$, the x axis the summarization level of $a_2$.}\label{fig:corr_a1}
\end{figure}

Looking in more detail into the previous results, two observations become apparent: In the NREL dataset (Figure \ref{fig:lerr_a1}) the privacy of $a_1$ increases slightly for lower summarization levels of $a_2$, this is probably an artifact of the reduced number of data points at high summarization levels (cf. Figure \ref{fig:data_counts}).
In the ECBT dataset (Figure \ref{fig:lerr_a1}) the privacy of $a_1$ increases if $a_2$ summarizes with a summarization level of one, which is a counterintuitive artifact of the privacy measure.
If one of the two members of a group summarizes with a summarization level of one, the group data is the average between this constant and the data of the other supplier.
Averaging a signal with a constant modifies the scale and the offset of a signal, but does not vary its shape, which could reveal highly privacy sensitive information (cf. Figure \ref{fig:pc}).

The measure of privacy-correlation behaves as expected in the case the summarized data of a supplier is a constant: if $a_2$ chooses a summarization level of one (first column of Figure \ref{fig:corr_a1}) the privacy-correlation between the data of $a_1$ and the group data is minimal.
Assuming that neither data suppliers choose a summarization level of one, the results produced by the privacy-correlation measure are similar to those produced by the standard privacy measure (Figure \ref{fig:lerr_a1}), thus the standard measure gives valid results for summarization levels other than one.


\section{Empirical group frequencies for different probability distributions}

\begin{figure}[!htb]
\begin{minipage}[b]{.5\linewidth}
\centering
\includegraphics[width=.9\linewidth]{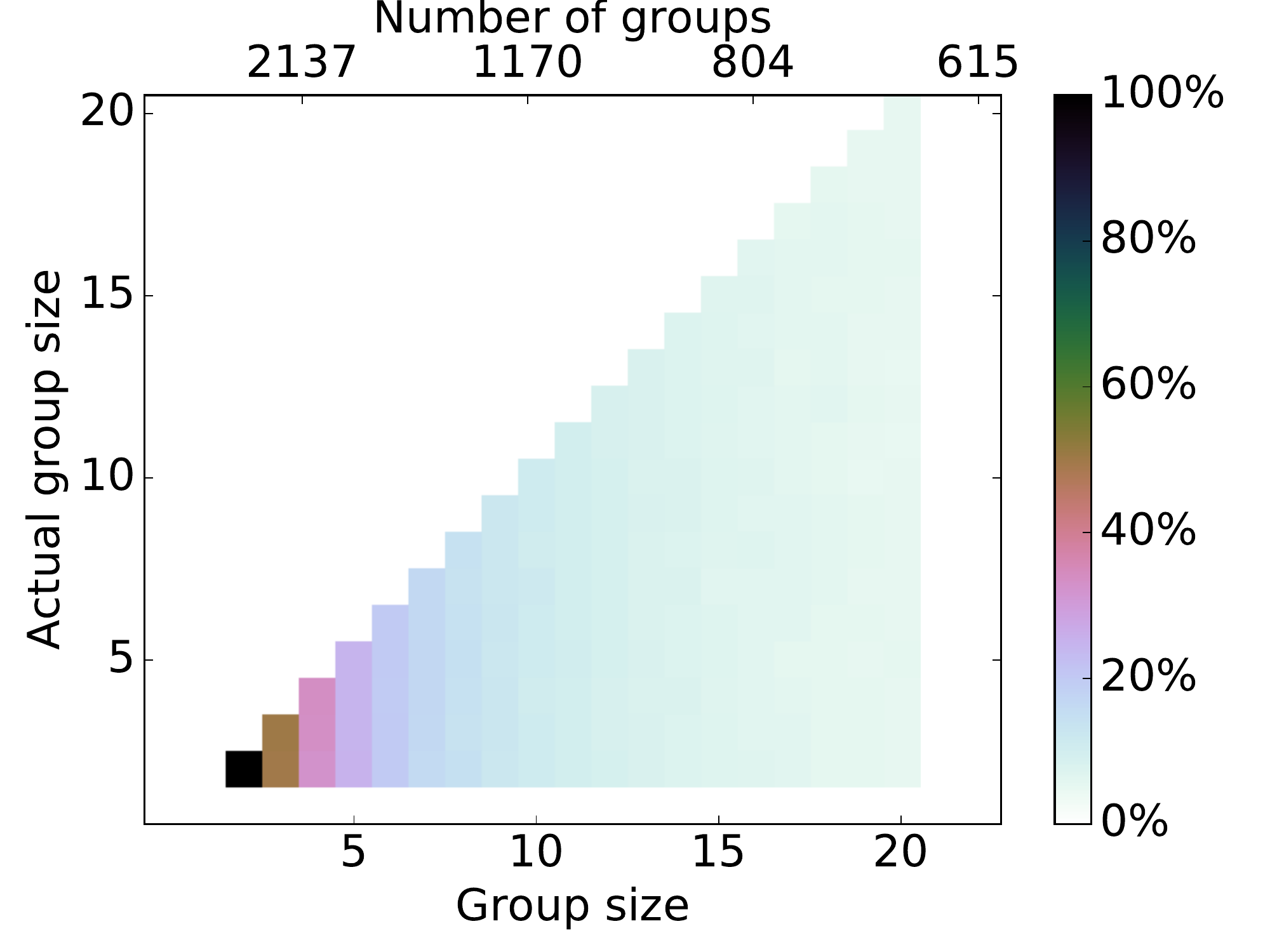}
\subcaption{ECBT dataset}\label{fig:3a}
\end{minipage}%
\begin{minipage}[b]{.5\linewidth}
\centering
\includegraphics[width=.9\linewidth]{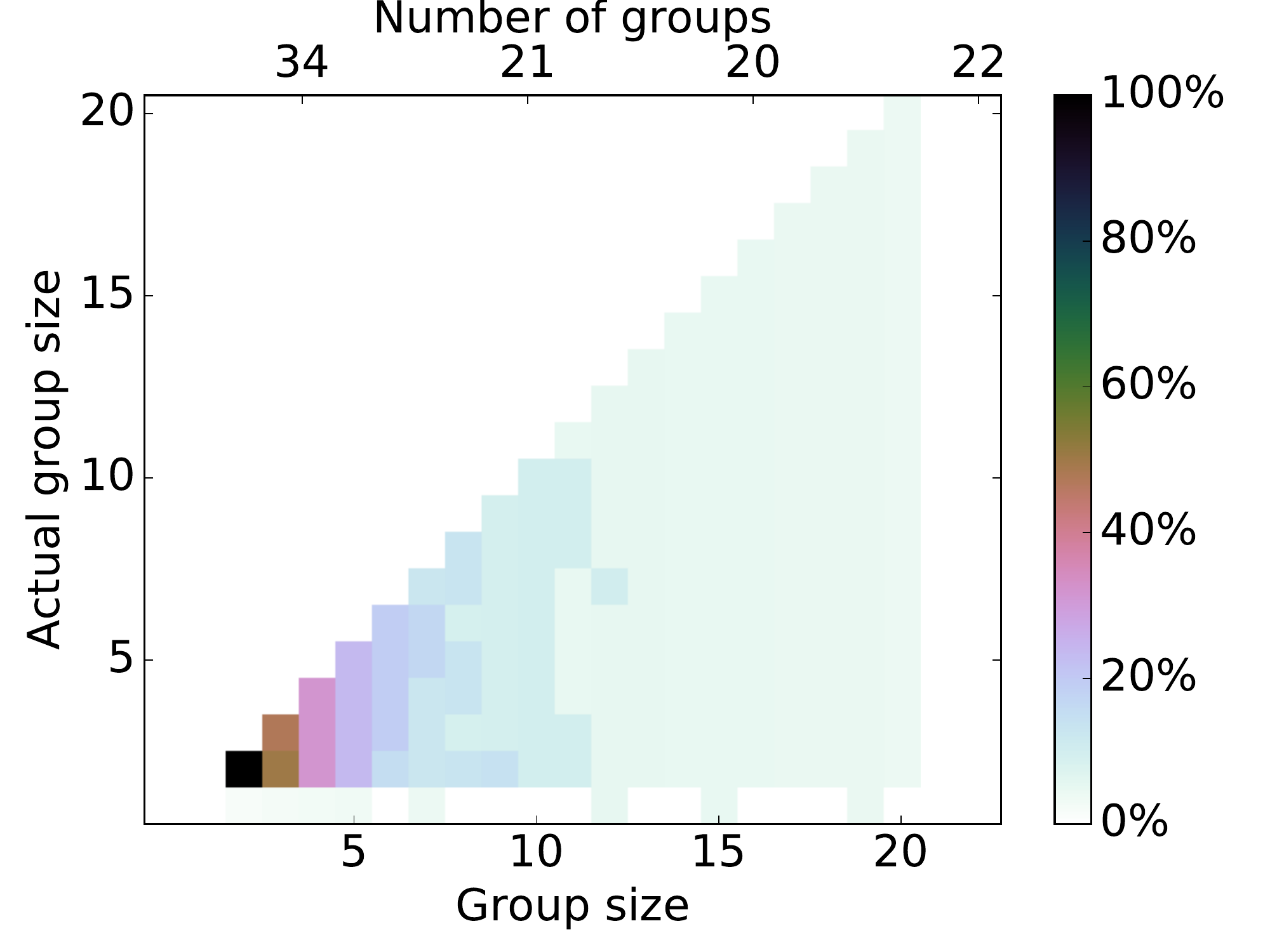}
\subcaption{NREL dataset}\label{fig:3b}
\end{minipage}
\caption{Empirical frequency of group sizes. Group sizes generated randomly by sampling from a uniform distribution.}\label{fig:3}
\end{figure}

\begin{figure}[!htb]
\begin{minipage}[b]{.5\linewidth}
\centering
\includegraphics[width=.9\linewidth]{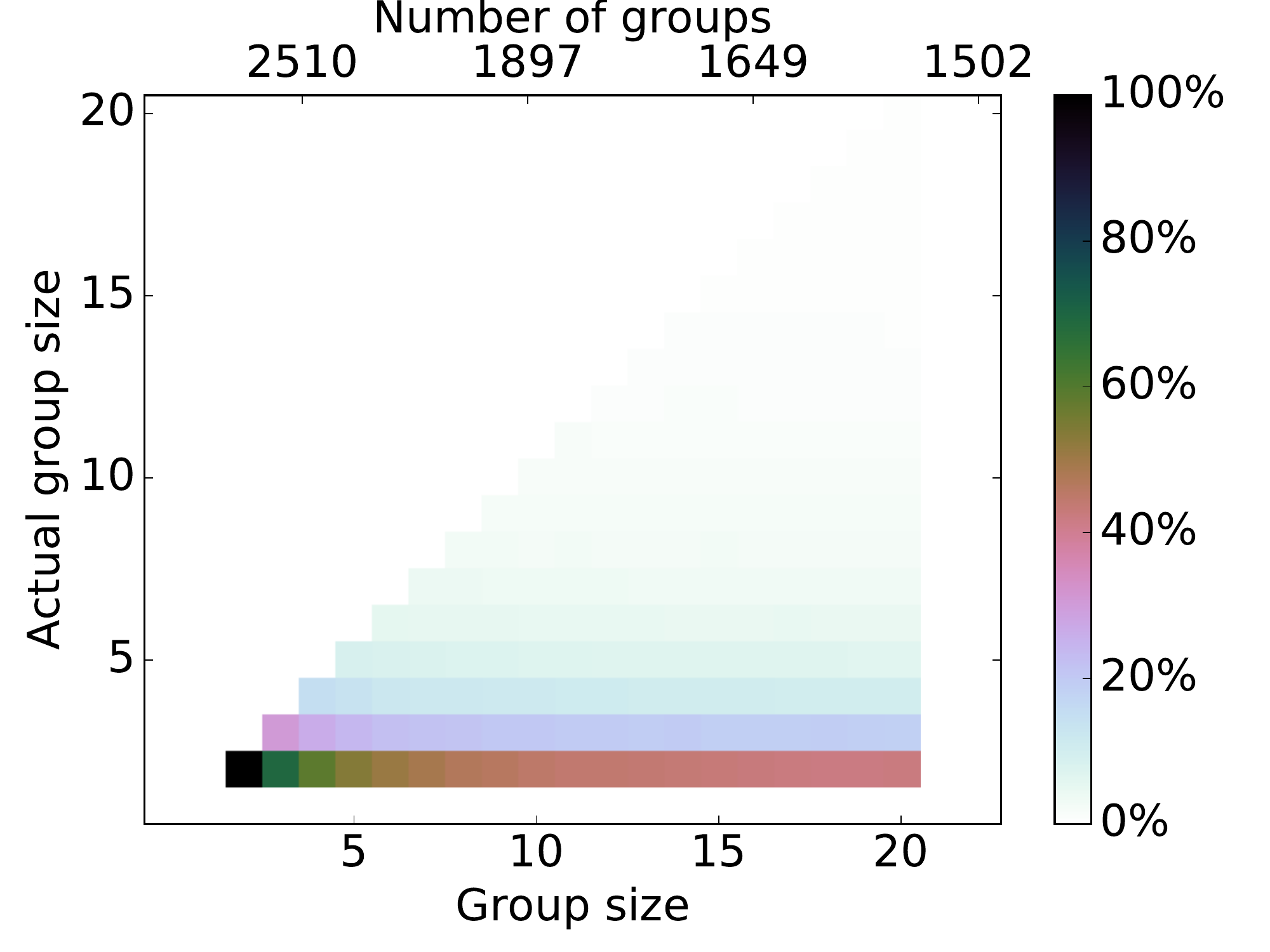}
\subcaption{ECBT dataset}\label{fig:6a}
\end{minipage}%
\begin{minipage}[b]{.5\linewidth}
\centering
\includegraphics[width=.9\linewidth]{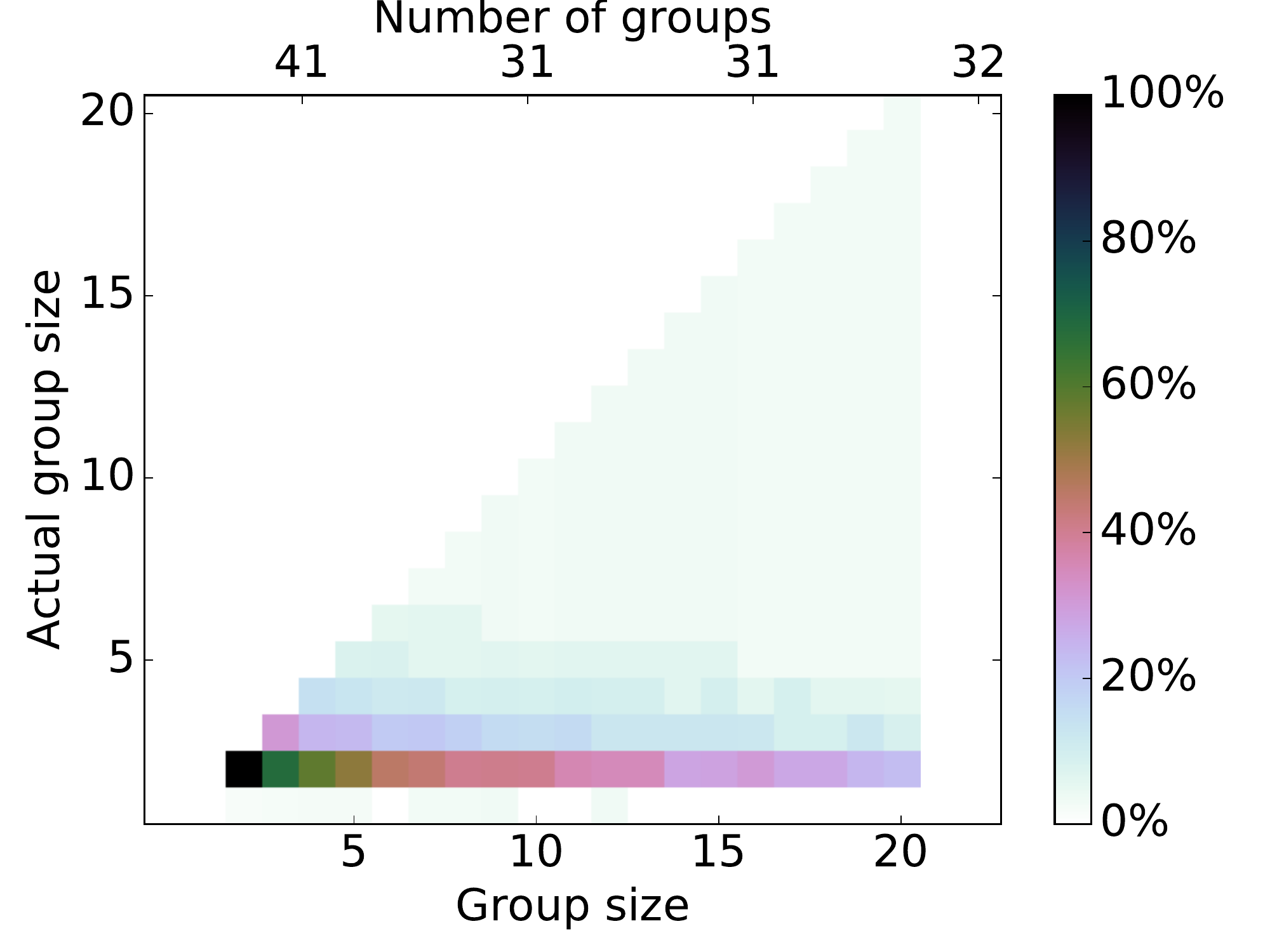}
\subcaption{NREL dataset}\label{fig:6b}
\end{minipage}
\caption{Empirical frequency of group sizes. Group sizes generated randomly by sampling from a power law distribution.}\label{fig:6}
\end{figure}

\begin{figure}[!htb]
\begin{minipage}[b]{.5\linewidth}
\centering
\includegraphics[width=.9\linewidth]{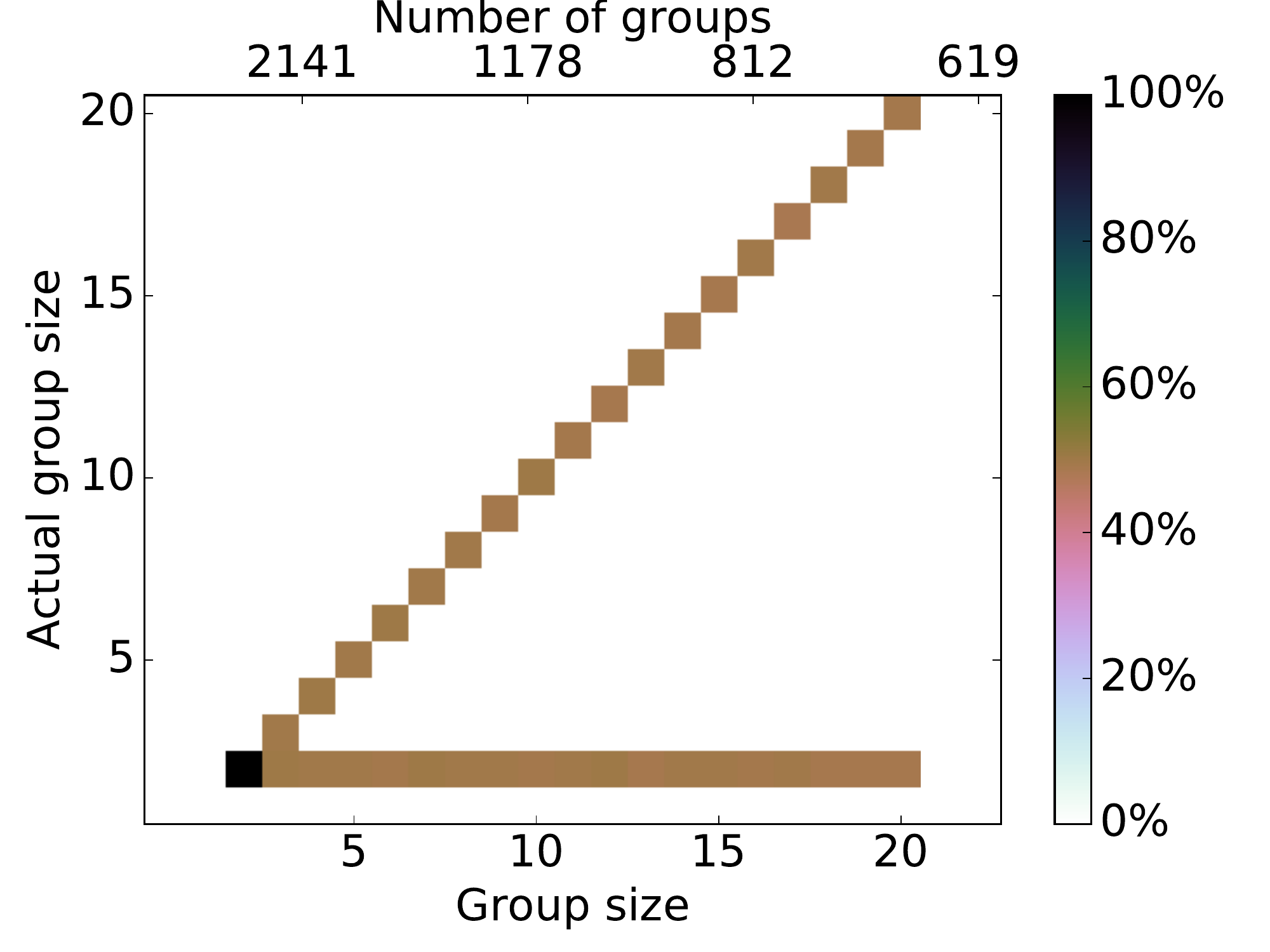}
\subcaption{ECBT dataset}\label{fig:7a}
\end{minipage}%
\begin{minipage}[b]{.5\linewidth}
\centering
\includegraphics[width=.9\linewidth]{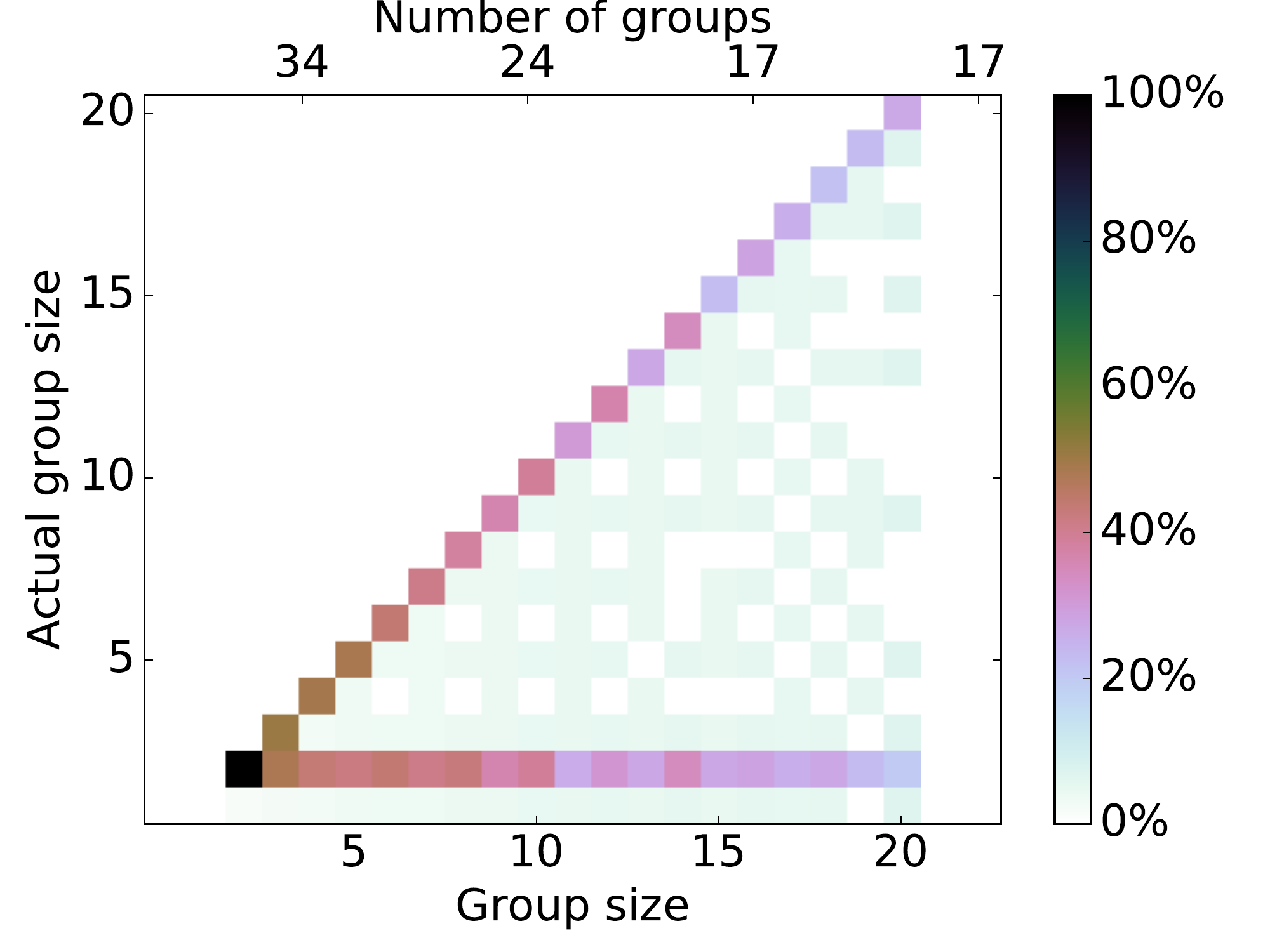}
\subcaption{NREL dataset}\label{fig:7b}
\end{minipage}
\caption{Empirical frequency of group sizes. Group sizes generated randomly by sampling from a step function.}\label{fig:7}
\end{figure}

Figures \ref{fig:3}, \ref{fig:6}, \ref{fig:7} show the empirical frequencies of group sizes for different probability distributions.
The grouping mechanism produces artifacts, representing groups of unexpected size: at first the group sizes are drawn from the probability distribution and then the population is divided accordingly in groups. Depending on the sampling, the sum of the group sizes and the size of the population might not be equal. The remaining data suppliers are grouped together.
This is apparent for a step function (Figure \ref{fig:7b}), where some groups have size between 2 and $N$, represented as squares in the middle of the frequency diagram.

\end{document}